\documentclass[aps,prx,longbibliography,twocolumn,superscriptaddress,floatfix]{revtex4-2}
\usepackage[table,xcdraw]{xcolor}
\usepackage{amsmath,amsthm,amssymb, color}
\usepackage{enumerate} 

\usepackage{color}
\usepackage{multirow}
\usepackage{mathtools}
\usepackage{braket}
\usepackage{tikz}
\usetikzlibrary{positioning}
\usepackage{caption}
\usepackage{subcaption}
\usepackage[compat=0.4]{yquant}
\useyquantlanguage{groups}
\usetikzlibrary{quotes}
\usepackage{diagbox}
 \usepackage{graphicx}
 \usepackage{tablefootnote}
 \usepackage[normalem]{ulem} 
 \usepackage{comment}
 \usepackage{makecell}
 \usepackage{algorithm}
 \usepackage[noend]{algpseudocode}
\usepackage{subcaption}
\usepackage{threeparttable}
\usepackage{booktabs} 
\usepackage{quantikz}
\usepackage[colorlinks=true, allcolors=blue, breaklinks=true]{hyperref}
\usepackage{xurl} 
\usepackage[capitalize]{cleveref}
\Crefname{appendix}{App.}{Apps.}
\Crefname{equation}{Eq.}{Eqs.}
\Crefname{figure}{Fig.}{Figs.}
\Crefname{section}{Sec.}{Secs.}
\tikzset{
  U/.style={
    draw,
    rectangle,
    minimum width=0.35cm,
    minimum height=0.35cm,
    fill=white
  },
  Uf/.style={
    draw,
    rectangle,
    minimum width=0.25cm,
    minimum height=0.25cm,
    fill=black!70
  },
  emptyctl/.style={       
    fill=none,          
    inner sep=0pt,      
    minimum size=0pt    
  }
}

\DeclareFontFamily{U}{MnSymbolC}{}
\DeclareFontShape{U}{MnSymbolC}{m}{n}{
  <-5.5> MnSymbolC5
  <5.5-6.5> MnSymbolC6
  <6.5-7.5> MnSymbolC7
  <7.5-8.5> MnSymbolC8
  <8.5-9.5> MnSymbolC9
  <9.5-11.5> MnSymbolC10
  <11.5-> MnSymbolCb12
}{}

\newcommand{\fc}[2]{{#1}^{\dagger}_{#2}}
\newcommand{\fan}[2]{{#1}_{#2}}
\newcommand{\wstat}{\ket{j}}
\newcommand{\ws}{j}

\newcommand{\oracle}[1]{\mathcal{O}_{}}

\def\qeq{\mathrel{%
    \mathchoice{\QEQ}{\QEQ}{\scriptstyle\QEQ}{\scriptscriptstyle\QEQ}%
}}
\def\QEQ{{%
    \setbox0\hbox{=}%
    \rlap{\hbox to \wd0{\hss?\hss}}\box0
}}

\newcommand{\spar}{\textcolor{black}{\ensuremath{O_C}}}

\newcommand{\sparnew}{\textcolor{black}{\textnormal{\textsc{sparsity }}}}

\newcommand{\ampnew}
{\textcolor{black}{\textnormal{\textsc{amplitude }}}}

\newcommand{\amp}{\textcolor{black}{\ensuremath{O_A }}}

\newcommand{\sel}{\textcolor{black}{\textsc{select }}}
\newcommand{\prep}{\textcolor{black}{\textsc{prepare }}}

\DeclareMathOperator{\polylog}{polylog}
\newtheorem{theorem}{Theorem}[section]

\newtheorem{definition}[theorem]{Definition}

\newcommand\norm[1]{\left\lVert#1\right\rVert}
\renewcommand{\H}{\mathcal{H}}

\newcommand{\occ}[2]{\operatorname{occ}_{#1}(#2)}
\newcommand{\occshort}[1]{\operatorname{occ}_{#1}}
\newcommand{\redrev}[1]{#1} 
\newcommand{\greenrev}[1]{#1} 

\begin{document}
\title{Low-gate-count block encodings for second-quantized fermionic Hamiltonians}
\author{Diyi Liu}
\affiliation{Applied Mathematics and Computational Research Division, Lawrence Berkeley National Laboratory, Berkeley, California 94720, USA}
\affiliation{Department of Mathematics, University of Minnesota, Minnesota 55455, USA}

\author{Shuchen Zhu}
\affiliation{Department of Mathematics, Duke University, Durham, North Carolina 27708, USA}
\affiliation{Duke Quantum Center, Duke University, Durham,  North Carolina 27701, USA}
\affiliation{Department of Computer Science, Georgetown University, Washington, DC 20057, USA}

\author{Lin Lin}
\affiliation{Applied Mathematics and Computational Research Division, Lawrence Berkeley National Laboratory, Berkeley, California 94720, USA}
\affiliation{Department of Mathematics, University of California, Berkeley, California 94720, USA}
\author{Guang Hao Low}
\affiliation{Google Quantum AI, Venice, California 90291, USA}
\author{Chao Yang}
\affiliation{Applied Mathematics and Computational Research Division, Lawrence Berkeley National Laboratory, Berkeley, California 94720, USA}

\date{\today}

\begin{abstract}
Efficient block encoding of many-body Hamiltonians is a central requirement for quantum algorithms in scientific computing, particularly in the early fault-tolerant era. In this work, we introduce new explicit constructions for block encoding second-quantized Hamiltonians that substantially reduce Clifford+$T$ gate complexity and ancilla overhead. Using a SWAP-based data-lookup strategy for the sparsity oracle $O_C$ and a SELECT--SWAP direct-sampling construction for the amplitude oracle $O_A$, we achieve a T count that scales as $\tilde{\mathcal{O}}(\sqrt{L})$ with respect to the number of interaction terms $L$ in general second-quantized Hamiltonians. We also reduce important Clifford-gate overheads in the oracle implementations. Furthermore, we design a block encoding that directly targets the $\eta$-particle subspace, thereby reducing the subnormalization factor from $\mathcal{O}(L)$ to $\mathcal{O}(\sqrt{L})$, and improving fault-tolerant efficiency when simulating systems with fixed particle numbers. Building on the block encoding framework developed for general many-body Hamiltonians, we extend our approach to electronic Hamiltonians whose coefficient tensors exhibit translation invariance or possess decaying structures. Our results provide a practical path toward early fault-tolerant quantum simulation of many-body systems, substantially lowering resource overheads compared to previous methods. 
\end{abstract}
\maketitle

\tableofcontents



\section{Introduction}

Quantum algorithms for simulating many-body systems rely on efficient schemes for encoding non-unitary operators or Hamiltonian data into quantum circuits~\cite{lloyd1996universal, abrams1997simulation, georgescu2014quantum}. These schemes, often called \textit{input models}, specify how classical data---such as sparse matrices~\cite{camps2022explicit,sunderhauf2024block,yang2025dictionary}, kernel matrices~\cite{nguyen2022block}, matrix product operators~\cite{nibbi2024block}, partial differential operators~\cite{hu2024quantum,kharazi2024explicit,guseynov2024efficient}, physical Hamiltonians~\cite{liu2025efficient,du2023multi,du2024hamiltonian,simon2025ladder,lamm2024block,babbush2017exponentially,su2021fault,berry2024quantum,kane2025block}, and pseudo-differential operators~\cite{li2023efficient}---are represented and accessed on quantum hardware~\cite{PRLHarrowQlinear, berry2015hamiltonian}. The choice of input model determines both the resources required for simulation and the class of quantum algorithms that can be applied.

One particularly general input model is \textit{block encoding}~\cite{low2019hamiltonian, gilyen2019quantum}. A block encoding embeds an operator $\mathcal{H}$, scaled by a subnormalization factor $\alpha$, into the top-left block of a larger unitary operator $U$. Equivalently, when the ancilla register is initialized in $\ket{0}$ and postselected on $\ket{0}$ after applying $U$, the induced operation on the system register is proportional to $\mathcal{H}$. Once such an encoding is available, quantum signal processing (QSP) and quantum singular value transformation (QSVT) can be used to implement matrix functions such as time evolution, eigenvalue filtering, and linear-system transformations with near-optimal query complexity~\cite{low2017optimal, gilyen2019quantum}.

A widely used constructive framework for block encoding, especially in quantum chemistry, is the \textit{linear combination of unitaries} (LCU) method~\cite{childs2012hamiltonian,babbush2018encoding,lin2022lecture}. In LCU, the target operator is represented as
\begin{equation}
\mathcal{H}=\sum_{i=0}^{L-1}  \alpha_i \mathcal{H}_{i},
\label{eq:lcueq}
\end{equation}
where each $\mathcal{H}_i$ is either unitary or is itself embedded into a unitary. The corresponding block encoding is implemented using a \prep oracle, which prepares amplitudes proportional to the coefficients, and a \sel oracle, which conditionally applies the terms $\mathcal{H}_i$. The unitary $W=(\prep^{\dagger}\otimes I)\sel(\prep\otimes I)$ then block encodes $\mathcal{H}$ with subnormalization factor $\norm{\alpha}_1=\sum_i |\alpha_i|$.

In this paper, we focus on input models for second-quantized fermionic Hamiltonians of the form
\begin{equation}
  \mathcal{H}=\sum_{p,q} h_{pq} \fc{a}{p}\fan{a}{q} + \sum_{p<q,\  r<s} h_{pqrs} \fc{a}{p}\fc{a}{q} \fan{a}{r} \fan{a}{s},
  \label{eq:2quant}
\end{equation}
where $\fc{a}{p}$ and $\fan{a}{q}$ are fermionic creation and annihilation operators satisfying
\begin{equation}
 \begin{split}
     \{\hat{a}_p, \hat{a}_q^\dagger\} &= \delta_{pq},\\
     \{\hat{a}_p, \hat{a}_q\} = \{\hat{a}_p^\dagger, \hat{a}_q^\dagger\} &= 0.
 \end{split}
\label{eq:comm}
\end{equation}
The coefficients $h_{pq}$ and $h_{pqrs}$ encode the one- and two-body interaction strengths. A standard LCU construction for~\eqref{eq:2quant} first maps the fermionic operators to Pauli strings, for example by the Jordan--Wigner or Bravyi--Kitaev transformation~\cite{seeley2012bravyi,tranter2018comparison}.

Although naive term-by-term LCU block encodings for~\eqref{eq:2quant} are well established, their oracle cost typically scales linearly with the number of Hamiltonian terms $L$~\cite{lin2022lecture}. For a general electronic-structure Hamiltonian, $L=\Theta(n^4)$, where $n$ is the number of spin orbitals. In addition, the \prep oracle must load the classical data $h_{pq}$ and $h_{pqrs}$ into a quantum circuit, for example through QROM and alias-sampling techniques~\cite{babbush2018encoding,berry2019qubitization,vose1991linear}. Circuit-level resource estimates and lower bounds have been developed for such classical-data access models~\cite{clader2022quantum,zhang2024circuit}. These data-loading circuits can have large depth and significant fault-tolerant overhead. Moreover, in a Jordan--Wigner representation, local fermionic operators can map to nonlocal Pauli strings, which introduces additional control overhead and large constants in the gate count~\cite{seeley2012bravyi,tranter2018comparison}.

We develop an alternative block-encoding construction that acts more directly on the occupation-number representation. Instead of first expanding $\fc{a}{p}\fan{a}{q}$ and $\fc{a}{p}\fc{a}{q}\fan{a}{r}\fan{a}{s}$ into Pauli strings, our construction implements their action through controlled bit flips, while encoding the associated fermionic signs using SWAP-UP networks~\cite{low2018trading,wan2021exponentially}, CNOT ladders, and targeted Pauli-$Z$ gates. These ingredients define the column-sparsity oracle $O_C$, which determines the locations of the nonzero matrix elements of $\mathcal{H}$. The coefficient values are encoded by an amplitude oracle $O_A$ based on data lookup and direct sampling. With the SELECT--SWAP architecture~\cite{low2018trading}, the dominant Clifford+T cost scales as $\tilde{\mathcal{O}}(\sqrt{L})$ rather than $\mathcal{O}(L)$ for the corresponding data-loading component.

A second contribution is a block encoding tailored to the fixed-$\eta$ particle sector. Since~\eqref{eq:2quant} preserves particle number, many applications require only the restriction of $\mathcal{H}$ to the subspace with exactly $\eta$ fermions. Standard constructions that diffuse over all formal two-body indices do not exploit this symmetry and therefore incur unnecessarily large subnormalization factors. Our construction uses state-dependent enumeration in the occupation basis to restrict the diffusion to occupied modes. For the general two-body case, this reduces the relevant subnormalization scaling from $\mathcal{O}(n^4)$ in the full space to $\mathcal{O}(n^2\eta^2)$ in the $\eta$-particle sector; equivalently, when $L=\Theta(n^4)$ and $\eta\ll n$, the fixed-particle block encoding gives a substantial reduction in qubitization cost. Indeed, related fixed-particle input models and symmetry-restricted normalization strategies have been studied~\cite{tong2021fast,carolan2024succinct,cortes2024assessing,loaiza2023block,loaiza2023reducing}. However, existing approaches either do not readily extend to arbitrary many-body Hamiltonians or do not provide a rigorous guarantee that the resulting subnormalization factor attains the optimal scaling achieved by our construction.

A complementary line of work changes the persistent state encoding itself, replacing the $n$-qubit occupation-number string by a list of occupied modes~\cite{kirby2021compactencoding} or by a succinct representation of the fixed-weight occupation string~\cite{carolan2024succinct}. In those encodings the occupied modes are addressable by construction, whereas here the same state dependence is obtained by computing an occupied-mode address transiently, leaving the persistent register in the occupation basis. We compare the two routes in \Cref{sec:compact-encoding-comparison}.

We organize the paper as follows. In \Cref{sec:prelim}, we introduce notation and the block-encoding framework used throughout the paper. In \Cref{sec:select}, we construct the sparsity oracle $O_C$ for one- and two-body fermionic interactions. In \Cref{sec:prepare}, we construct the amplitude oracle $O_A$ using SELECT--SWAP data lookup and direct sampling. In \Cref{sec:eta}, we describe the fixed-$\eta$ particle construction and its resource advantages. In \Cref{sec:structuredBE}, we discuss additional reductions for Hamiltonians with translation invariance, locality, or coefficient decay. In \Cref{sec:compare}, we compare the resulting resource estimates with existing input models and related simulation approaches. We conclude in \Cref{sec:conclusion}.

\section{Preliminaries}
\label{sec:prelim}
We use Dirac notation $\bra{\cdot}$ and $\ket{\cdot}$ to represent row and column vectors, respectively. Specifically, $\ket{0}$ and $\ket{1}$ are used to denote the unit vectors $e_{0} = [1 \ 0]^T$ and $e_{1} = [0 \ 1]^T$. The tensor product of $m$ $\ket{0}$'s is written as $\ket{0^m}$. We use $\ket{x,y}$ to represent the Kronecker product of $\ket{x}$ and $\ket{y}$, which can also be written as $\ket{x}\ket{y}$ or $\ket{xy}$. The $N \times N$ identity matrix is denoted by $I_N$, and we may omit the subscript $N$ when the dimension is clear from the context.

We use standard asymptotic notation throughout. For non-negative functions $f$ and $g$ of the problem parameters, $f=\mathcal{O}(g)$ denotes an upper bound up to a constant factor, $f=\Theta(g)$ denotes matching upper and lower bounds up to constant factors, and $f=o(g)$ means that $f/g\to 0$ in the asymptotic regime under consideration. We use $\tilde{\mathcal{O}}$ to suppress factors polylogarithmic in $n$, $\eta$, and $1/\epsilon$.

We map the binary representation of an integer $l \in \mathbb{N}$ where $0 \leq l \leq 2^{n} - 1$, 
\begin{equation*}
    l = l_0 \cdot 2^{0} + l_1 \cdot 2^{1} + \dots + l_{n-1} \cdot 2^{n-1},
\end{equation*}
with $l_i \in \{0, 1\}$ for $i = 0, 1, 2, \dots, n-1$, to a quantum state $\ket{l}$, often written as $\ket{l_{0}, \dots, l_{n-1}}$. The superposition of these states can be encoded using $n$ qubits, where $\ket{l_i}$ corresponds to the $i$-th qubit.

\begin{figure}[htbp]
  \centering
  \begin{quantikz} 
    \lstick[3]{ $\ket{l}$} \ket{l_0} &  \qw & \gate[3]{U} & \qw & \qw & \qw \\
     \ket{l_1} &  \qw & \qw   & \qw & \qw &\qw \\ 
     \ket{l_2} & \qw  & \qw  &\qw & \qw & \qw 
\end{quantikz}
\caption{Illustration of the circuit convention.}
\label{fig:democirc}
\end{figure}
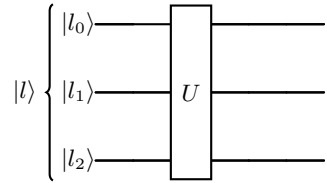

We use $H$, $X$, $Y$, and $Z$ to denote the Hadamard, Pauli-$X$, Pauli-$Y$, and Pauli-$Z$ matrices, respectively. 
We adhere to the standard convention for drawing quantum circuits, which feature multiple parallel lines intersected by several layers of rectangular boxes. Each line represents either a single qubit or multiple qubits, depending on its label, and each box represents a single-qubit or multi-qubit gate, depending on how many qubit lines pass through it. Qubits in a circuit diagram are numbered in increasing order from top to bottom, as shown in the 3-qubit circuit $U$ in \Cref{fig:democirc}. An integer $l = [l_{n-1} \cdots l_1 l_0]$ input to a circuit is prepared as a quantum state $\ket{l_{0} \ldots l_{n-1}}$, with $\ket{l_0}$ mapped to the $0$-th qubit and $\ket{l_{n-1}}$ mapped to the $(n-1)$-th qubit.

\subsection{Block encoding}

Block encoding~\cite{GSLW18} is a technique for embedding a properly scaled non-unitary matrix $A \in \mathbb{C}^{N \times N}$ into a unitary matrix $U_A$ of the form
\begin{equation}
U_A = \begin{bmatrix}
A & \ast  \\
\ast & \ast
\end{bmatrix}.
\end{equation}
Applying $U_A$ to a vector of the form
\begin{equation}
v = \begin{bmatrix}
x \\
0
\end{bmatrix} = \ket{0}\ket{x},
\end{equation}
yields
\begin{equation}
    w = U_A v = 
    \begin{bmatrix}
    Ax \\
    *
    \end{bmatrix} = \ket{0}(A\ket{x}) + \ket{1}\ket{\ast}.
\end{equation}

If a measurement of the first qubit yields $\ket{0}$, the post-measurement state of the second register is $A\ket{x}$, and this occurs with probability $\|A\ket{x}\|^2$.
A formal definition of block encoding is as follows.

\begin{definition}[Block encoding] 
Given an $n$-qubit matrix $A \in \mathbb{C}^{2^n\times 2^n}$, if there exist $\alpha, \varepsilon \in \mathbb{R}_+$, and an $(m+n)$-qubit unitary matrix $U_{A}$ so that 
\begin{equation}
  || A-\alpha(\bra{0^m} \otimes I_{2^n}) U_{A} (\ket{0^m} \otimes I_{2^n}) ||\leq \varepsilon ,
  \label{eq:bedef}
\end{equation}
then $U_A$ is called an $(\alpha,m,\varepsilon)$-block-encoding of $A$. When the block encoding is exact with $\varepsilon=0$, $U_A$ is called an $(\alpha, m)$-block-encoding of $A$. 
\end{definition}

Assuming an efficient circuit implementation of $U_A$, measuring the ancilla register prepares the state $\ket{Ax}$ in the system register.  We should note that the unitary $W$ constructed in the LCU-based block encoding of $\mathcal{H}$ in \eqref{eq:2quant} is a $(\norm{\alpha}_1,\log L,0)$-block-encoding.

The following theorem outlines a general method for constructing a block encoding of an $s$-sparse matrix, which is defined as a matrix with no more than $s$ nonzero elements in each column (see~\cite{camps2022explicit,gilyen2019quantum,liu2025efficient}). 

\begin{theorem} \label{thm:Usparse}
Let $c(j,\ell)$ be a function that gives the row index of the $\ell$-th (among a list of $s$) non-zero matrix elements in the $j$-th column of an $s$-sparse matrix $A \in \mathbb{C}^{N\times N}$ with $N=2^n$, where $s=2^m$. We denote the $(i,j)$ entry of $A$ by $A_{i,j}$. If there exists a unitary $O_C$ (\sparnew oracle) such that 
\begin{equation}
O_C \ket{\ell}\ket{j} = \ket{\ell}\ket{c(j,\ell)},
\label{eq:oc}
\end{equation}
and a unitary $O_A$ (\ampnew oracle) such that
\begin{equation}
O_A \ket{0}\ket{\ell}\ket{j} = \left(
A_{c(j,\ell),j} \ket{0} + \sqrt{1-|A_{c(j,\ell),j}|^2} \ket{1}
\right)
\ket{\ell}\ket{j},
\label{eq:oa}
\end{equation}
then 
\begin{equation}
U_A = \left(I_2 \otimes D_s\otimes I_N \right) 
\left(I_2 \otimes O_C \right) 
O_A
\left(I_2 \otimes D_s \otimes I_N\right),
\label{eq:uafact}
\end{equation}
block encodes $A/s$.
Here $D_s$ is called a diffusion operator and is defined as 
\begin{equation}
D_s \equiv \underbrace{H\otimes H \otimes \cdots \otimes H}_{m},
\end{equation}
\end{theorem}

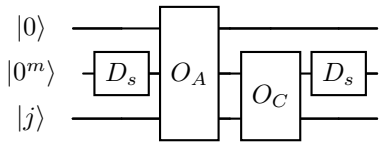
\begin{figure}
  \begin{center}
    \begin{quantikz}
    &\ket{0} \quad & \qw            & \gate[3]{O_{A}} &\qw  &\qw &\qw &\qw \\ 
    &\ket{0^m} \quad & \gate{D_s}     & \qw         &\qw &\gate[2]{O_C} & \gate{D_{s}} &\qw\\
    &\ket j \quad & \qw            & \qw         &\qw & & \qw &\qw
  \end{quantikz}
  \end{center}
  \caption{A schematic illustration of a quantum circuit for the block encoding of an $s$-sparse matrix.}
\label{fig:complete_circ}
\end{figure}

The basic structure of the block encoding circuit is shown in \Cref{fig:complete_circ}.
Note that the input to the circuit consists of three sets of qubits. The top qubit is used to encode nonzero matrix elements.  The second set of qubits that takes $\ket{0^m}$ as the input is used to encode the position of the nonzero elements. The bottom set of qubits that take $\ket{j}$ as the input are the qubits that encode the column indices of the matrix to be block encoded.

The circuit consists of several blocks. The $O_C$ block is used to encode the positions of the nonzero elements, which is defined by the mapping $c(j,\ell)$ in  \Cref{thm:Usparse}. The $O_A$ block is used to encode the nonzero matrix elements. 

The circuit structure of $O_C$ depends on the mapping $c(j,\ell)$, which describes the sparsity structure of $A$. Ref.~\cite{camps2022explicit} showed how $c(j,\ell)$ is defined for tridiagonal Toeplitz matrices and sparse matrices whose sparsity structure can be described by a binary tree, as well as how the $O_C$ circuit can be constructed efficiently for those types of matrices. 

Once $O_C$ is specified, we can construct $O_A$, which encodes the numerical value of each non-zero matrix element.

\subsection{Quantum signal processing and quantum singular value transformation}
In this section, we introduce quantum signal processing and quantum singular value transformation to show how block encoding can be used for solving linear algebra problems and scientific computing. 
To solve the linear algebra problems involving $A$ or compute physical observables, one needs to block encode a matrix function of $A$. 
Quantum signal processing~\cite{LC17} (QSP) for scalar polynomials and its extension to matrix polynomials through the quantum singular value transformation~\cite{GSLW18} (QSVT) enable access to the block encoding of a matrix polynomial of $A$ based on the block encoding of $A$.

\begin{theorem}[Quantum signal processing] \label{thm:qsp}
Let 
\begin{equation}
 U(t) = 
 \begin{bmatrix}
 t & \sqrt{1-t^2} \\
 \sqrt{1-t^2} & -t
 \end{bmatrix},
\label{eq:Ut}
\end{equation}
where $|t|\leq 1$.
There exists a set of phase angles $\Phi_d \equiv \{\phi_0,...,\phi_d\} \in \mathbb{R}^{d+1}$ so that 
\begin{align}
U_{\Phi_d}(t) &\equiv (-i)^{d} e^{i\phi_0 Z} \Pi_{j=1}^d \left[U(t)e^{i\phi_j Z} \right] \nonumber\\
 &=
 \begin{bmatrix}
 p(t) & -q(t)\sqrt{1-t^2}\\
 q^*(t)\sqrt{1-t^2} & p^*(t)
 \end{bmatrix},
 \label{eq:Uphi}
\end{align}
if and only if $p(t)$ and $q(t)$ are complex valued polynomials in $t$ and satisfy
\begin{enumerate}
\item $\mathrm{deg}(p) \leq d$, $\mathrm{deg}(q) \leq d-1$;
\item $p$ has parity $d$ mod $2$ , and $q$ has parity  $d-1$ mod $2$; 
\item $|p(t)|^2 + (1-t^2)|q(t)|^2 = 1$, $\forall t\in [-1,1]$.
\end{enumerate}
When $d=0$, $\deg(q)\le-1$ should be interpreted as $q=0$.
\end{theorem}

Note that $U_{\Phi_d}(t)$ is a block encoding of the function $p$ of a scalar $t$ in \Cref{thm:qsp}. This theorem can be extended to a properly scaled matrix, which is the theory behind QSVT. 
\begin{figure*}[htbp]
\centering
  \begin{quantikz} 
    \ket{0} &\gate{H}&\gate[2]{CR_{\phi_{d}}}& \qw & \gate[2]{CR_{\phi_{d-1}}}& \qw &\qw \dots & \qw &\gate[2]{CR_{\phi_{0}}} & \gate{H} &\qw \\ 
     \ket{0^{m}} & \qw &\qw &\gate[2]{U_{A}} & \qw & \gate[2]{U_{A}^{\dagger}} & \qw \dots & \gate[2]{U_{A}} & \qw & \qw &\qw \\
    \ket{\psi}  &  \qw & \qw & \qw &\qw & \qw &\dots & \qw &\qw &\qw \\
\end{quantikz}
\caption{Illustration: the schematic circuit design of quantum singular value transformation (for an odd $d$; for an even $d$ the last $U_A$ is replaced by $U_A^{\dag}$). The additional Hadamard gate selects only the real part of the polynomial $p$. 
}
\label{fig:QSP}
\end{figure*}
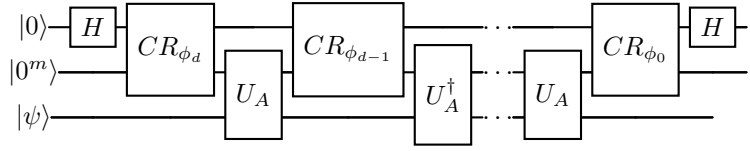

It follows from QSVT that the block encoding of $p(A)$ for some polynomial $p$ can be expressed in terms of the block encoding of $A$, as illustrated in \Cref{fig:QSP}. 
In the circuit representation, $\phi_i$ is the same phase angle appearing in \Cref{thm:qsp}. Each controlled-rotation gate $CR_{\phi_{i}}$ can be implemented as a single-qubit gate placed on the top ancilla qubit that performs $e^{-i\phi_{i}Z}$, preceded and followed by a multi-qubit CNOT gate. 

While \Cref{thm:qsp} asserts the existence of these phase angles for quantum signal processing, they can be effectively computed through solving a numerical optimization problem as detailed in Refs.~\cite{dong2022infinite,Dongefficient}.

\section{Sparsity oracle \texorpdfstring{$O_C$}{OC}}
\label{sec:select}
We start this section with a brief description of the occupation (qubit) and operator representations of a quantum many-body state and how the representation changes after $a_i^\dagger a_j$ and $a_p^\dagger a_q^\dagger a_r a_s$ are applied to such a state. In particular, we show that the new state can be represented by flipping certain qubits of the input state and introducing a phase factor determined by the indices $i$, $j$, or $p$, $q$, $r$, and $s$.  
We then show how $a_i^\dagger a_j$ and $a_p^\dagger a_q^\dagger a_r a_s$ can be implemented as controlled flip operations and how the phase factor can be encoded efficiently in a column sparsity oracle $O_C$.


\subsection{Representation of many-body state and many-body Hamiltonian}

We introduce the formalism we use to represent a quantum many-body basis state and the matrix representation of a second-quantized many-body Hamiltonian in a set of such basis states. 

Because the Hamiltonian we consider in \Cref{eq:2quant} is particle number conserved, i.e., $\mathcal{H}$ commutes with number operator $\mathcal{N}=\sum_{i} \fc{a}{i}\fan{a}{i}$,
we can consider the invariant subspace of $\mathcal{H}$ associated with $\eta$-particle states $\ket{j}$ that satisfies 
$\mathcal{N}\ket{j}=\eta \ket{j}$ for some $\eta \in \mathbb{N}$. The occupation representation of such a state $\ket{j} \equiv \ket{j_0 j_1 ...j_{n-1}}$, where $n$ is the total number of single-particle states from which the $\eta$ particles are chosen and $j_0 j_1 ... j_{n-1}$, with $j_i \in \{0,1\}$, is the binary representation of $j$, satisfies 
\begin{equation}
\left\{
\begin{array}{cl}
j_i = 1 & \mbox{for integers $i = f_0, f_1, \dots, f_{\eta-1}$}, \\
j_i = 0 & \mbox{for other integers $i \in [0,n-1]$},
\end{array}
\right.
\end{equation}
where $0 \leq f_0 < f_1 < \cdots < f_{\eta-1} \leq n-1$ are a set of integers between 0 and $n-1$. Such an $\eta$-particle state can alternatively be represented as 
 \begin{equation}
      \fc{a}{f_0} \fc{a}{f_{1}} \dots \fc{a}{f_{\eta-1}}\ket{vac},
  \label{eq:binaryRepresentationOfFockState}
\end{equation}
where $\ket{vac}$ represents the empty or vacuum state. Note that \Cref{eq:binaryRepresentationOfFockState} is only valid when $f_0$, $f_1$, \ldots, $f_{\eta-1}$ form an increasing integer sequence. When this is not true, we can use the anti-commutation identity \Cref{eq:comm} to repeatedly interchange the order of two adjacent out-of-order creation operators indexed by different $f_k$'s until the indices of the reordered creation operators become an increasing sequence.  Every time we invoke the anti-commutation identity, we introduce a negative sign. The final sign of the created many-body basis state depends on the number of reorderings performed, i.e., \Cref{eq:binaryRepresentationOfFockState} may produce $-\ket{j}$ for some many-body basis state $\ket{j}$. For example, $\fc{a}{2} \fc{a}{1} \ket{vac}$ is not a valid many-body basis state. However, it is a many-body basis state multiplied by $-1$ because
 \begin{equation}
 	\fc{a}{2} \fc{a}{1} \ket{vac}=-\fc{a}{1} \fc{a}{2}   \ket{vac}:= - \ket{011}.
  \label{eq:a2a1vac}
 \end{equation}
 
The occupation representation of the $\ket{j}$ basis state can be mapped directly to a $n$-qubit register with $\ket{j_p}$ mapped to the $p$-th qubit.  The representation defined by \Cref{eq:binaryRepresentationOfFockState} is convenient when we examine the state resulting from the application of a set of creation and annihilation operators to $\ket{j}$. 

%
The matrix representation of $\mathcal{H}$ can be obtained by evaluating   $\H_{i,j} \equiv \bra{i} \mathcal{H}\ket{j}$ for all $i,j$ pairs, where $i,j \in \{0,1,\ldots,2^n-1\}$. For the two-body $\mathcal{H}$ defined in \Cref{eq:2quant}, many of these matrix elements are zero. In particular, $\bra{i}\mathcal{H}\ket{j}=0$, if $\ket{i}$ and $\ket{j}$ differ by more than two single particle states.

The nonzero structure of the matrix
representation of $\mathcal{H}$ can be
ascertained by examining the basis states that appear in $\mathcal{H}\ket{j}$ for each $j$. Because the two-body Hamiltonian $\mathcal{H}$ defined in \Cref{eq:2quant} is a linear combination of one-body operator 
$\fc{a}{p}\fan{a}{q}$ and two-body operators $\fc{a}{p}\fc{a}{q}\fan{a}{r}\fan{a}{s}$, for $p,q,r,s\in \{0,1,\ldots,n-1\}$, it is sufficient to examine basis states produced when each one of these operators is applied to $\ket{j}$ to deduce the position $i$ of the nonzero matrix in column $j$ of the matrix $\H$. The value of the nonzero matrix element depends on the coefficients $h_{pq}$ and $h_{pqrs}$ in \Cref{eq:2quant}, which we will discuss later.    %
%
%

For simplicity, let us first look at basis states produced from applying $\fc{a}{p}\fan{a}{q}$ to $\ket{j}$. We note that
\begin{itemize}
\item $\fc{a}{p}\fan{a}{q}$ may not contribute to a nonzero matrix element in the $j$-th column of $\H$. This is because $\fc{a}{p}\fan{a}{q} \ket{j} = 0 \ket{vac}=0$ if  $\ket{j}_p = 1$ or $\ket{j}_q = 0$, where $\ket{j}_p$ denotes the occupation number of the $p$-th single-particle state in $\ket{j}$. %

\item $\fc{a}{p}\fan{a}{q}$ contributes to a nonzero element in the $j$-th column of $\H$ if $\ket{j}_p = 0$ and $\ket{j}_q  = 1$.

 \item When $\ket{j}_p = 0$ and $\ket{j}_q  = 1$, 
 \begin{equation}
 \fc{a}{p}\fan{a}{q} \ket{j} = (-1)^{d_{j;p,q}} \ket{\mbox{FLIP}(j;p,q)},
 \label{eq:signedswap}
 \end{equation}
 where \mbox{FLIP($j;p,q$)} denotes a binary string obtained from flipping the $p$-th and $q$-th bits of the binary representation of $j$, and $d_{j;p,q}$ is an integer to be described below.
\end{itemize}

We will refer to the factor $(-1)^{d_{j;p,q}}$ as the phase factor below. To see where this phase factor comes from, let us examine a simple example in which a one-body operator $\fc{a}{2}\fan{a}{0}$ is applied to the state $\ket{j} = \ket{110}$.
Using the alternative representation $\ket{j} = \ket{110}=\fc{a}{0}\fc{a}{1}\ket{vac}$, we obtain
 \begin{equation}
   \fc{a}{2}\fan{a}{0} \ket{110}=\fc{a}{2}\fan{a}{0}\fc{a}{0}\fc{a}{1}\ket{vac}=\fc{a}{2} \fc{a}{1} \ket{vac}=-\ket{011}.
   \label{eq:onebodyexample}
 \end{equation}
 However, the right-hand side of \Cref{eq:onebodyexample} is not a valid basis state because the subscripts of the creation operators are not in increasing order. It follows from \Cref{eq:a2a1vac} that we can rearrange the creation operators by using the anti-commutation identity, which introduces a negative sign.  As a result, we obtain
\begin{equation}
    \fc{a}{2}\fan{a}{0}\ket{110} = - \ket{011}.
\end{equation}

For a general one-body interaction term, the phase factor resulting from applying $\fc{a}{p}\fan{a}{q}$ to $\ket{j}\equiv\ket{j_0j_1\dots j_{n-1}}$, with assumption that $p<q$, $\ket{j}_p=\ket{0}$ and $\ket{j}_q=\ket{1}$, can be derived as follows,
\begin{equation}
    \begin{split}
      &\fc{a}{p}\fan{a}{q} \fc{a}{f_0} \fc{a}{f_{1}}  \dots  \fc{a}{q} \dots \fc{a}{f_{\eta-1}}\ket{vac}\\
    =&(-1)^{j_0+j_1+\cdots+j_{q-1}}\fc{a}{p} \fc{a}{f_0} \fc{a}{f_{1}}  \dots \fan{a}{q} \fc{a}{q} \dots \fc{a}{f_{\eta-1}}\ket{vac}\\
    =& (-1)^{j_0+j_1+\cdots+j_{q-1}}\fc{a}{p} \fc{a}{f_0} \fc{a}{f_{1}}  \dots [1-\fc{a}{q} \fan{a}{q} ] \dots \fc{a}{f_{\eta-1}}\ket{vac}\\
    =& (-1)^{j_0+j_1+\cdots+j_{q-1}}\fc{a}{p} \fc{a}{f_0} \fc{a}{f_{1}}  \dots \times 1 \times \dots \fc{a}{f_{\eta-1}}\ket{vac}\\
    =& (-1)^{j_{p+1}+j_{p+2}+\cdots+j_{q-1}} \fc{a}{f_0} \fc{a}{f_{1}}  \dots \fc{a}{p} \dots \fc{a}{f_{\eta-1}}\ket{vac}.
    \end{split}
\end{equation}
In the first step of the derivation, we move $\fan{a}{q}$ to the right using the anti-commutation identity \Cref{eq:comm} until it is just to the left of $\fc{a}{q}$. Each time the anti-commutation identity is invoked, we pick up a negative sign. The total number of negative signs we pick up is the number of creation operators to the left of $\fc{a}{q}$. This number is the same as the number of 1's in the binary representation of $j$, i.e., $j_0 j_1\ldots j_{q-1}j_q\ldots j_{n-1}$, to the left of the $q$-th bit, which is identical to $j_0+j_1+\cdots +j_{q-1}$.
We then replace $\fan{a}{q}\fc{a}{q}$ with $1-\fc{a}{q}\fan{a}{q}$ using the anti-commutation identity in \Cref{eq:comm}. This effectively splits $\fc{a}{p}\fan{a}{q}\ket{j}$ into two terms. The second term vanishes because applying $\fan{a}{q}$ to a basis state in which the $q$-th single-particle state is unoccupied results in 0.  In the next step, we again use the anti-commutation identity in \Cref{eq:comm} to move $\fc{a}{p}$ to the right place in the ordered integer sequence $f_0, f_1,\ldots,f_{\eta-1}$. Such movements produce a multiplicative factor $(-1)^{j_0+j_1+\cdots +j_{p-1}}$. Combining it with the multiplicative factor $(-1)^{j_0+j_1+\cdots +j_{q-1}}$ produced from the first step of the derivation, we obtain the final phase factor 
\begin{equation}
\begin{split}
(-1)^{d_{j;p,q}}&=(-1)^{j_0+\cdots+j_{p-1}} \times (-1)^{j_0+\cdots+j_{p-1}+j_{p}+\cdots + j_{q-1}} \\
&=(-1)^{j_{p+1} +j_{p+2} + \cdots + j_{q-1}}.
\end{split}
\label{eq:phasepq1}
\end{equation}
The derivation of such a phase factor only relies on the fact that $p\neq q$. The expression for the phase factor $d_{j;p,q}$ is similar for $p>q$,
\begin{equation}
\begin{split}
(-1)^{d_{j;p,q}}&=(-1)^{j_0+\cdots+j_{q-1}} \times (-1)^{j_0+\cdots+j_{q-1}+0+j_{q+1}+\cdots + j_{p-1}} \\
&=(-1)^{j_{q+1} +j_{q+2} + \cdots + j_{p-1}}.
\end{split}
\label{eq:phasepq2}
\end{equation}
The derivation is similar except for the phase factor generated by $\fc{a}{p}$. Instead of $(-1)^{j_0+\dots +j_{p-1}}$, the $j_q$ in the phase factor should be replaced by zero as $\fan{a}{q}$ has been applied to $\ket{j}$. In the special case when $p=q$ or $|p-q|=1$, it can be seen that the phase factor is $1$ or equivalently $d_{j;p,q}=0$. 

For a general two-body interaction term, the phase factor resulting from applying $\fc{a}{p}\fc{a}{q}\fan{a}{r}\fan{a}{s}$ to $\ket{j}\equiv\ket{j_0j_1\dots j_{n-1}}$ can be derived similarly. Clearly, not all combinations of $\fc{a}{p}$,$\fc{a}{q}$,$\fan{a}{r}$,$\fan{a}{s}$ contribute to a nonzero matrix element in $\H$. For example, when $p=q$ or $r=s$, $\fc{a}{p}\fc{a}{q}\fan{a}{r}\fan{a}{s}\ket{j}=0$ for all $j$ because we cannot create or annihilate the same state twice. 

In general, whether a two-body operator contributes a non-zero matrix element to the $j$-th column of $\H$ depends on the value of $p$, $q$, $r$ and $s$ as well as the state of the $i$-th qubit in $\ket{j}$ for $i=p,q,r,s$.

When 1) $p=r$ and $q\neq s$, 2) $p=s$ and $q\neq r$, 3) $q=r$ and $p\neq s$ and 4) $q=s$ and $p\neq r$, the corresponding two-body interaction term becomes  $\fc{a}{p}\fc{a}{q}\fan{a}{p}\fan{a}{s}$, $\fc{a}{p}\fc{a}{q}\fan{a}{r}\fan{a}{p}$, $\fc{a}{p}\fc{a}{q}\fan{a}{q}\fan{a}{s}$ or $\fc{a}{p}\fc{a}{q}\fan{a}{r}\fan{a}{q}$ respectively. Using the anticommutation identity, we can rewrite these terms as
$-n_p \fc{a}{q}\fan{a}{s}$, $n_p \fc{a}{q}\fan{a}{r}$, $n_q\fc{a}{p}\fan{a}{s}$ and $-\fc{a}{p}\fan{a}{r}n_q$ %
respectively. Because $n_q \ket{j}$ is either 0 or $\ket{j}$, these special instances of the two-body term effectively reduce to one-body terms.
In the case $p=r$ and $q=s$, the two-body term becomes $\fc{a}{p}\fc{a}{q}\fan{a}{p}\fan{a}{q}=-n_p n_q$, i.e., the signed product of two number operators; as in the four cases above, the sign can be absorbed into the corresponding coefficient.

Note that there cannot be three or more identical indices in the creation and annihilation operators contained in a two-body term because, in this case, two creation operators or two annihilation operators will be identical. As we discussed earlier, this type of two-body term yields 0 when it is applied to $\ket{j}$ for any $j$.

Therefore, it suffices to examine the phase factors produced by the following four types of operators:  $n_{p} \fc{a}{q} \fan{a}{s}$ with $q\neq s$, $ \fc{a}{p} \fan{a}{r}n_{q}$ with $p\neq r$, $n_pn_q$ and $\fc{a}{p}\fc{a}{q}\fan{a}{r}\fan{a}{s}$ with distinct $p$, $q$, $r$, $s$. The first two cases reduce to the one-body case discussed earlier in which the phase factor is defined by \Cref{eq:phasepq1} or \Cref{eq:phasepq2}. The third case does not introduce a phase factor at all.

For the last case in which $p,q,r,s$ are all different, we assume, without loss of generality, that $p<q$ and $r>s$. If the indices are given in a different order, we can bring them to this canonical ordering using the anticommutation identity in \Cref{eq:comm} and absorb the resulting sign into the coefficient $h_{pqrs}$, which reconciles the assumption $r>s$ with the convention $r<s$ used in \Cref{eq:2quant}.
In this case, applying the two-body operator to $\ket{j}$ yields a different basis state 
 if $\ket{j}_p=\ket{0}$, $\ket{j}_q=\ket{0}$, $\ket{j}_r=\ket{1}$ and $\ket{j}_s=\ket{1}$.
 The phase factor resulting from this  operation can be derived in two steps. In the first step, we examine the phase factor resulting from applying $\fan{a}{r}\fan{a}{s}$ to $\ket{j}$. It can be easily verified that such a phase factor is in fact
\begin{equation}
	(-1)^{d_{j;r,s}}=(-1)^{j_{s+1}+\cdots +j_{r-1}}.
\end{equation} 
The product of this phase factor and the state %
\begin{equation}
	\ket{j'} = \ket{\mbox{FLIP}(j;r,s)}
    \label{eq:jprime}
\end{equation}
is the outcome of $\fan{a}{r}\fan{a}{s}\ket{j}$.

In the second step, we derive the phase factor associated with applying $\fc{a}{p}\fc{a}{q}$ to $\ket{j'}$ defined by \Cref{eq:jprime}. It can be easily shown that such a factor is 
\begin{equation}
 (-1)^{d_{j';p,q}}=(-1)^{j'_{p+1}+\dots +j'_{q-1}}.
\end{equation}

As a result, the total phase factor associated with applying the two-body operator $\fc{a}{p}\fc{a}{q}\fan{a}{r}\fan{a}{s}$ to $\ket{j}$ is 
\begin{equation}
	 (-1)^{d_{j';p,q}+d_{j;r,s}}.
\end{equation}
The resulting state is 
 \begin{equation}
 \begin{split}
     &\fc{a}{p}\fc{a}{q}\fan{a}{r}\fan{a}{s} \ket{j} \\=& (-1)^{d_{j';p,q}+d_{j;r,s}} \ket{\mbox{FLIP}(\mbox{FLIP}(j;r,s);p,q)}.
 \end{split}
 \label{eq:2bodyj}
 \end{equation}

\subsection{Oracle construction}
\label{sec:oc}
In this subsection, we detail the construction of an efficient circuit for the \spar \  oracle (column \sparnew oracle) for one-body and two-body interactions. 

\subsubsection{One-body interaction}
To begin the discussion, we consider a one-body Hamiltonian given by:
\begin{equation}
\label{equ:onebodypq}
    \mathcal{H}= \sum_{p \neq q} \fc{a}{p}\fan{a}{q}.
\end{equation}
The \spar \ oracle can be constructed as the product of two oracles: the $O_{s}$ and $O_{\phi}$.  The $O_s$ oracle is defined as:
\begin{align}
  &O_{s} \ket{1}  \ket{p}\ket{q} \ket{\ws} \nonumber \\
  &= \begin{cases}
    \ket{0}  \ket{p}\ket{q} \ket{\mbox{FLIP}(j;p,q)} &\text{if $\ket{j}_p = 0$ and $\ket{j}_q  = 1$},\\
    \ket{1} \ket{p} \ket{q} \ket{j} &\text{otherwise}.
  \end{cases}
\end{align}
The $O_{\phi}$ oracle can be defined as 
\begin{equation}
    O_{\phi} \ket{1}\ket{p}\ket{q} \ket{ j} = (-1)^{d_{j;p,q}}\ket{1}\ket{p}\ket{q} \ket{j}.
\end{equation}
Applying $O_C$ to $\ket{1}\ket{p}\ket{q}\ket{j}$ yields
\begin{equation}
	\begin{split}
		 &O_C \ket{1}\ket{p} \ket{q} \ket{j} \\ =& O_s O_{\phi} \ket{1}\ket{p} \ket{q} \ket{j} \\
		=& (-1)^{d_{j;p,q}} O_s  \ket{1}\ket{p} \ket{q} \ket{j} \\
		=& (-1)^{d_{j;p,q}}   \ket{o(j,p,q)}\ket{p} \ket{q} \ket{\mbox{FLIP}(j;p,q)},
	\end{split}
\end{equation}
where
\begin{equation}
	\ket{o(j,p,q)} = \begin{cases}
		\ket{0} &\text{if $\ket{j}_p = 0$ and $\ket{j}_q  = 1$},\\
		\ket{1} & \text{otherwise.}
	\end{cases}
\end{equation}

\begin{figure}[htbp!]
    \centering
    \includegraphics[width=0.5\textwidth]{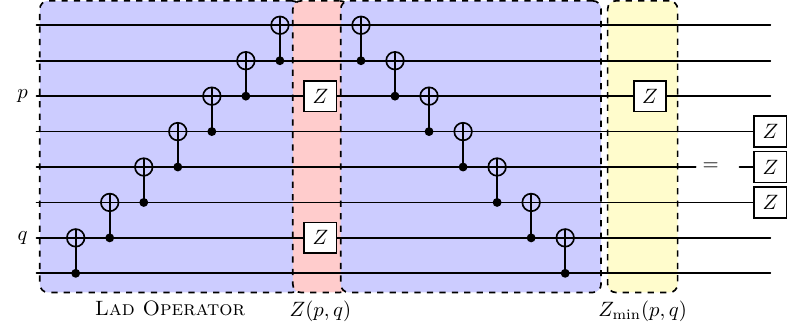}
    \caption{Illustration of a circuit identity for the construction of $O_{\phi}$. Note that in the circuit, the two Pauli $Z$ gates are placed on the $p$-th and $q$-th qubits where $p<q$.  }
    \label{fig:quantum_phase}
\end{figure}

For a fixed $(p, q)$ pair, the corresponding phase oracle $O_{\phi}$ can be constructed using the circuit shown on the right-hand side of the equation in \Cref{fig:quantum_phase}. Each $Z$ gate in the circuit accounts for the $(-1)^{j_k}$ factor in \Cref{eq:phasepq1} for $k = p+1, \ldots, q-1$. To construct the phase oracle for the sum $\sum_{p,q} \fc{a}{p}\fan{a}{q}$, controls can be placed on the qubits representing $\ket{p}$ and $\ket{q}$ to select the appropriate circuit for each one-body term. This approach, as outlined in ~\cite{liu2025efficient}, requires $\mathcal{O}(n^2)$ circuit blocks and at least $\mathcal{O}(n^2)$ gates.

Using previously developed techniques~\cite{Wan2021exponentiallyfaster,kivlichan2018quantum}, the number of circuit blocks and gates can be significantly reduced. 
One of the key things to observe is that the circuit on the right hand side of the equation in  \Cref{fig:quantum_phase} is equivalent to the circuit on the left hand side which consists of a set of staggered CNOTs starting from the least significant qubit and ending at the most significant qubit used to encode $\ket{j}$.  We refer to this set of staggered CNOTs as a ladder circuit. The ladder circuit is followed by two $Z$ gates placed on the $p$-th and $q$-th qubits respectively. The inverse of the ladder operator is then applied to undo the ladder operation.%
Readers are referred to~\cite{Wan2021exponentiallyfaster} for verification that the equation in \Cref{fig:quantum_phase} holds.
 

At first glance, the circuit on the left-hand side of the equation in \Cref{fig:quantum_phase} may appear to use more gates than the circuit on the right. However, when combined with a circuit that performs the so-called ``SWAP-UP'' operation, which moves the state of a specific qubit in the register encoding $\ket{j}$ to the first (least significant) qubit of this register, the operation on the right-hand side of \Cref{fig:quantum_phase} can be performed simultaneously for all combinations of $p$ and $q$.

Formally, the SWAP-UP (denoted by SW($p$) below) is defined as \cite{low2018trading} \begin{equation}
    \mbox{SW}(p) : \ket{p}\ket{j_0}\ket{j_1} \cdots \ket{j_{n-1}} \to \ket{p}  \ket{j_p} \ket{*}\cdots\ket{*}.
    \label{eq:swapup}
\end{equation}
Note that the operation can be generalized such that $j_i$ is of length $m_b$. The SWAP-UP operation requires $m_bn+\lceil \log_2(n) \rceil$ qubits, and $8m_bn+\mathcal{O}(\log(n))$ $T$ gate count with $\log(n)$ T depth. 


\begin{figure}[htbp!]
    \centering
   \includegraphics[width=0.5\textwidth]{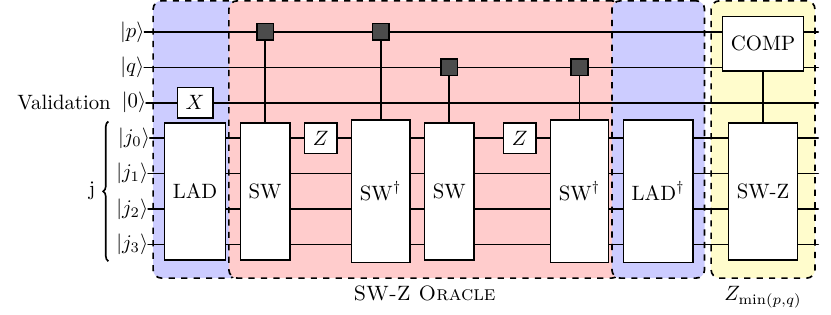}
    \caption{Illustration of quantum circuit for $O_{\phi}$ for one-body interaction $\fc{a}{p}\fan{a}{q}, \fan{a}{p}\fan{a}{q}, \fc{a}{p}\fc{a}{q}$ when $p\neq q$.}
    \label{fig:QSelect1}
\end{figure}

It is not difficult to verify that the circuit block in the shaded region of \Cref{fig:QSelect1}, which implements
$\mbox{SW}(p) (Z\otimes I) \mbox{SW}^{\dagger}(p)$
effectively places a $Z$ gate on the $p$-th qubit of the target register that encodes $\ket{j}$.  
Similarly, $\mbox{SW}(q) (Z\otimes I) \mbox{SW}^{\dagger}(q)$ places a $Z$ gate on the $q$-th qubit of the target register. We refer to the circuit block that excludes the controlling qubits at the top of the circuit (for encoding $p$ and $q$) as a SW-$Z$ block. Note that this block is identical in SW($p$) and SW($q$).

By using Hadamard gates applied to $\ket{0}$ to generate a superposition of $\ket{p}$ and $\ket{q}$, we can effectively generate a superposition of SW($q$)$\cdot$SW($p$) circuits that apply appropriate phase factors associated with different $\fc{a}{p}\fan{a}{q}$ terms.

The last layer of the circuit consists of the COMP and SW-Z circuit blocks. The COMP block, described in~\cite{oliveira2007quantum}, compares $\ket{p}$ and $\ket{q}$. Based on the output from \text{COMP}, \text{SW-Z} applies $Z_{\min(p,q)}$ to the corresponding qubits in the circuit. We emphasize that the $O_{\phi}$ constructed here is designed for the case $p\neq q$. It can be easily extended to $p=q$ by adding additional control on $p=q$.

\begin{figure}[htbp!]
  \centering
  \begin{subfigure}{0.5\textwidth}
    \centering
    \includegraphics[width=1\textwidth]{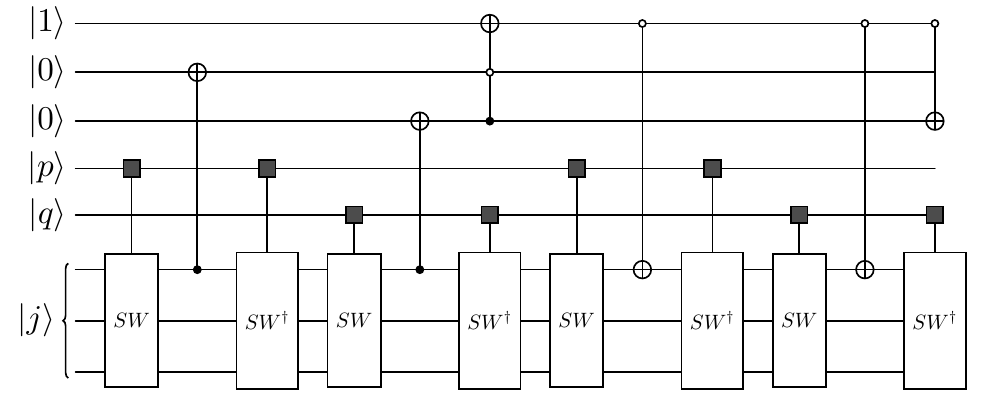}
    \caption{$\fc{a}{p}\fan{a}{q}$.}
    \label{fig:ocfast}
  \end{subfigure}

  \vspace{1em}

  \begin{subfigure}{0.5\textwidth}
    \centering
    \includegraphics[width=1\textwidth]{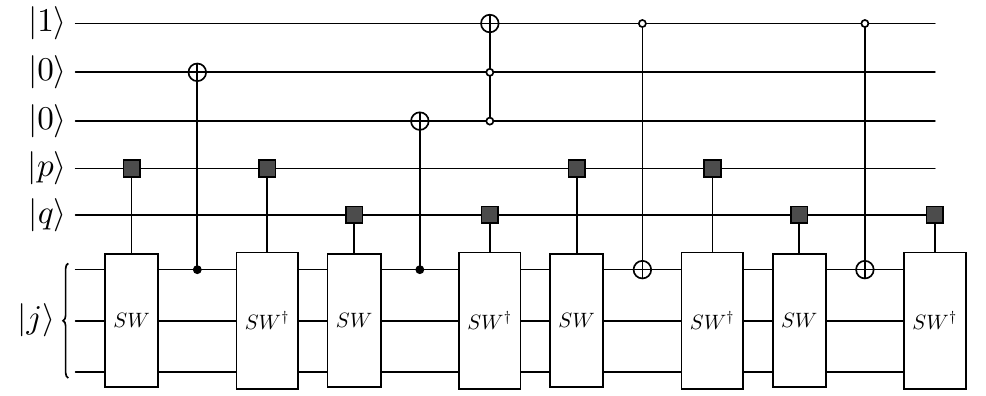}
    \caption{$\fc{a}{p}\fc{a}{q}$.}
    \label{fig:ocfast2}
  \end{subfigure}

  \caption{Illustrations of the $O_s$ oracle for one-body interactions, where $SW$ denotes the SWAP-UP gate used in \cite{low2018trading,wan2021exponentially}.}
  \label{fig:os-onebody}
\end{figure}

The $O_{s}$ oracle is based on a similar approach to the \spar \ oracle introduced in \cite{liu2025efficient}.  We use the $SW$ gate to reduce the complexity of looking up the classical data stored in the $p$-th qubit, illustrated in \Cref{fig:ocfast}. The first ancilla qubit is used to mark whether the computation is valid. Its value is initially set to $\ket{1}$. By checking $\ket{j}_p$ and $\ket{j}_q$ with the SW gate and recording them in two ancilla qubits, we can make the computation valid when $\ket{j}$ satisfies $\ket{j}_p=0$ and $\ket{j}_q=1$. Based on whether the computation is valid (the control from the first ancilla qubit), we use the SW gate to swap up the $p$-th qubit and apply a NOT gate. The same process is applied to the $q$-th qubit, and an additional CNOT gate is applied to uncompute the classical data stored on the third qubit.

To estimate the cost of the phase oracle $O_\phi$, we add up the number of gates required to implement all circuit blocks in \Cref{fig:QSelect1,fig:os-onebody}. Each ladder oracle (LAD) in $O_{\phi}$ requires $n$ CNOT gates, and the SWAP-UP block uses $\mathcal{O}(n)$ gates to implement $Z_{p}$ (a precise accounting of all gate costs can be found in Appendix A and B of \cite{wan2021exponentially}). For the $O_s$ oracle, a total of eight SWAP-UP operations are used, as illustrated in~\Cref{fig:ocfast,fig:ocfast2}. Consequently, for the one-body interaction, the \spar \ oracle has a T-gate count of $\mathcal{O}(n)$ and a $T$ depth of $\mathcal{O}(\log n)$, following the resource estimates in~\cite{low2018trading,wan2021exponentially}. %

\subsubsection{Number operator}
We consider a Hamiltonian 
\begin{equation}
\label{equ:onebodypp}
    \mathcal{H}= \sum_{p} \fc{a}{p}\fan{a}{p},
\end{equation}
and discuss the construction of the \spar \ oracle for this Hamiltonian. Specifically, the goal is to design two oracles, $O_s$ and $O_{\phi}$. For number operators, no non-trivial phase is generated, and $O_\phi$ acts as the identity operator. Therefore, it suffices to construct the $O_s$ oracle,%
\begin{equation}
\begin{split}
       O_{s} \ket{1}  \ket{p} \ket{\ws} 
      =  \begin{cases}
    \ket{0}  \ket{p}\ket{j} &\text{if $\ket{j}_p = 1$ },\\
    \ket{1} \ket{p}  \ket{j} &\text{otherwise}.
  \end{cases}
\end{split}
\end{equation}

\begin{figure}[htbp!]
    \centering
    \includegraphics[width=0.5\textwidth]{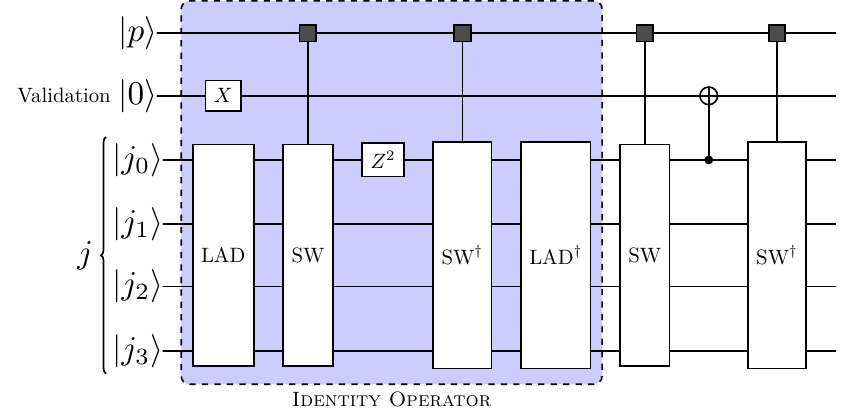}
    \caption{Illustration of the \spar \ oracle for a Hamiltonian expressed as a summation of number operators. It is worth noting that the first part of the circuit, located between the two Lad oracles, is equivalent to an identity operation where $Z^2=Z\cdot Z=I$. This structure is used to facilitate comparison with the \spar \ oracle for general one-body interactions. %
    }
    \label{fig:QSelect2}
\end{figure}

Using the SW gate, we can determine whether $\ket{j}_p$ equals $\ket{0}$ or $\ket{1}$. \Cref{fig:QSelect2} illustrates the construction of the \spar \ oracle, maintaining a structure comparable to \Cref{fig:QSelect1} to highlight that the phase is always $1$ for number operators. 

\begin{figure}[htbp!]
  \centering
 \includegraphics[width=0.45\textwidth]{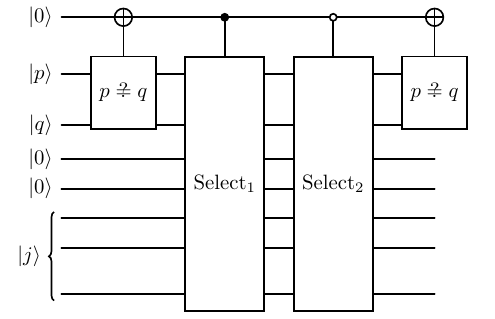}
\caption{Illustration of \spar \ oracle for one-body interaction $\fc{a}{p}\fan{a}{q}$. We use $\text{Select}_1$ and $\text{Select}_2$ to denote the different \spar \ oracles designed in \Cref{fig:QSelect1} and \Cref{fig:QSelect2}. The gate marked with $p \qeq q$ serves as the control for the CNOT gate: the CNOT gate is applied only when $p = q$. %
}
\label{fig:QSelect_onebody}
\end{figure}

We have discussed the \spar \ oracle for both cases, $p = q$ and $p \neq q$. By incorporating an additional control to distinguish whether $p = q$, we can construct an \spar \ oracle for general one-body interactions, as illustrated in \Cref{fig:QSelect_onebody}.

\subsubsection{Two-body interactions}

\begin{figure}[htbp!]
  \centering
  \includegraphics[width=0.5\textwidth]{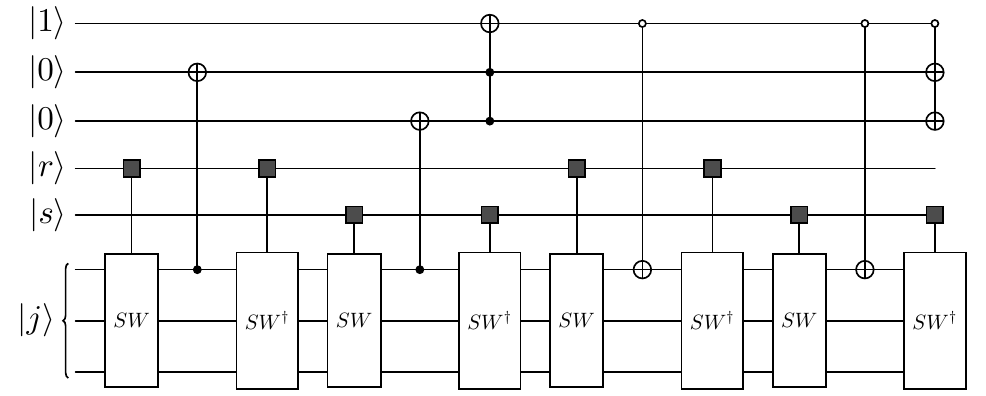}
\caption{Illustration of $O_s$ oracle for one-body interaction $\fan{a}{r}\fan{a}{s}$ where $SW$ is used to denote the SWAP-UP gate used in \cite{low2018trading,wan2021exponentially}.}
\label{fig:ocfast3}
\end{figure}

In this section, we discuss the \spar \ oracle for two-body interactions. Before delving into the details, it is important to note that the \spar \ oracle described here has only minor variations across different types of interactions, including $\fc{a}{p}\fc{a}{q}\fan{a}{r}\fan{a}{s}$, $n_p n_q$ and $n_r\fc{a}{p}\fan{a}{q}$. By block encoding each type of interaction, LCU techniques can ultimately be applied to provide the input model for the full Hamiltonian.

For the two-body interaction, there are three distinct types. The first type of interaction is given by:%
\begin{equation}
\fc{a}{p}\fc{a}{q}\fan{a}{r}\fan{a}{s}, 
\end{equation}
where $p < q$ and $r > s$. For simplicity, we initially assume that $p$, $q$, $r$, and $s$ are four distinct indices. The general form can be extended by adding controls to handle cases where $p = q$ or $r = s$. Notably, the \spar \ oracle is only required to function for the specific ordering of $p$, $q$, $r$, and $s$, as the \spar \ oracle can be set to assign zero coefficients to all other orderings.
We aim to design a \sparnew oracle $O_C$ such that
\begin{equation}
\label{equ:twobody_Oc_phase_flip}
\begin{split}
    &O_C \ket{1} \ket{1} \ket{p} \ket{q} \ket{r} \ket{s} \ket{j}\\
    &= (-1)^{d_{j';p,q}+d_{j;r,s}} \ket{o_1(j;p,q)}\ket{o_2(j;r,s)} \\&\times\ket{p} \ket{q} \ket{r} \ket{s}\ket{\mbox{FLIP}(j';p,q)},
\end{split}
\end{equation}
where $j'=\mbox{FLIP}(j;r,s)$. 
The $o_1$ and $o_2$ indicate whether $\fc{a}{p}\fc{a}{q}$ and $\fan{a}{r}\fan{a}{s}$ are valid operations for the state $\ket{j}$. To be specific,
\begin{equation}
	\ket{o_1(j;p,q)} = \begin{cases}
		\ket{0} &\text{if $\ket{j}_p = 0$ and $\ket{j}_q  = 0$},\\
		\ket{1} & \text{otherwise},
	\end{cases}
\end{equation}
and
\begin{equation}
	\ket{o_2(j;r,s)} = \begin{cases}
		\ket{0} &\text{if $\ket{j}_r = 1$ and $\ket{j}_s  = 1$},\\
		\ket{1} & \text{otherwise}.
	\end{cases}
\end{equation}

\begin{figure}[htbp!]
  \centering
  \includegraphics[width=0.275\textwidth]{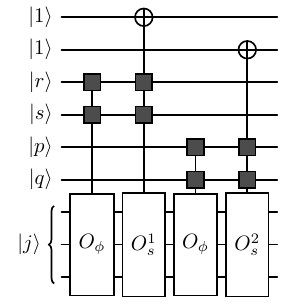}
\caption{Illustration of \spar \ oracle for two-body interaction $\fc{a}{p}\fc{a}{q}\fan{a}{r}\fan{a}{s}$ where $SW$ is used to denote the SWAP-UP gate used in \cite{low2018trading,wan2021exponentially}. The $O_s^{1}$ and $O_s^{2}$ are part of the \spar \ oracle designed for $\fc{a}{p}\fc{a}{q}$ and $\fan{a}{r}\fan{a}{s}$ respectively. We note that we abuse the notation $O_{\phi}$, $O_{s}^1$ and $O_{s}^2$ in the circuit illustration as we do not include the control on $p,q,r,s$ or the validation qubits in the two oracles, aiming to clarify the structure of the circuit.}
\label{fig:QSelect_twobody1}
\end{figure}

The design of such an \spar \ oracle is derived directly from the \spar \ oracle for the two one-body operators $\fc{a}{p}\fc{a}{q}$ and $\fan{a}{r}\fan{a}{s}$. The $O_s$ oracles for these two types of interactions are shown in \Cref{fig:ocfast2} and \Cref{fig:ocfast3}. The phase oracle for both operators is identical to the one illustrated in \Cref{fig:QSelect1}. By stacking these two \spar \ oracles together, the desired \spar \ oracle is obtained, as illustrated in \Cref{fig:QSelect_twobody1}. To account for cases where $p = q$ or $r = s$, the same procedure as in \Cref{fig:QSelect_onebody} is applied.

\begin{figure}[htbp!]
  \centering
  \includegraphics[width=0.5\textwidth]{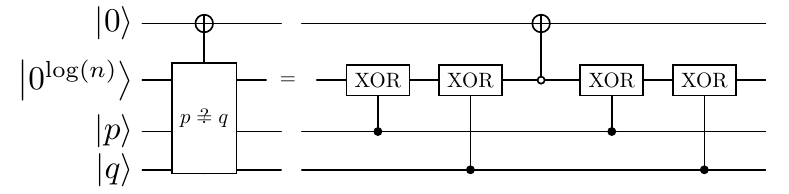}
\caption{Illustration of implementation of condition $p=q$ in \spar \ oracle constructed in~\Cref{fig:QSelect_onebody}. If $p=q$, the first qubit is in state $\ket{0}$; otherwise it is in state $\ket{1}$. 
}
\label{fig:QSelect_identitycontrol}
\end{figure}

The \spar \ oracle for two-body interaction can be designed with less cost if the interaction has special structures, for example,
\begin{equation}
    n_p n_q \ \text{or} \ n_r\fc{a}{p}\fan{a}{q}.
\end{equation}
To construct the \spar \ oracle for these interactions, the \spar \ oracle (including $O_{\phi}$ and $O_{s}$) for $\fc{a}{p}\fc{a}{q}$ and $\fan{a}{r}\fan{a}{s}$ is replaced with an \spar \ oracle for $\fc{a}{p}\fan{a}{q}$ and the number operator $n$. This process is illustrated in \Cref{fig:QSelect1} and \Cref{fig:QSelect2}.
We note that additional control of $p=q$ is needed for the \spar \ oracle to include the designs for the \spar \ oracle in both the case of $p=q$ and $p\neq q$, as illustrated in \Cref{fig:QSelect_onebody} and \Cref{fig:QSelect_identitycontrol}.

\subsubsection{Beyond two-body interaction}

The generalization of the \spar \ oracle to three-body interactions and beyond follows a similar approach to the \spar \ oracles for one-body and two-body interactions. The additional cost primarily arises from the increased use of SWAP circuits~\cite{low2018trading}. It is also worth noting that the \spar \ oracle can be extended to handle Hamiltonians that do not conserve particle number. However, the specific details of the circuit design for such Hamiltonians are beyond the scope of this paper and are left for future work.

\section{Amplitude oracle \texorpdfstring{$O_A$}{OA}}
\label{sec:prepare}

In previous work~\cite{liu2025efficient,du2023multi,du2024hamiltonian}, the \amp \ oracle, which encodes numerical values of all matrix elements of the Hamiltonian, is composed of a sequence of controlled rotations with the rotation angle determined by the values of the coefficients $h_{ijkl}$ in \Cref{eq:2quant}.

The number of controlled rotations is the number of terms in one and two-body operators which is in general $n^2$ and $n^4$ respectively.  It is desirable to reduce the number of controlled gates in a block encoding circuit. This can be done using a table lookup input model described below.

Another potential problem with using controlled rotation is that the unitary 
\begin{equation}
R(\theta) = \begin{bmatrix}
\cos(\theta)  & \sin(\theta) \\
-\sin(\theta) & \cos(\theta)
    \end{bmatrix}
\end{equation}
required to perform the rotation may not be a native gate, therefore may need to be decomposed further into products of other simpler unitaries to yield an approximate rotation.

The data lookup input model, which makes use of a previously developed SELECT-SWAP circuit, provides an efficient mechanism to encode the coefficients as probability amplitudes of a basis state in a superposition state.




To simplify the discussion, we assume that the Hamiltonian to be block-encoded contains $L_e$ distinct coefficients corresponding to the one-body and two-body terms. Note that $L_e$ may be smaller than the total number of terms $L$ in certain structured Hamiltonians in~\Cref{sec:translation_invariance}. We denote each coefficient by $\rho_l$ for $l=0,1,\ldots,L-1$. We seek to construct a circuit for \amp \ oracle that yields %
\begin{widetext}
    \begin{equation}
  \frac{1}{\sqrt{L}} \sum_{l \in [L]} \ket{0^{m_b-1}}\left( \rho_l \ket{0} + g(\rho_l) \ket{1}\right) \ket{l}+\ket{*},
  \label{eq:suprho}
\end{equation}
\end{widetext}
 when it is applied to the superposition state
$
\frac{1}{\sqrt{L}} \sum_{l \in [L]} \ket{0} \ket{l},
$
produced by applying $I\otimes D_s$ to $\ket{0}\ket{0^m}$ where $m = \log (L)$,  and $D_s$ is a Kronecker product of Hadamard matrices. Here $\mathrm{sgn}(x)$ is a sign function that yields $\mathrm{sgn}(x) = -1$ if $x < 0$, and $\mathrm{sgn}(x) = 1$ otherwise. To simplify the notation, we define $g(x) \equiv \mathrm{sgn}(x)-x$, which denotes the amplitude of the $\ket{1}$ branch that is discarded upon post-selection; this notation is used throughout the paper. The integer $m_b$ in \Cref{eq:suprho} is determined by the accuracy of the block encoding and is discussed further in \Cref{sec:sampling}. Also, $\ket{*}$ denotes a quantum state that contains at least one $\ket{1}$ in the first $m_b$ qubits, and will thus vanish when the first $m_b$ qubits are measured with $\ket{0^{m_b}}\bra{0^{m_b}}$.

This circuit is constructed in two steps. In the first step,
we construct an efficient circuit $O_B$ as a data-lookup oracle that can map $l$ to a binary representation of a finite precision approximation of $\rho_l$ which we denote by $b_l$, i.e.,
\begin{equation}
	O_B: \ket{l} \ket{0^{m_b}} \to  \ket{l}\ket{b_l}.
    \label{eq:ob}
\end{equation}

In the second step, we construct a \textit{sampling} circuit $O_S$ that can turn $\ket{b_l}$ into an amplitude factor of the state $\ket{0}\ket{l}$, i.e.,
\begin{equation}
\begin{split}
    O_S:& \ket{b_l}\ket{0^{m_b-1}}\ket{0} \\ \to&
\ket{b_l}\ket{0^{m_b-1}} \left( \rho_l \ket{0} 
+ g(\rho_l) \ket{1} \right)+ \ket{*}.
\label{eq:os}
\end{split}
\end{equation}
We apply another $O_B$ oracle to uncompute $\ket{b_l}$. In particular, we note that when the data-lookup oracle is written abstractly as in \Cref{eq:ob}, only the address register $\ket{l}$ and the output register $\ket{b_l}$ appear explicitly, and the ancilla qubits required by its circuit implementation are suppressed. In our construction, each application of $O_A$ implicitly invokes the SELECT-SWAP circuit (described in \Cref{sec:datalookup}) twice in the worst case: once to compute $\ket{b_l}$ and once to uncompute it. Each invocation employs an additional ancilla register, which can consist of either clean ancilla qubits initialized to $\ket{0}$ or dirty ancilla qubits borrowed in an arbitrary state~\cite{low2018trading}. Since these ancilla qubits are restored to their initial state at the end of each invocation, they do not appear in \Cref{eq:ob}, and their count is accounted for separately in the resource estimates of \Cref{sec:datalookup}. 


%

\subsection{Data-lookup oracle}
\label{sec:datalookup}

\begin{figure}

    \centering
    \includegraphics[width=1\linewidth]{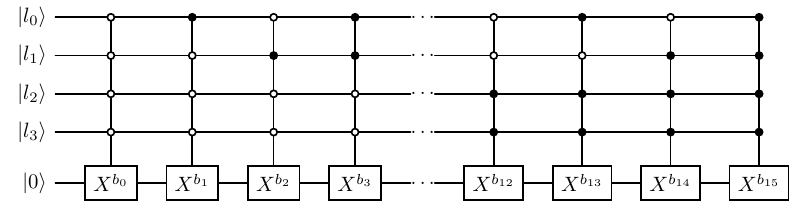}
    \caption{Illustration of quantum read-only memory (QROM) circuit.}
    \label{fig:QROM}
\end{figure}

%
An $m_b$-bit binary representation of a finite precision approximation to the coefficient $\rho_l$ can be generated by applying a sequence of $X$ gates to turn some of the qubits in $\ket{0^{m_b}}$ from $\ket{0}$ to $\ket{1}$ to yield $\ket{b_l}$.  We will use $X^{b_l}\ket{0^{m_b}}$ to represent such an operation.  A superposition of $X^{b_l}\ket{0}$ can be generated using a selection circuit shown in \Cref{fig:QROM}.  The application of $X^{b_l}$ is activated by a single $\ket{l}$-controlled NOT that turns a controlling ancilla qubit from $\ket{0}$ to $\ket{1}$. The activated controlling qubit is reset to $\ket{0}$ once $X^{b_l}$ is applied to $\ket{0^{m_b}}$.  In this case, the number of $T$ gates required to implement multi-qubit CNOT gates is $\mathcal{O}(L)$.

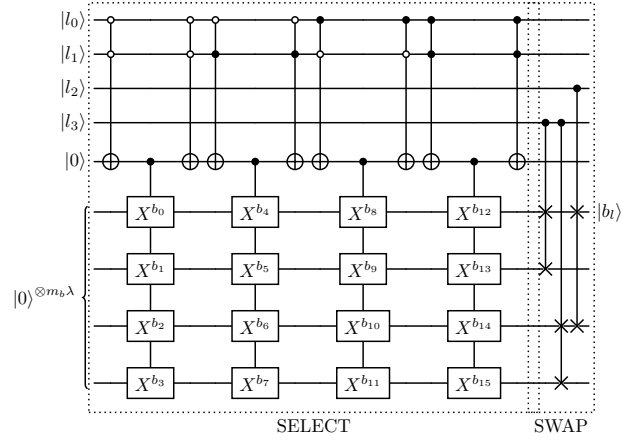
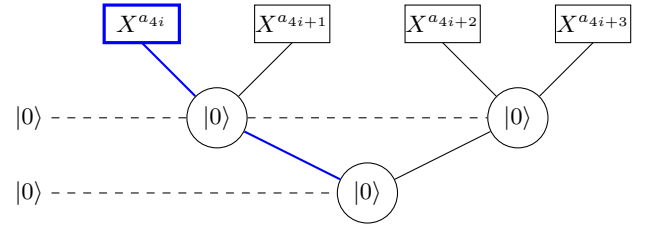
\begin{figure}[htbp]
\centering

\begin{subfigure}{\linewidth}
  \centering
  \includegraphics[width=\linewidth]{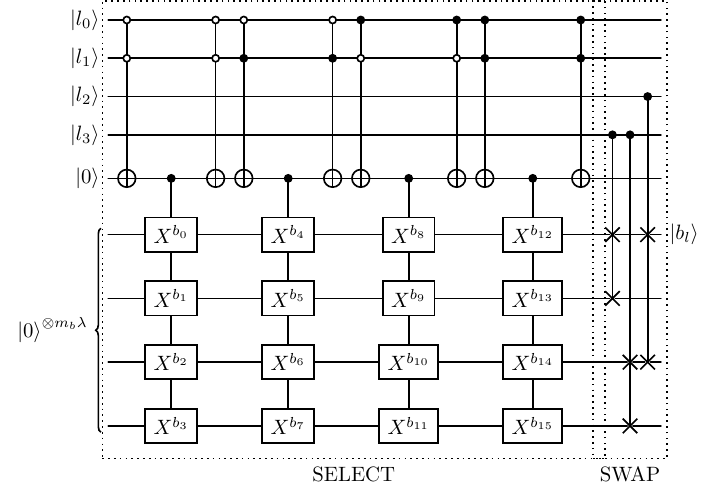}
  \begin{tikzpicture}
  \end{tikzpicture}
  \caption{The SELECT--SWAP circuit introduced in Ref.~\cite{low2018trading}, for word length $m_b=1$, tuning parameter $\lambda=4$, and $L=16$. If the stored bit at address $b_i$ is $0$, then $X^{b_i}$ is the identity; otherwise, it is a Pauli $X$. The upper half of the address bits is used by the SELECT subcircuit to pick the column of $X^{b_i}$, and the SWAP subcircuit uses the lower half to extract the row. The classical bit is then routed to $\ket{b_x}$. The optimal version is given in Ref.~\cite[Fig.~1(d)]{low2018trading}.} 
  \label{fig:select_swap}
\end{subfigure}

\vspace{1em}

\begin{subfigure}{\linewidth}
  \centering
  \begin{tikzpicture}

    \draw[very thick, blue] (-0.5, -0.5)  rectangle node{\textcolor{black}{$X^{a_{4i}}$}} ++(1,0.5);
    \draw (1.5, -0.5)  rectangle node{$X^{a_{4i+1}}$} ++(1,0.5);
    \draw (3.5, -0.5)  rectangle node{$X^{a_{4i+2}}$} ++(1,0.5);
    \draw (5.5, -0.5)  rectangle node{$X^{a_{4i+3}}$} ++(1,0.5);

    \node[draw, circle,minimum size=0.8cm,inner sep=0pt] (t1p) at (1,-1.5){$\ket{0}$};
    \node[draw, circle,minimum size=0.8cm,inner sep=0pt] (t2p) at (5,-1.5){$\ket{0}$};
    \node[draw, circle,minimum size=0.8cm,inner sep=0pt] (t0p) at (3,-2.5){$\ket{0}$};

    \draw[blue, thick] (t1p) to (0,-0.5);
    \draw (t1p) to (2,-.5);
    \draw (t2p) to (4,-.5);
    \draw (t2p) to (6,-.5);
    \draw[blue, thick] (t0p) to (t1p);
    \draw (t0p) to (t2p);

    \node at (-1.5,-1.5) {$\ket{0}$};
    \node at (-1.5,-2.5) {$\ket{0}$};
    \draw[dashed] (-1.2,-1.5) to (t1p);
    \draw[dashed] (t1p) to (t2p);
    \draw[dashed] (-1.2,-2.5) to (t0p);

  \end{tikzpicture}
  \caption{High-level routing scheme of the SWAP subcircuit for $N=16$. The index $i \in \{0,1,2,3\}$ represents the column activated by the SELECT subcircuit. The controlled-SWAPs (circles) act as a router, carving a path (blue) through the binary tree to retrieve the classical bit.}
  \label{fig:swap_scheme}
\end{subfigure}

\caption{SELECT-SWAP circuit and its routing scheme through the binary tree.}
\label{fig:select_swap_combined}
\end{figure}

The number of multi-qubit CNOT gates can be significantly reduced by using the SELECT-SWAP technique proposed in \cite{low2018trading}. In this approach, a set of $\mathcal{O}(\log(\frac{L}{\lambda}))$  ancilla qubits are used to group all $X^{b_l}\ket{0^{m_b}}$ operations into $\mathcal{O}(\frac{L}{\lambda})$ groups as shown in \Cref{fig:select_swap} where $L=16, \lambda=4$ and $m_b=1$. 

%
Within each group, there are $\lambda$ data items and each data item has an $m_b$-bit binary representation. The most significant subset of qubits used to encode $l$ is used to identify the group $X^{b_l}$ is in for a given $l$. For example, in~\Cref{fig:select_swap}, the qubits labeled by $\ket{l_0}$ and $\ket{l_1}$ are used to locate the group of data being inquired.  The remaining qubits are used to control SWAP operations used to move $X^{b_l}\ket{0^{m_b}}$ to the leading $m_b$ qubits of the ancilla within that group. This process can be viewed as a binary-tree routing procedure, where the controlled-SWAPs steer the active path determined by the SELECT output to load the corresponding classical bit, as illustrated in \Cref{fig:swap_scheme}. In this approach, the number of $T$ gates used in multi-qubit controls is $\mathcal{O}(\frac{L}{\lambda})$ and the number of $T$ gates used in control SWAP operations is $\mathcal{O}(\lambda m_b)$.
%

We refer readers to~\cite{low2018trading} for a comprehensive resource estimation of SELECT-SWAP circuit. In total, the data lookup oracle requires 
\begin{equation}
	\label{equ:selectswap_qubit}
	\lambda m_b+2\lceil \log_2 (L) \rceil
\end{equation}
 ancilla qubits.  It contains 
 \begin{equation}
 	\label{equ:selectswapTgate}
 	4\left\lceil\frac{L}{\lambda}\right\rceil+8 \lambda m_b 
 \end{equation}
 $T$ gates  and yields a circuit that has a $T$ depth of 
 \begin{equation}
 	\label{equ:selectswapTdepth}
 	\frac{L}{\lambda}+\log(\lambda).
 \end{equation}
 
  The parameter $\lambda$ is used to divide all $X$ operations into multiple groups and determine the number of classical data items within each group. It can be easily shown from \Cref{equ:selectswapTgate} and \Cref{equ:selectswapTdepth} that we can set $\lambda$ to $\lfloor \sqrt{\frac{L}{2m_b}} \rfloor$ to minimize the T-gate count which attains $\mathcal{O}(\sqrt{L m_b})$.

A majority of the Clifford gates within SELECT-SWAP are used to implement the controlled multi-NOT logic, which is used for loading classical data. For a quantum device with all-to-all connectivity, one way to implement such a logic is by using $Lm_b+o(L)$ CNOT gates with a circuit depth of the same scaling. However, we note that with mid-circuit measurement and qubit reset, every quantum circuit in the Clifford group can be implemented with $\mathcal{O}(1)$ depth with measurement-based quantum computation (MBQC)~\cite{raussendorf2001computational,gottesman1999quantum} and similar Clifford gate cost. Such a measurement-based approach can also achieve constant depth, even when the hardware lacks all-to-all connectivity~\cite{baumer2025measurement}.

The SELECT-SWAP circuit is a framework for a quantum data lookup input model that encodes classical data on a quantum device. It is not necessary to optimize toward lowering the $T$ gate count. Indeed, the previously developed PREPARE oracle in~\cite{babbush2018encoding} employs QROM, which can be regarded as using a SELECT-SWAP circuit with $\lambda=1$, i.e, optimizes qubit usage but incurs a linear $T$ gate count. A comprehensive review of different data lookup schemes within a unified architecture is presented in Ref.~\cite{zhu2024unified}, which also demonstrates how SELECT-SWAP can be tuned for 2D local connectivity of qubits, exhibiting noise resilience while achieving sublinear scaling across all resource measures.

\subsection{Direct sampling method}
\label{sec:sampling}

To prepare a desired quantum state, an additional oracle is required to convert the binary representation of a coefficient into probability amplitude. For a binary string $b$ with first bit $b_0$ representing sign of the number it represents and last $m_b-1$ bits (denoted by $b_1$) representing a real number $0 \leq r_b \leq 1$, a sampling oracle is constructed to map $\ket{b}\ket{0^{m_b-1}} \ket{0}$ to the state given by the right-hand side of \Cref{eq:os}
%

Related sampling-based approaches to quantum state preparation and matrix block encoding include quantum rejection sampling and its coefficient-structured extensions~\cite{ozols2012quantum,lemieux2024quantum}, while black-box state preparation without coherent arithmetic provides another alternative~\cite{sanders2019black}.

When the number of coefficients $L$ in the Hamiltonian greatly exceeds the $\ell_1$ norm of the coefficients (for instance, for two-body interactions, the $\ell_1$ norm of the coefficients $h_{pqrs}$ is defined as $\sum_{p,q,r,s=0}^{n-1}|h_{pqrs}|$), alias sampling~\cite{babbush2018encoding} has been proposed to reduce the subnormalization factor from $L$ to the $\ell_1$ norm of the coefficients. However, this is unnecessary if each coefficient is $\mathcal{O}(1)$ or if the coefficients are known to decrease with respect to their indices (or index differences). In such cases, a direct sampling approach to be presented below reduces the additional subnormalization introduced by diffusion operators.

\begin{figure}
  \centering
  \includegraphics[width=0.45\textwidth]{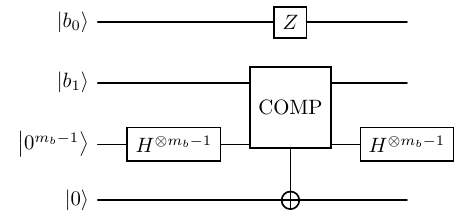}
\caption{Illustration of sampling approach to transform classical data from data-lookup into the coefficient factors.}
\label{fig:sampling}
\end{figure}

The construction of the sampling circuit is illustrated in \Cref{fig:sampling}. Initially, $\ket{0}^{m_b-1}$ is diffused to produce the superposition
\begin{equation}	
\sum_{i=0}^{2^{m_b-1}-1} \frac{1}{\sqrt{2^{m_b-1}}}\underbrace{\ket{b_0}\ket{b_1}}_{\ket{b}} \ket{i} \ket{0}.
\label{eq:diffusei}
\end{equation}
%
Next, $\ket{b_1}$ and $\ket{i}$ are compared using a comparison oracle~\cite{berry2018improved}. If $b_1 \leq i$,  the $\ket{0}$ in last qubit is flipped to $\ket{1}$. This step effectively splits \Cref{eq:diffusei} into the sum of two terms
\begin{equation}
\label{equ:direct1}
\begin{split}
    &\frac{1}{\sqrt{2^{m_b-1}}}\sum_{i=0}^{b_1-1}\ket{b_0}\ket{b_1} \ket{i} \ket{0}  \\
    +& \frac{1}{\sqrt{2^{m_b-1}}}\sum_{i=b_1}^{2^{m_b-1}-1}\ket{b_0}\ket{b_1}\ket{i} \ket{1}.
\end{split}
\end{equation}
The second set of Hadamard gates in \Cref{fig:sampling} is applied to $m_b-1$ qubits holding superpositions of $\ket{i}$ to turn the state $\ket{i}$ into
\begin{equation}
    \frac{1}{\sqrt{2^{m_b-1}}}\ket{0^{m_b-1}}+*
\end{equation}
where $*$ denotes linear combination of states orthogonal to $\ket{0^{m_b-1}}$ and \Cref{equ:direct1} becomes 
\begin{equation}
    \begin{split}
&\ket{b_0}\ket{b_1}\ket{0^{m_b-1}} \left( \frac{b_1}{2^{m_b-1}} \ket{0} + \frac{2^{m_b-1}-b_1}{2^{m_b-1}} \ket{1} \right)\\+&\ket{b_0}\ket{b_1}\ket{\Psi},
    \end{split}
\end{equation}
where $\ket{\Psi}$ denotes an unnormalized state orthogonal to $\ket{0^{m_b-1}}\ket{0}$ and  $\ket{0^{m_b-1}}\ket{1}$.
A Pauli-$Z$ gate is applied to the qubit holding $\ket{b_0}$ to add a phase $(-1)^{b_0}$ to the state. These operations yield %
\begin{equation}
	\ket{b}\ket{0^{m_b-1}} \left( r_b \ket{0} + g(r_b) \ket{1} \right)+ \ket{b}\ket{\Psi},
    \label{eq:rb}
\end{equation}
where $r_b=(-1)^{b_0} \cdot\frac{b_1}{2^{m_b-1}}$ represents the amplitude of state $\ket{b}\ket{0^{m_b}}$, and we recall that $g(r_b)=\mathrm{sgn}(r_b)-r_b$ as defined in \Cref{sec:prepare}. The second term in \Cref{eq:rb} above vanishes in post-selection when $\ket{0^{m_b-1}}$ is measured. Note that $m_b-1$ scales as $\mathcal{O}(\polylog(\frac{1}{\epsilon}))$ for input model with one unique coefficient and at most $\epsilon$ error. The actual format of polynomial dependence relies on the structure of the coefficients tensor in the second-quantized Hamiltonian. The cost of direct sampling comes from $2m_b-2$ Hadamard gates and one compare oracle. A single call to the compare oracle requires $\mathcal{O}(m_b)$ $T$ gate count and generates a circuit of $\mathcal{O}(m_b)$ $T$ gate depth. 

\subsection{Oracle construction}

There are three main steps in constructing the amplitude oracle using the data-lookup oracle and direct sampling, illustrated in the $O_A$ part within \cref{fig: be_general_1body}. First, the classical data is loaded via a data-lookup oracle. Next, the direct sampling oracle maps the binary representation of the coefficient to an amplitude in an auxiliary register. Finally, the data-lookup oracle is applied again to uncompute the classical data loaded in the first step.


For data lookup based methods, sampling techniques are required to convert the binary representation of coefficients into their actual coefficient values. For example, alias sampling~\cite{vose1991linear} mentioned earlier has been employed with QROM to prepare quantum states with reduced subnormalization factor~\cite{babbush2018encoding}.

\section{The complete circuit for general Hamiltonians}
\label{sec:full_encoding}

In this section, we show how the circuit gadgets developed for both the \sparnew oracle $O_C$ and \ampnew oracle $O_A$ in previous sections are assembled together according to the general block encoding circuit diagram shown in~\Cref{fig:complete_circ}. In particular, we specify the additional ancilla qubits required in such a circuit and give an estimation of the gate resources required to implement such a circuit.  To simplify our discussion, we continue to use the Hamiltonian defined in~\Cref{equ:onebodypq} which contains only one-body interactions. The extension of a more general Hamiltonian consisting of both one and two-body interactions is relatively straightforward.

\begin{figure*}[htbp!]
\centering
\includegraphics[width=0.95\textwidth]{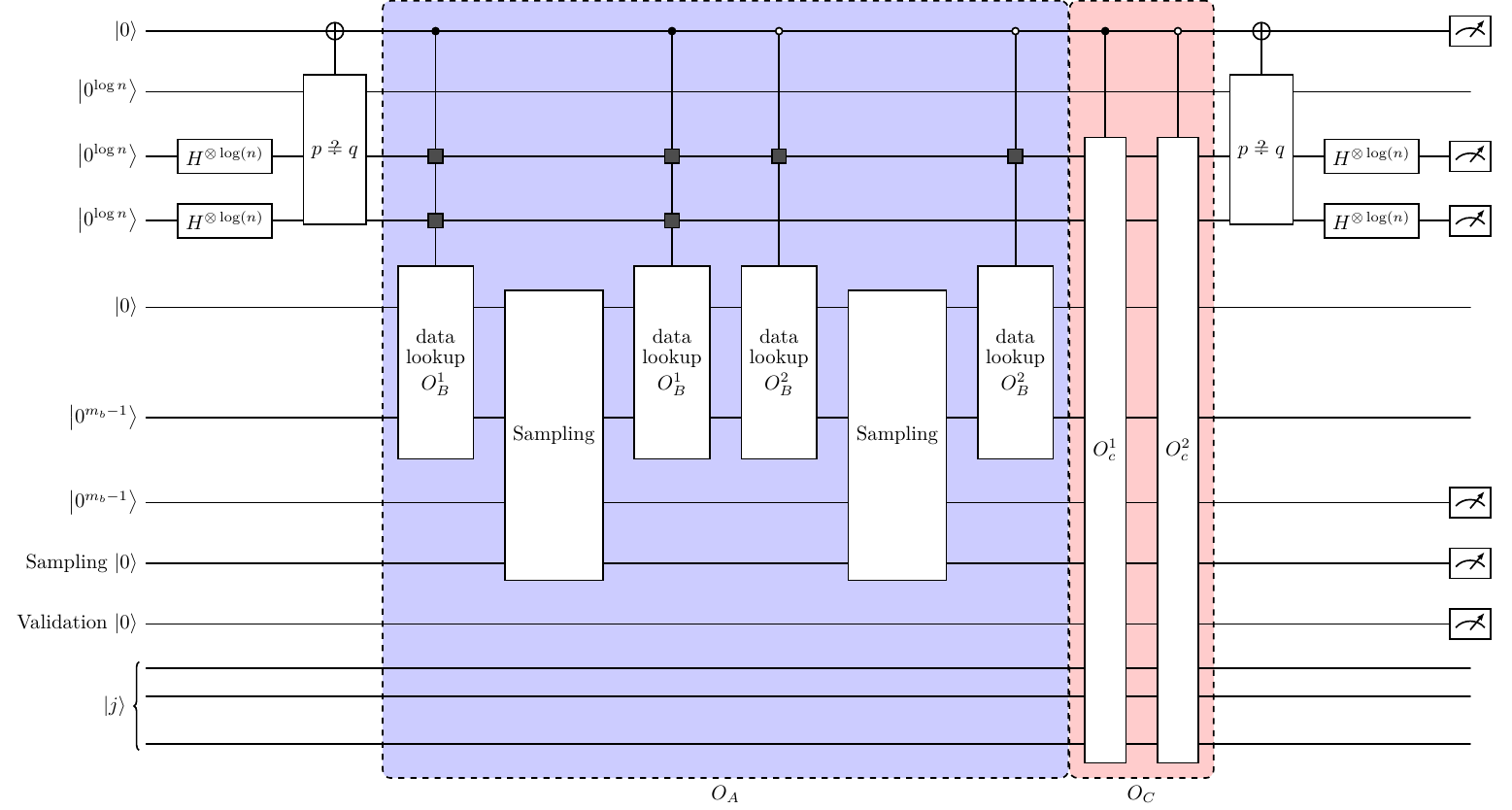}
    \caption{The quantum circuit for the block encoding of a one-body Hamiltonian.}
    \label{fig: be_general_1body}
\end{figure*}

The overall structure of the block encoding circuit for $\mathcal{H}$ is shown in \Cref{fig: be_general_1body}. The $O_A$ and $O_C$ components of the circuit are shaded in red and blue, respectively. Except for the qubits labeled by $\ket{j}$ at the bottom of the circuit, all other qubits are ancilla. The first qubit that takes the input $\ket{0}$ is used to distinguish one-body terms with $p=q$, which is a number operator, from those with $p\neq q$ using the circuit gadget shown in~\Cref{fig:QSelect_identitycontrol}. 
The $O_C$ oracles required for these two cases are slightly different. They are implemented by the $O_C^1$ and $O_C^2$ circuit blocks shown towards the end of the circuit. Only one of them is activated depending on the outcome of the comparison between $p$ and $q$, which are generated from the diffusion operator (Hadamard gates) applied to the two sets of qubits that take  $\ket{0^{\log (n)}}$ as the input. 

A few sets of ancilla qubits are used as the input to the $O_A$ oracle. The qubit that takes $\ket{0}$ as the input and the qubits that take the first $\ket{0^{m_b-1}}$ as the input are used for the SELECT-SWAP based data lookup circuit ($O_B^1$ or $O_B^2$) shown in \Cref{fig:select_swap}. We note that within \Cref{fig: be_general_1body}, there are $m_b \lambda$ ancilla qubits used for the data-lookup circuit but not included within the circuit. These qubits are used to construct the data-lookup circuit shown in \Cref{fig:select_swap}. A superposition of the binary representations of $h_{pq}$ is kept in these qubits after a sequence of controlled $X$ operations is applied to these qubits in \Cref{fig: be_general_1body}.  The next set of qubits that take the second $\ket{0^{m_b-1}}$ as the input and the following qubit that takes $\ket{0}$ as the input are used in the direct sampling oracle designed to encode a finite precision approximation to $h_{pq}$ for the corresponding $p$ and $q$ generated from the diffusion operator applied to $\ket{0^{\log (n)}}$ at the beginning of the circuit.

Note that we again treat the number operator ($p=q$) separately from the case in which $p\neq q$. The leading controlling qubit is used to select one of the $O_B$ oracle ($O_B^1$ or $O_B^2$) circuits. The output states of the $O_B^1$ or $O_B^2$ circuit are used as the $\ket{b_0}$ and $\ket{b_1}$ input to the direct sampling circuit discussed earlier and shown in~\Cref{fig:sampling}. In addition to these inputs, the qubits initialized with the second $\ket{0^{m_b-1}}$ and $\ket{0}$ are also taken as the input to the direct sampling oracle.

The ancilla qubits that are used as the input to both the data lookup and direct sampling oracles are uncomputed to $\ket{0}$ and $\ket{0^{m_b-1}}$ states after they are used in the first data lookup circuit and direct sampling oracle with another data lookup oracle $O_B^{1}$. The controlling qubit at the top of the circuit is reset to $\ket{0}$ by an additional application of the $p\qeq q$ check oracle applied at the beginning of the circuit. We restore the validation qubit to $\ket{0}$ in order to construct a linear combination of non-vanishing one-body interactions. As usual, the qubits used to encode the indices of the interaction terms are post-selected to $\ket{0}$. To be clear, all ancilla qubits that require post-selection are measured at the end of the circuit, and all other ancilla qubits are uncomputed to $\ket{0}$.

In general, the complexity of the block encoding circuit for a Hamiltonian of the form
\begin{equation}
    H = \sum_{\ell=0}^{L-1} \rho_\ell F_\ell ,
    \label{eq:general-term-H}
\end{equation}
where each $F_\ell$ is a fermionic monomial, i.e.,
\begin{equation}
    F_\ell \in
    \left\{
    a_p^\dagger a_q,\;
    n_p,\;
    a_p^\dagger a_q^\dagger a_r a_s,\;
    n_p n_q,\;
    n_r a_p^\dagger a_q
    \right\},
\end{equation}
depends on $L$ in the second quantized Hamiltonian and the precision $\epsilon$ required to encode all coefficients of the Hamiltonian. 

To obtain an $\epsilon$-accurate block encoding, we approximate each
coefficient $\rho_\ell$ using $m_b$ bits.  Since there are $L$ terms, it
is sufficient to choose
\begin{equation}
    m_b = \Theta\!\left(\log\frac{L}{\epsilon}\right).
    \label{eq:mb-general}
\end{equation}
The amplitude oracle $O_A$ uses the SELECT-SWAP data-lookup construction
with tunable block size $\lambda$.  As discussed in \Cref{sec:prepare}, the lookup
part requires
\begin{equation}
    O(\lambda m_b+\log L+m_b)
    \label{eq:qubit-general-lambda}
\end{equation}
ancilla qubits, where the term $\lambda m_b$ is the workspace used to
store a block of $\lambda$ coefficient values, $\log L$ qubits are used
for the term index, and the remaining $m_b$-dependent registers are used
for sampling and uncomputation.  The corresponding T-count is
\begin{equation}
    O\!\left(\frac{L}{\lambda}+\lambda m_b+m_b\right),
    \label{eq:T-general-lambda}
\end{equation}
and the corresponding $T$ depth is
\begin{equation}
    O\!\left(\frac{L}{\lambda}+\log \lambda+m_b\right).
    \label{eq:Tdepth-general-lambda}
\end{equation}
Here the term $L/\lambda$ comes from the SELECT part of the lookup
oracle, while the term $\lambda m_b$ comes from the SWAP part acting on
a block of $\lambda$ data items, each represented using $m_b$ bits.

The sparsity oracle $O_C$ contributes the cost of the SWAP-UP, ladder,
validation, and controlled bit-flip circuits described in \Cref{sec:select}.  For
one- and two-body fermionic terms, this contribution scales at most
linearly in $n$, up to logarithmic factors, and is therefore lower order
than the data-lookup cost in the general unstructured case
$L=\Theta(n^4)$.  Consequently, the dominant asymptotic dependence of
the full-Hilbert-space construction is determined by
Eqs.~\eqref{eq:T-general-lambda} and \eqref{eq:Tdepth-general-lambda}.

Optimizing the tunable parameter $\lambda$ in
Eq.~\eqref{eq:T-general-lambda} to reduce T-count gives
\begin{equation}
    \lambda
    =
    \Theta\!\left(\sqrt{\frac{L}{m_b}}\right).
    \label{eq:lambda-opt-general}
\end{equation}
With this choice,
\begin{equation}
    \frac{L}{\lambda}
    =
    \Theta(\sqrt{L m_b}),
    \qquad
    \lambda m_b
    =
    \Theta(\sqrt{L m_b}).
\end{equation}
Thus the leading contribution to the qubit count, T-count, and $T$ depth 
is
\begin{equation}
\label{eq:equ:generalBEn_OptimalTgate_depth_Lmb}
    O(\sqrt{L m_b}).
\end{equation}
Substituting Eq.~\eqref{eq:mb-general} yields
\begin{equation}
    O\!\left(
    \sqrt{
    L \log\frac{L}{\epsilon}
    }
    \right).
    \label{equ:generalBEn_OptimalTgate_depth}
\end{equation}

For the general electronic-structure Hamiltonian in \Cref{eq:2quant},
$L=\Theta(n^4)$.  Therefore~\Cref{equ:generalBEn_OptimalTgate_depth} becomes
\begin{equation}
    O\!\left(
    n^2
    \sqrt{
    \log\frac{n^4}{\epsilon}
    }
    \right).
    \label{eq:general_es_complexity}
\end{equation}
These are the scalings reported in the row labeled ``General, This
work'' in \Cref{tab:cost1-full} for qubit count, T count and T depth.

Finally, we emphasize that the improvement in this full-Hilbert-space
construction is a reduction in the cost of implementing the coefficient
loading and amplitude oracle.  The subnormalization factor is not
improved in the general full-space setting.  Because the construction
uses a standard diffusion over all $L$ term labels, the
subnormalization factor is
\begin{equation}
    \alpha = O(L)=O(n^4).
    \label{eq:general-alpha-final}
\end{equation}
The reduction of the subnormalization factor requires the
state-dependent, fixed-particle construction introduced in \Cref{sec:eta}.

\section{Block encoding for $\eta$-particle Hamiltonian}
\label{sec:eta}
The block encoding scheme discussed above in \Cref{sec:select} and \Cref{sec:prepare} for the Hamiltonian given by \Cref{eq:2quant} introduces a subnormalization factor $L$ that is proportional to $n^4$ where $n$ is the number of orbitals required to represent the Hamiltonian. Such a subnormalization factor results from using the diffusion operator to generate superpositions of all possible $ijkl$ combinations. Because \Cref{eq:2quant} can be block diagonalized by basis states that preserve particle number, one would expect that the block encoding for a diagonal block of the Hamiltonian associated with a fixed particle number $\eta$ will have a small subnormalization factor.

In this section, we show that this is indeed the case. We present an efficient quantum circuit for block encoding the $\eta$-particle sector of \Cref{eq:2quant}. We show that the subnormalization factor associated with this block encoding circuit is $\eta^2 n^2$, which can be much smaller than $n^4$ when $\eta \ll n$.

\subsection{Indirect diffusion}
The key idea we use to reduce the subnormalization factor is to use an indirect diffusion oracle to generate a subset of indices associated with an annihilation operator in \Cref{eq:2quant}.  Note that, if $\ket{j}$ is an $\eta$-particle state, i.e., $\eta$ of the $n$ orbitals are occupied, $\eta$ qubits of the $n$-qubit representation of $\ket{j}$ are in the $\ket{1}$ state while others are in the $\ket{0}$ state. The only relevant annihilation operators are those that annihilate the occupied orbitals.  Therefore, the indirect diffusion oracle is devised to indirectly generate a superposition of the indices of the occupied orbitals associated with $\ket{j}$.

For a computational basis state $\ket{j}$, let $\redrev{\occ{i}{j}}$ denote the bit index of its $i$-th occupied qubit. For example, if $n=4$, $\eta=2$, and $\ket{j}=\ket{0101}$, then $\redrev{\occ{0}{j}}=1$ and $\redrev{\occ{1}{j}}=3$. The implementation of an indirect diffusion oracle hinges on constructing an oracle that maps $\ket{i}$ and $\ket{j}$ to $\redrev{\occ{i}{j}}$. We refer to this oracle as the \textit{occupation detection oracle} and denote it by $O_{occ}$ throughout the paper.
The operation performed by $O_{occ}$ can formally be expressed as
\begin{equation}
\ket{i}\ket{0^{\log(n)}}\ket{j} \xrightarrow[]{O_{occ}} \ket{i} \ket{\redrev{\occ{i}{j}}} \ket{j},
    \label{eq:occ}
\end{equation}
and 
\begin{equation}
    \ket{i} \ket{\redrev{\occ{i}{j}}} \ket{j} \xrightarrow[]{O_{occ}} \ket{i}\ket{0^{\log(n)}}\ket{j}.
    \label{eq:occback}
\end{equation}

We emphasize that $O_{occ}$ does not compact the persistent system register: $\ket{j}$ remains the standard $n$-qubit occupation string, and the occupied-mode address $\occ{i}{j}$ is computed only in an ancilla register and later uncomputed. Because the system register keeps this representation throughout, the fixed-$\eta$ construction composes directly with occupation-basis state preparation, orbital rotations by Givens networks~\cite{kivlichan2018quantum}, and computational-basis measurement, with no conversion overhead. Encodings that instead store occupied-mode addresses as the system representation are discussed in \Cref{sec:compact-encoding-comparison}.

By composing a standard Hadamard diffusion operator that performs the following operation
\begin{equation}
    \begin{split}
       &\ket{0^{\log(\eta)}} \ket{0^{\log(n)}}\ket{j}\\
       \xrightarrow[]{H^{\otimes \log(\eta)}}&\frac{1}{\sqrt{\eta}}\sum_{i=0}^{\eta-1}\ket{i}\ket{0^{\log(n)}}\ket{j},
    \end{split}
\end{equation}
to generate a superposition of $\ket{i}$'s for $i=0,1,2,...,\eta-1$ with the oracle shown in \Cref{eq:occ},  we obtain the indirect diffusion oracle $O_{\rm IDF}$ that yields the following mapping,
\begin{equation}
\begin{split}
&\ket{0^{\log(\eta)}}\ket{0^{\log(n)}}\ket{j} \\ \xrightarrow[]{O_{\rm IDF}} &\frac{1}{\sqrt{\eta}}\sum_{i=0}^{\eta-1}\ket{i}\ket{\redrev{\occ{i}{j}}}\ket{j}.
\end{split}
\end{equation}
Note that, unlike the standard diffusion operator that takes $\ket{0^{\log(\eta)}}$ as the only input, the input to $O_{\rm IDF}$ consists of both $\ket{0^{\log(\eta)}}$ and $\ket{j}$, i.e., it depends on $\ket{j}$. The $\ket{j}$ dependency of the diffusion has a direct consequence on how the block encoding of the $\eta$-particle Hamiltonian is constructed.  
The indirect diffusion operator can be easily generalized to include multiple qubit registers as the input.
For example,
\begin{equation}
\begin{split}
&\ket{0^{\log(\eta)}}\ket{0^{\log(\eta)}}\ket{0^{\log(n)}}\ket{0^{\log(n)}}\ket{j} \\ \xrightarrow[]{O_{\rm IDF}} &\frac{1}{\eta}\sum_{i_1=0}^{\eta-1}\sum_{i_2=0}^{\eta-1}\ket{i_1}\ket{i_2}\ket{\redrev{\occ{i_1}{j}}}\ket{\redrev{\occ{i_2}{j}}}\ket{j}
\end{split}
\end{equation}
performs an indirect diffusion to generate a superposition of pairs of quantum states representing indices of creation and annihilation pairs that preserve the $\eta$-particle number in $\ket{j}$. 

\subsection{State-dependent \texorpdfstring{$O_C$}{OC} and \texorpdfstring{$O_A$}{OA} oracles}
\label{sec:state_select_prepare}
The superposition of states representing indices of the annihilation operators that preserve the particle number of an $\eta$-particle basis state $\ket{j}$ can be used as the input to a pair of $O_A$ and $O_C$ oracles to block encode an $\eta$-particle Hamiltonian.


To illustrate how these oracles are constructed, we first use the block encoding of the one-body term of the Hamiltonian \Cref{equ:onebodypq} as an example.


To construct the \ampnew oracle $O_A$, we follow the same technique presented in \Cref{sec:prepare} to use a $O_B$ oracle first to create a finite precision binary representation of each one-body matrix element $h_{i_1,\redrev{\occ{i_2}{j}}}$. Instead of applying $O_B$ to $\ket{i_1}\ket{i_2}$ and $\ket{0^r}$ for a general one-body term, we apply $O_B$ to $\ket{i_1}\ket{\redrev{\occ{i_2}{j}}}\ket{0^r}$ using the SELECT-SWAP circuit presented in \Cref{sec:prepare}. %
Due to the dependence of $\ket{\redrev{\occ{i_2}{j}}}$ on $\ket{j}$, the output of such an oracle is state dependent. Similar to what is used in the \amp \ oracle for the general Hamiltonian, the binary representation of selected one-body matrix elements is then turned into amplitude factors through a direct sampling oracle $O_S$ discussed in \Cref{sec:sampling}.





To start, apply $O_B$ and $O_S$ to the state
\begin{equation}
\label{equ:indirect}
\frac{1} {\sqrt{n\eta}} \sum_{i_1=0}^{n-1} \sum_{i_2=0}^{\eta-1}  \ket{i_1}\ket{i_2}\ket{\redrev{\occ{i_2}{j}}} \ket{0^{m_b}} \ket{0^{m_b-1}} \ket{0}
        \ket{j}.
\end{equation}
which is produced by the indirect diffusion operator $O_{\rm IDF}$ applied to the input
\begin{equation}
\ket{0^{\log(n)}}\ket{0^{\log(\eta)}} \ket{0^{ \log(n)}} \ket{0^{m_b}}  \ket{0^{m_b-1}} \ket{0} \ket{j},
\end{equation}
yields 
\begin{widetext}
\begin{equation}
\frac{1} {\sqrt{n\eta}} \sum_{i_1=0}^{n-1} \sum_{i_2=0}^{\eta-1}  \ket{i_1}\ket{i_2}\ket{\redrev{\occ{i_2}{j}}} \ket{0^{m_b}} \ket{0^{m_b-1}} \left(h_{i_1,\redrev{\occ{i_2}{j}}}\ket{0}+g(h_{i_1,\redrev{\occ{i_2}{j}}})\ket{1} \right) \ket{j} +\ket{*}.
\label{eq:postprep}
\end{equation}
\end{widetext}

Here, as in \Cref{eq:suprho}, $\ket{*}$ denotes a quantum state that contains at least one $\ket{1}$ in the $m_b$ qubits holding $\ket{0^{m_b-1}}\ket{0}$ in \Cref{equ:indirect}. As a result, not all of the weight in \Cref{eq:postprep} is used: only the branch in which these qubits remain in $\ket{0^{m_b}}$ carries the desired coefficient $h_{i_1,\redrev{\occ{i_2}{j}}}$, and post-selection on this register (i.e., measuring it with $\ket{0^{m_b}}\bra{0^{m_b}}$) is needed to remove both $\ket{*}$ and the branch weighted by $g(h_{i_1,\redrev{\occ{i_2}{j}}})$.




%



We can now apply the \sparnew oracle $O_C$ introduced in \Cref{sec:select} to 
qubits holding $\ket{i_1} \ket{\redrev{\occshort{i_2}}}\ket{j}$ in \Cref{eq:postprep} to obtain
\begin{widetext}
\begin{equation}
\frac{1}{\sqrt{n\eta}}
\sum_{i_1,i_2}(-1)^{d_{j;i_1,\redrev{\occshort{i_2}}}}
\ket{o(j,i_1,\redrev{\occshort{i_2}})}
\ket{i_1}
\ket{i_2}
\ket{\redrev{\occshort{i_2}}}
\ket{0^{m_b}}
\ket{0^{m_b-1}}
\left(h_{i_1,\redrev{\occshort{i_2}}}
\ket{0}
+g(h_{i_1,\redrev{\occshort{i_2}}})
\ket{1} \right)
\ket{{\rm FLIP}(j;i_1,\redrev{\occshort{i_2}})} +\ket{*}.
\label{eq:postsel}
\end{equation}
\end{widetext}


In order to ensure \Cref{eq:bedef} is satisfied, we need to use an uncompute oracle to turn $\ket{i_1}$, $\ket{i_2}$, $\ket{\redrev{\occshort{i_2}}}$ to $\ket{0^{\log(n)}}$, $\ket{0^{\log{(\eta)}}}$ and $\ket{0^{\log(n)}}$ respectively. For a general one-body term, the uncompute can be accomplished by the same Hadamard diffusion operator used to generate a superposition of indices $i_1$ and $i_2$ of the creation and annihilation operators. However, because the indirect diffusion operator $O_{\rm IDF}$ is $j$ dependent, we cannot use it to turn $\ket{\redrev{\occshort{i_2}}}$ back to $\ket{0^{\log (n)}}$ (uncompute) once $\ket{j}$ has been changed to $\ket{{\rm FLIP}(j;i_1,\redrev{\occshort{i_2}})}$ in \Cref{eq:postsel}.


Therefore, additional effort is needed to construct a desired uncompute operator. One way to overcome this difficulty is to use another register to hold a copy of $\ket{j}$ that will not be altered by the $O_C$ oracle and use this register to uncompute $\ket{\redrev{\occshort{i_2}}}$ according to \Cref{eq:occback}.  We should also note that making an additional copy of $\ket{j}$, which is a computational basis that can be generated by using a sequence of CNOT gates placed on matching qubits in $\ket{j}$ and $\ket{0^{\log (n)}}$ to yield
\begin{equation}
    \ket{j}\ket{0^{\log (n)}} \to \ket{j}\ket{j},
    \label{eq:copyj}
\end{equation}
does not violate the well-known \textit{no-cloning theorem}. Consider an example where $j=01$ (a 2-bit string). Our goal is to produce: \begin{equation} \ket{0}\ket{1}\ket{0}\ket{0} \rightarrow \ket{0}\ket{1}\ket{0}\ket{1}, \end{equation} where the first two qubits are initialized as $\ket{0}\ket{1}$ (representing $j$), and the last two start as $\ket{0}\ket{0}$. To accomplish this, we apply two CNOT gates: the first with the first qubit as control and the third as target, and the second with the second qubit as control and the fourth as target. After these operations, the third and fourth qubits will match the first and second, respectively, successfully copying the state of $\ket{j}$ onto a second register. We note that the copy operation described here is, in fact, an application of the exclusive OR (XOR) operator, as will be discussed in the next paragraph. We further emphasize that, although \Cref{eq:copyj} is stated for a single computational basis state $\ket{j}$, the copy operation is a unitary composed of CNOT gates and therefore extends by linearity to a coherent linear combination of basis states, mapping $\sum_j c_j \ket{j}\ket{0^{\log (n)}}$ to the entangled state $\sum_j c_j \ket{j}\ket{j}$. This is consistent with the no-cloning theorem: the output is not a product of two independent copies of the superposition, i.e., $\sum_j c_j \ket{j}\ket{j} \neq \left(\sum_j c_j \ket{j}\right)\otimes\left(\sum_j c_j \ket{j}\right)$. 


%

We could apply the $O_{occ}$ oracle to qubits holding $\ket{i_2}$, $\ket{\redrev{\occ{i_2}{j}}}$ and the first $\ket{j}$ according to \Cref{eq:occback} to turn $\ket{\redrev{\occ{i_2}{j}}}$ to $\ket{0^{\log (n)}}$. The qubits holding $\ket{i_1}$ and $\ket{i_2}$ can be restored to $\ket{0^{\log (n)}}$ and $\ket{0^{\log (\eta)}}$ respectively by applying the standard Hadamard diffusion operator.

However, we still need to turn the qubits holding the copied $\ket{j}$ back to $\ket{0^{\log (n)}}$. This is not easy to do without additional resources. To overcome this difficulty, we use CNOTs applied to matching qubits in $\ket{{\rm FLIP}(j;i_1,\redrev{\occshort{i_2}})}$ and $\ket{j}$ to set all qubits of the register holding $\ket{j}$ to $\ket{0}$ except the $i_1$-th and the $\redrev{\occshort{i_2}}$-th qubits, which are set to the $\ket{1}$ state because $\ket{{\rm FLIP}(j;i_1,\redrev{\occshort{i_2}})}$ and $\ket{j}$ differ only in these qubits, i.e., a circuit is constructed to perform
\begin{widetext}
\begin{equation}
\ket{{\rm FLIP}(j;i_1,\redrev{\occshort{i_2}})}\ket{j} \rightarrow \ket{{\rm FLIP}(j;i_1,\redrev{\occshort{i_2}})}\ket{{\rm FLIP}(j;i_1,\redrev{\occshort{i_2}})\oplus j},
\label{eq:xorj}
\end{equation}
\end{widetext}
where $\oplus$ is the exclusive OR operator (denoted as XOR in our quantum circuits).

The remaining qubits to be uncomputed are now in
\begin{equation}
\ket{i_1} \ket{\redrev{\occ{i_2}{j}}}\ket{{\rm FLIP}(j;i_1,\redrev{\occshort{i_2}})\oplus j}
\label{eq:remainingqbits}
\end{equation}
To turn the $i_1$-th qubit of $\ket{{\rm FLIP}(j;i_1,\redrev{\occshort{i_2}})\oplus j}$ to $\ket{0}$, we can use the SWAP-UP circuit (SW) defined by \Cref{eq:swapup} and shown in \Cref{fig:QSelect1} to move the $\ket{1}$ state in the $i_1$-th qubit to the leading qubit of the register. Applying an $X$ gate on this qubit followed by using ${\rm SW}^\dagger$ to move the resulting $\ket{0}$ back to the $i_1$-th qubit of the register achieves our goal.

The remaining $\ket{1}$ in the $\redrev{\occ{i_2}{j}}$-th qubit of the register holding $\ket{{\rm FLIP}(j;i_1,\redrev{\occshort{i_2}})\oplus j}$ as well as the ones in $\ket{\redrev{\occ{i_2}{j}}}$ can be turned to $\ket{0}$ by using a slight modification of the $O_{occ}$ circuit that we will describe in the next section. We will refer to this special uncompute oracle circuit as $O_{uocc}$.

\subsection{The implementation of $O_{occ}$ and $O_{uocc}$}
\label{sec:Occ}
%



In this section, we describe how to construct quantum circuits to implement the $O_{occ}$ oracle defined in \Cref{eq:occ}. We also show how such an oracle can be easily modified to uncompute the qubits holding $\ket{\redrev{\occ{i_2}{j}}}$ and the $\redrev{\occ{i_2}{j}}$-th qubit in the register holding $\ket{{\rm FLIP}(j;i_1,\redrev{\occshort{i_2}})\oplus j}$ as we discussed in the previous section.

\begin{figure}[htbp!]
  \centering

  \begin{subfigure}{1\linewidth}
    \centering
    \includegraphics[width=\linewidth]{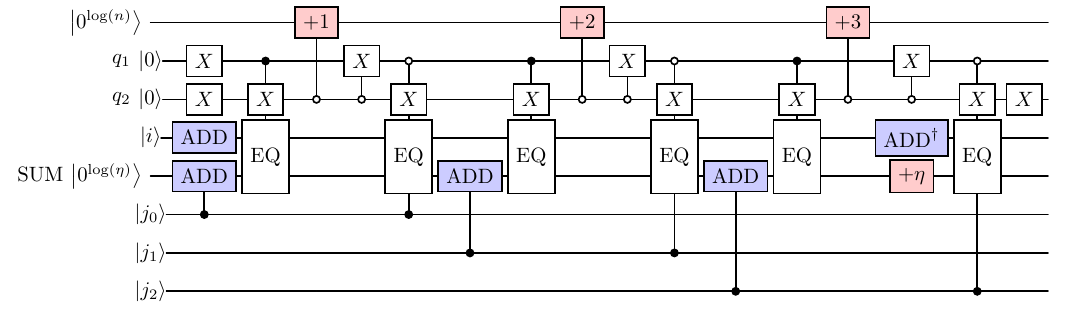}
    \caption{Occupation detection oracle $O_{occ}$.}
    \label{fig:Occnew}
  \end{subfigure}

  \vspace{0.8em}

  \begin{subfigure}{1\linewidth}
    \centering
    \includegraphics[width=\linewidth]{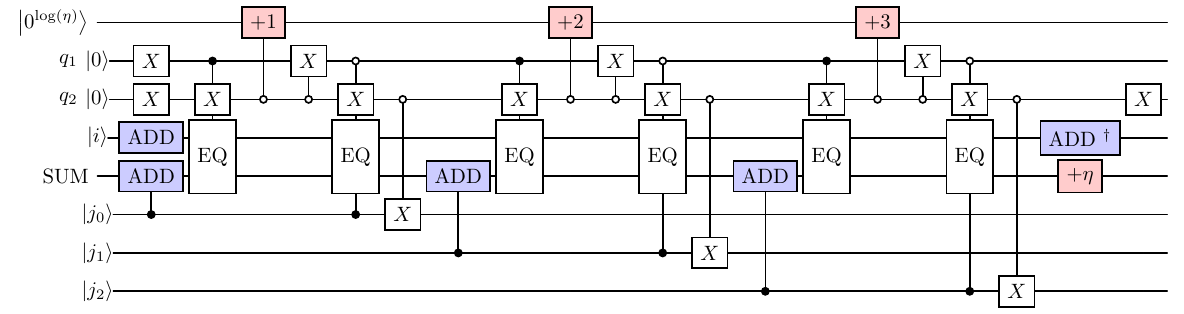}
    \caption{The $O_{uocc}$ circuit used to uncompute the remaining $\ket{1}$ ancilla qubits in \Cref{eq:remainingqbits}.}
    \label{fig:UOccnew}
  \end{subfigure}

  \caption{Occupation detection oracle and its uncompute for the computational basis $\ket{j}$ as a $\eta$-particle state.}
  \label{fig:occAll}
\end{figure}

The basic idea is to construct a circuit to scan each qubit in the register holding $\ket{j}$ in order by using a sequence of controlled adder gates to increment a separate register (labeled as SUM in \Cref{fig:Occnew}) holding $\log(\eta)$ ancilla qubits that serve to record the number of qubits that have been identified to hold a $\ket{1}$ state in $\ket{j}$. When the integer value represented by the SUM register matches the integer value represented by $\ket{i+1}$, which indicates that the qubit currently being scanned is the $i$-th occupied qubit in $\ket{j}$, i.e., it holds the $i$-th $\ket{1}$ (counting from zero) rather than being the $i$-th qubit of $\ket{j}$, we can then return the position of this $\ket{1}$ state in $\ket{j}$ (from the register at the top of the circuit in \Cref{fig:Occnew}), which is equivalent to the number of qubits in $\ket{j}$ that have been scanned. To keep track of the number of qubits that have been scanned, we use a separate controlled adder to change the register initialized to $\ket{0^{\log (n)}}$ to $\ket{\redrev{\occ{i}{j}}}$. Two additional ancilla qubits (labeled by $q_1$ and $q_2$ in \Cref{fig:Occnew}) are used to set proper controls so that only one of the adders is applied for each $i$ when the value represented by the SUM register equals $i+1$. The latter condition is checked by a comparison circuit labeled by EQ in \Cref{fig:Occnew}. Both the $q_1$ and $q_2$ qubits are initially set to $\ket{1}$. The $q_2$ qubit is flipped to $\ket{0}$ when EQ returns $\ket{1}$ and $q_1$ is in $\ket{1}$. It is then used to activate the adder on the leading qubits and to reset the $q_1$ qubit back to $\ket{0}$. At the end of each period, the $q_2$ qubit is flipped back into state $\ket{1}$ by a second EQ circuit, subject to the additional condition on the EQ output and the state $\ket{j}$: this EQ circuit is activated only when $q_1$ is in $\ket{0}$ (open control) and the qubit of $\ket{j}$ just scanned is in $\ket{1}$ (closed control). Restoring $q_2$ to $\ket{1}$ ensures that the adder on the leading qubits, which is activated by $q_2$ being in $\ket{0}$, does not fire again in subsequent periods. At the very end of the circuit, an $X$ gate flips $q_2$ into $\ket{0}$. We also note that the first flip of $q_2$, which activates the adder, occurs precisely when the qubit of $\ket{j}$ just scanned is in the $\ket{1}$ state and increments the SUM register to the value $i+1$. Although EQ continues to return $\ket{1}$ on subsequent scans (the SUM register retains the value $i+1$ until the next $\ket{1}$ in $\ket{j}$ is encountered), the $q_1$ and $q_2$ qubits ensure that the adder on the leading qubits is applied exactly once, recording the position of the $i$-th $\ket{1}$ in $\ket{j}$. 

Suppose the input $i$ of the $O_{occ}$ oracle satisfies $i+1 > \eta$, i.e., the state $\ket{j}$ contains fewer than $i+1$ occupied qubits. In this case, the SUM register never reaches the value $i+1$, the EQ circuit never returns $\ket{1}$, and the adder on the leading qubits is not activated in any scanning period. As a result, $q_1$ ends up in the state $\ket{1}$ and $q_2$ ends up in the state $\ket{0}$. Post-selection on the $q_1$ qubit is therefore needed, since not every ancilla qubit can be uncomputed back to the state $\ket{0}$ within the oracle in this case.



The $O_{occ}$ circuit shown in \Cref{fig:Occnew}, which implements the following transformation
\begin{equation}
    \ket{0^{\log(n)}}\ket{0^2}\ket{i}\ket{0^{\log(\eta)}}\ket{j} \rightarrow \ket{\redrev{\occ{i}{j}}}\ket{0^2}\ket{i}\ket{0^{\log(\eta)}}\ket{j}
    \label{eq:occmap}
\end{equation}
is reversible, i.e., the left-hand side of \Cref{eq:occmap} is produced if the input to the circuit is the right-hand side of \Cref{eq:occmap}, provided that there are at least $i+1$ $\ket{1}$'s in the state $\ket{j}$, i.e., $\eta \geq i+1$. When this condition fails, the oracle cannot locate the position of the $i$-th $\ket{1}$ in $\ket{j}$, and the qubit $q_1$ is left in the state $\ket{1}$ instead of $\ket{0}$, as discussed below.

In particular, if the input to $O_{occ}$ is $\ket{\redrev{\occ{0}{j'}}}\ket{0^2}\ket{0^{\log(\eta)}}\ket{j'}$, where
$\ket{j'}$ is the state produced in the register holding $\ket{{\rm FLIP}(j;i_1,\redrev{\occshort{i_2}})\oplus j}$ after the $i_1$-th qubit of this register has been set to $\ket{0}$, leaving only one qubit of this register, which was the $\redrev{\occ{i_2}{j}}$-th qubit of this register when it held $\ket{j}$, in $\ket{1}$, the output of the $O_{occ}$ circuit
is $\ket{0^{(\log n)}}\ket{0^2}\ket{i_2}\ket{\log^{(\log \eta)}}\ket{j'}$. Note that $\redrev{\occ{0}{j'}}=\redrev{\occ{i_2}{j}}$ because the $0$-th occupied qubit in the register holding $j'$ is the same as the $i_2$-th occupied qubit of the register when it held $\ket{j}$.

This transformation does not completely accomplish the goal of resetting all ancilla qubits to $\ket{0}$ even after a set of Haddamard gates are applied to $\ket{i_2}$ to turn it to $\ket{0^{(\log \eta)}}$ because the $\redrev{\occ{0}{j'}}$-th qubit in the register holding $\ket{j'}$ (or equivalently, the $\redrev{\occ{i_2}{j}}$-th qubit in the same register holding the original $\ket{j}$), is still in $\ket{1}$. However, this qubit can be set to $\ket{0}$ by modifing
the $O_{occ}$ circuitly slightly to insert a CNOT at the end of each period to clear the remaining $\ket{1}$ qubit.  
We denote the modified circuit by $O_{uocc}$ which is shown in \Cref{fig:UOccnew}. 

When $\eta$ is not a power of $2$, the circuit necessarily takes some inputs $i$ with $i+1 > \eta$. For such inputs the $O_{occ}$ oracle fails to find the address of the $i$-th $\ket{1}$ in the state $\ket{j}$, and the qubit $q_1$ in both the $O_{occ}$ and $O_{uocc}$ circuits is left in the state $\ket{1}$ instead of $\ket{0}$. In that case \Cref{eq:occmap} no longer holds. However, post-selecting on $q_1$ being in $\ket{0}$ allows us to keep the computation result precisely on those branches where the address is found; this post-selection also serves as a consistency check that the protocol works correctly. We note that in this case two distinct ancilla qubits require post-selection on the state $\ket{0}$, used respectively as qubit $q_1$ in the $O_{occ}$ and $O_{uocc}$ circuits.

The discussion extends straightforwardly to two-body interactions, where the register to be uncomputed contains two occupied qubits whose positions are held in two separate address registers. Since every two-body operator acts on the occupation-number register as a product of two one-body operators, the same procedure is simply applied twice, once for each occupied qubit and its address register.

We now estimate the number of $T$ gates required in the circuit implementation of $O_{occ}$ and $O_{uocc}$ and their T depth. Each ADD subcircuit requires $\mathcal{O}(\log (\eta))$ $T$ gates and has a T depth of $\mathcal{O}(\log (\eta))$~\cite{gidney2018halving}. The EQ operation can be implemented by using a sequence of CNOT gates to check whether each qubit in the register holding $\ket{i}$ and the register labeled by $\text{SUM}$ is in the same state. When that is the case and the state on $q_1$ qubit equals to $\ket{0}$, a NOT is applied to qubit $q_2$. This operation can be implemented with $\mathcal{O}(\log (\eta))$ $T$ gates in a circuit that has a T depth of $\mathcal{O}(\log \log (\eta))$. One clean ancilla is used in such a circuit. In total, these two oracles require $\mathcal{O}(n \log (\eta))$ $T$ gates and have a T depth of $\mathcal{O}(n\log (\eta))$.

\begin{figure}
\centering
\includegraphics[width=0.95\linewidth]{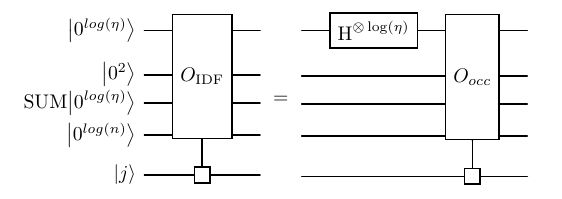}
    \caption{Quantum circuit of indirect diffusion oracle.}
    \label{fig:IDF}
\end{figure}

\subsection{The complete circuit for $\eta$-particle Hamiltonian}
\label{sec:input}
We now assemble the indirect diffusion operator, the state-dependent
amplitude oracle, and the state-dependent sparsity oracle into a
complete block-encoding circuit for the restriction of a
particle-number-conserving Hamiltonian to the $\eta$-particle subspace.
The key distinction from the full-Hilbert-space construction in
\Cref{sec:full_encoding} is that the diffusion over annihilation indices is now
state-dependent: for an input $\eta$-particle basis state $|j\rangle$,
the indirect diffusion oracle generates only indices corresponding to
occupied orbitals in $|j\rangle$.  Thus, the block encoding does not
diffuse over all formal monomials in the Hamiltonian, but only over
those monomials that can contribute a nonzero matrix element in the
column labeled by $|j\rangle$.

\begin{figure*}[htbp]
\centering
\includegraphics[width=1\textwidth]{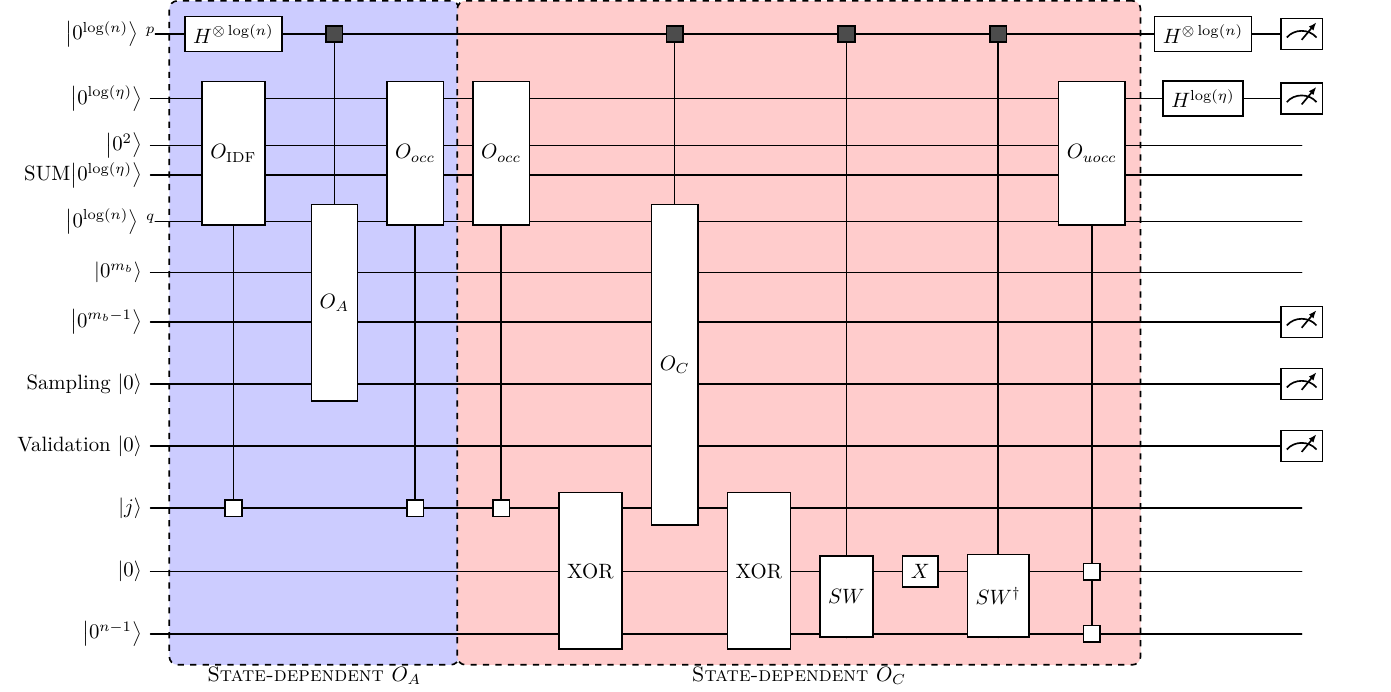}
	\caption{The block encoding quantum circuit of an $\eta$-particle one-body operator.}
    \label{fig:onebodyBEeta}
\end{figure*}

The complete circuit shown in~\Cref{fig:onebodyBEeta} has the same high-level form as the sparse-matrix
block encoding in~\Cref{fig: be_general_1body}, but with the standard diffusion operator
replaced by an indirect diffusion operator. 

In the circuit shown in~\Cref{fig:onebodyBEeta}, Hadamard gates are used to generate a superposition of the indices ($p$) for the creation operator $a_p^\dagger$. The indirect diffusion oracle $O_{\rm IDF}$, which takes $\ket{0^{\log(n)}}$ and $\ket{0^{\log(\eta)}}$ as the input in addition to $\ket{j}$, is used to generate the superposition of selected $q$ indices associated with valid annihilation operators for the state $\ket{j}$, i.e., the $q$'s that satisfy $a_q \ket{j} \neq 0$. The $O_A$ circuit block contains both the SELECT-SWAP based data lookup circuit gadgets and direct sampling based circuit gadgets to encode the numerical value of $h_{pq}$ for selected $q$'s. (We note that the data lookup circuit itself still stores all the classical data encoding the coefficients for all $p,q$ pairs.) Note that the inputs to $O_A$ include the superposition of $p$'s at the top circuit and selected $q$'s returned from $O_{\rm IDF}$ as well as the qubits initialized with $\ket{0^{m_b}}$ and $\ket{0^{m_b-1}}$, the ancilla qubit labeled by `Sampling'. 
To the right of the $O_A$ circuit block is an $O_{occ}$ circuit used to restore the qubits holding the superposition of a subset of indices ($q$) of valid annihilation operators for the state $\ket{j}$ to $\ket{0^{\log(n)}}$. However, such an uncomputing operation is not necessary in the overall circuit because the superposition of selected $q$'s is needed again in the $O_{C}$ oracle circuit. Instead of using another $O_{occ}$ to regenerate such a superposition from the uncomputed qubits, we can simply remove the two $O_{occ}$ circuit blocks together between the $O_A$ and $O_C$ in~\Cref{fig:onebodyBEeta}.  We display them in~\Cref{fig:onebodyBEeta} merely to show how the $O_A$ and $O_C$ circuit blocks can be assembled in the overall circuit. 

The XOR circuit block to the left of the $O_C$ circuit block is used to duplicate $\ket{j}$ onto the bottom $n$-qubits according to \Cref{eq:copyj}.  Recall that the duplicated $\ket{j}$ is to be used to facilitate the reset of ancilla qubits to $\ket{0}$ (uncompute). The $O_{C}$ circuit block takes the valid index pair $\{(i_1,\redrev{\occ{i_2}{j}}) | \  0\leq i_1 \leq n-1, 0\leq i_2\leq \eta-1 \}$ on qubits labeled with $p,q$ as the input. It turns the state $\ket{j}$ into $\ket{{\rm FLIP}(j;i_1,\redrev{\occshort{i_2}})}$.
The second XOR circuit block starts the uncompute process by setting all qubits holding $\ket{j}$ at the bottom of the circuit to $\ket{0}$ except the $i_1$th and the $\redrev{\occshort{i_2}}$-th qubits, i.e., it perform the operation defined by \Cref{eq:xorj}.

This operation is followed by the SW-X circuit block (a SWAP-UP and its inverse with an X gate in between) which is used to turn the $i_1$-th qubit in the register holding $\ket{{\rm FLIP}(j;i_1,\redrev{\occshort{i_2}})\oplus j}$ to $\ket{0}$, as discussed in~\Cref{sec:state_select_prepare}. Finally,
the $O_{uocc}$ circuit block turns the $\redrev{\occ{i_2}{j}}$-th qubit in the same register to $\ket{0}$ and effectively sets the entire register previously holding $\ket{{\rm FLIP}(j;i_1,\redrev{\occshort{i_2}})\oplus j}$ to $\ket{0^{\log (n)}}$.

At the end of the circuit, Hadamard gates are used to reset the two registers at the top of the circuit to $\ket{0^{\log (n)}}$ and $\ket{0^{\log (\eta)}}$ respectively.

\begin{figure}[htbp]
\centering
\includegraphics[width=\linewidth]{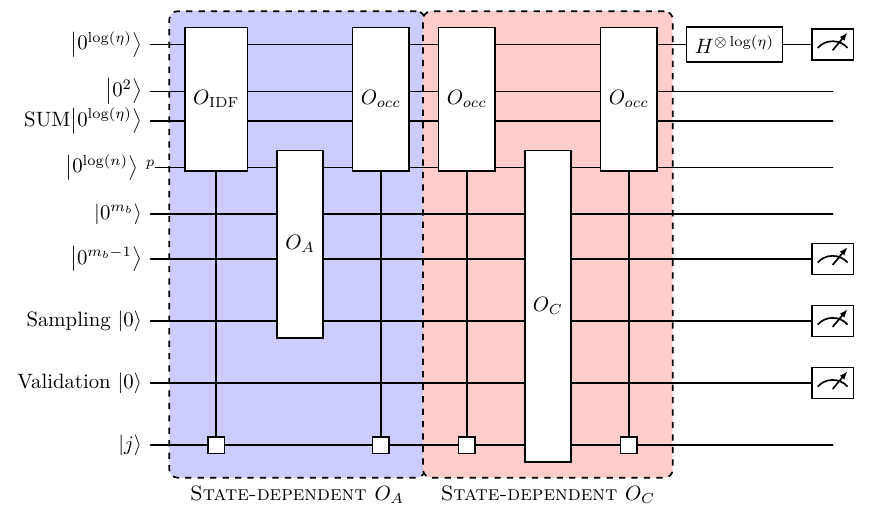}
    \caption{Block encoding of a diagonal one-body Hamiltonian
    $H_{\mathrm{diag}}=\sum_p h_{pp}n_p$ in the $\eta$-particle subspace.}
    \label{fig:numberBEeta}
\end{figure}

For the diagonal one-body Hamiltonian
$H_{\mathrm{diag}}=\sum_{p=0}^{n-1}h_{pp}n_p$, where
$n_p=a_p^\dagger a_p$, the circuit simplifies as shown in
\Cref{fig:numberBEeta}. Indirect diffusion selects only occupied
indices $p=\occ{i}{j}$, without a separate diffusion over a creation
index. On these valid branches, $n_p\ket{j}=\ket{j}$ and no fermionic
phase is generated. Since the state $\ket{j}$ is unchanged, the
address register can be uncomputed with $O_{occ}$ acting on this original
$\ket{j}$; the copied occupation register, XOR and SW-X cleanup blocks,
and $O_{uocc}$ are unnecessary. The two adjacent $O_{occ}$ blocks in
\Cref{fig:numberBEeta} only separate the state-dependent $O_A$ and
$O_C$ components and may be cancelled. In this diagram, the dependence
of $O_A$ and $O_C$ on the occupied index $p$ is included in the
corresponding blocks, and measurement symbols denote post-selection on
zero.

In the notation used below, this diagonal construction has lookup size
$L=n$ and state-dependent label count $S_\eta=\Theta(\eta)$. With the
coefficient normalization used for $O_A$, its subnormalization factor is
$\alpha_\eta=O(\eta)$: both the initial diffusion and the final
post-selection involve only the valid indices. The diagram
shows the power-of-two case. When $\eta$ is not a power of two, the
index register is padded to the next power
of two and a retained validity flag rejects indices $i\geq\eta$ at
post-selection. This changes the subnormalization by less than
a factor of two.

We now estimate the resource requirements.
We consider again a Hamiltonian of the form \eqref{eq:general-term-H}
where each $F_\ell$ is now  a particle-number-conserving fermionic monomial.
The integer $L$ denotes the number of coefficient entries that must be stored in the
data-lookup oracle.  For an unstructured electronic Hamiltonian with
general two-body coefficients, $L=\Theta(n^4)$.

For a fixed $\eta$-particle basis state $|j\rangle$, however, not all
of these $L$ monomials can produce nonzero matrix elements.  We denote
by $S_\eta$ the number of state-dependent labels generated by the indirect
diffusion procedure for each column of the Hamiltonian restricted to
the $\eta$-particle sector.  Equivalently, $S_\eta$ is the number of
candidate monomials that can contribute to the action of $H$ on a
fixed $\eta$-particle basis state.  For example, a general one-body
Hamiltonian has
\begin{equation}
    S_\eta = O(n\eta),
\end{equation}
because the annihilation index must be one of the $\eta$ occupied
orbitals while the creation index can be any of the $n$ orbitals.  A
general two-body Hamiltonian has
\begin{equation}
    S_\eta = O(n^2\eta^2),
    \label{eq:Seta-general-two-body}
\end{equation}
because the two annihilation indices are chosen from occupied orbitals
and the two creation indices are chosen from all orbitals.

To obtain an
$\epsilon$-accurate block encoding, it is sufficient to approximate
each coefficient to precision $O(\epsilon/S_\eta)$, since at most
$S_\eta$ state-dependent labels contribute in each column of the
$\eta$-particle Hamiltonian.  Thus we choose
\begin{equation}
    m_b
    =
    \Theta\!\left(
    \log\frac{S_\eta}{\epsilon}
    \right).
    \label{eq:eta-mb-general}
\end{equation}
The SELECT-SWAP data-lookup oracle stores $L$ coefficient entries and
uses a tunable block size $\lambda$.  Up to lower-order logarithmic
terms, the number of additional qubits required for data lookup is
\begin{equation}
    O(\lambda m_b).
\end{equation}
Including the occupation register and the auxiliary registers used by
the indirect diffusion and validation circuits, the total qubit count is
\begin{equation}
    O\!\left(
    n + \lambda m_b + \operatorname{polylog}(n,S_\eta)
    \right).
    \label{eq:eta-qubit-lambda}
\end{equation}
In the asymptotic estimates below, the lower-order logarithmic terms
are suppressed.

The T-gate complexity has two main contributions.  The first comes from
the indirect diffusion and occupied-orbital lookup circuits, which
require
\begin{equation}
    O(n\log\eta)
\end{equation}
$T$ gates and T depth.  The second comes from the SELECT-SWAP
data-lookup and direct-sampling part of $O_A$.  With block size
$\lambda$, this contribution amounts to 
\begin{equation}
    O\!\left(
    \frac{L}{\lambda}
    +
    \lambda m_b
    \right)
\end{equation}
$T$ gates.  Therefore the total T-count is

\begin{equation}
    O\!\left(
    n\log\eta
    +
    \frac{L}{\lambda}
    +
    \lambda
    \log\frac{S_\eta}{\epsilon}
    \right).
    \label{eq:eta-T-lambda}
\end{equation}
Similarly, the $T$ depth is
\begin{equation}
    O\!\left(
    n\log\eta
    +
    \frac{L}{\lambda}
    +
    2\log\lambda
    \right),
    \label{eq:eta-Tdepth-lambda}
\end{equation}
where the $2\log\lambda$ term comes from the SWAP routing and
uncomputation in the SELECT-SWAP construction.

The tunable parameter $\lambda$ may be chosen to reduce the T-count.
Using
\begin{equation}
    \lambda
    =
    \Theta\!\left(
    \sqrt{
    \frac{L}{
    \log(S_\eta/\epsilon)
    }}
    \right)
    \label{eq:eta-lambda-opt}
\end{equation}
gives
\begin{equation}
    \frac{L}{\lambda}
    =
    \Theta\!\left(
    \sqrt{
    L\log\frac{S_\eta}{\epsilon}
    }
    \right),
    \qquad
    \lambda
    \log\frac{S_\eta}{\epsilon}
    =
    \Theta\!\left(
    \sqrt{
    L\log\frac{S_\eta}{\epsilon}
    }
    \right).
\end{equation}
Thus, after optimizing $\lambda$, the leading resource estimates are
\begin{equation}
    O\!\left(
    n
    +
    \sqrt{
    L\log\frac{S_\eta}{\epsilon}
    }
    \right),
    \label{eq:eta-Q-opt-general}
    \end{equation}
    for qubit count,
\begin{equation} 
    O\!\left(
    n\log\eta
    +
    \sqrt{
    L\log\frac{S_\eta}{\epsilon}
    }
    \right),
    \label{eq:eta-T-opt-general}
\end{equation} 
for T-gate count, and
\begin{equation}
    O\!\left(
    n\log\eta
    +
    \sqrt{
    L\log\frac{S_\eta}{\epsilon}
    }
    \right).
    \label{eq:eta-DT-opt-general}
\end{equation} 
for T-depth.

The corresponding subnormalization factor is controlled by the
state-dependent diffusion size,
\begin{equation}
    \alpha_\eta
    =
    O(S_\eta).
    \label{eq:eta-alpha-general}
\end{equation}

We now specialize these estimates to the general two-body electronic
Hamiltonian in Eq.~(2), restricted to the $\eta$-particle subspace.  In
the unstructured case, the data-lookup table must store all two-body
coefficient entries, and therefore
\begin{equation}
    L = \Theta(n^4).
    \label{eq:eta-general-two-body-L}
\end{equation}
However, for each fixed $\eta$-particle basis state, the two
annihilation indices are restricted to occupied orbitals.  Hence the
number of state-dependent labels is
\begin{equation}
    S_\eta
    =
    \Theta(n^2\eta^2).
    \label{eq:eta-general-two-body-S}
\end{equation}
Consequently,
\begin{equation}
    m_b
    =
    \Theta\!\left(
    \log\frac{n^2\eta^2}{\epsilon}
    \right).
    \label{eq:eta-general-two-body-mb}
\end{equation}
Substituting Eqs.~\eqref{eq:eta-general-two-body-L} and
\eqref{eq:eta-general-two-body-S} into
Eqs.~\eqref{eq:eta-Q-opt-general}--\eqref{eq:eta-DT-opt-general}
yields
\begin{align}
    O\!\left(
    n
    +
    n^2
    \sqrt{
    \log\frac{n^2\eta^2}{\epsilon}
    }
    \right),
    \label{eq:eta-general-two-body-Q}
    \\
    O\!\left(
    n\log\eta
    +
    n^2
    \sqrt{
    \log\frac{n^2\eta^2}{\epsilon}
    }
    \right),
    \label{eq:eta-general-two-body-T}
    \\
    O\!\left(
    n\log\eta
    +
    n^2
    \sqrt{
    \log\frac{n^2\eta^2}{\epsilon}
    }
    \right)
    \label{eq:eta-general-two-body-DT}
\end{align}
for qubit count, T-gate count and T-depth respectively.

Suppressing the lower-order $n\log\eta$ contribution gives the simpler asymptotic form
\begin{equation}
    O\!\left(
    n^2
    \sqrt{
    \log\frac{n^2\eta^2}{\epsilon}
    }
    \right)
    \label{eq:eta-general-two-body-complexity}
\end{equation}
for qubit count, T-gate count and T-depth as shown in \Cref{tab:cost-combined-eta} in the row labeled by ``General". This is lower than the complexity of the block encoding circuit for a Hamiltonian represented in a full Hilbert space given by \Cref{eq:general_es_complexity}.

The subnormalization factor for the $\eta$-particle block encoding is
\begin{equation}
    \alpha_\eta
    =
    O(n^2\eta^2).
    \label{eq:eta-general-two-body-alpha}
\end{equation}

This is smaller than the full-Hilbert-space subnormalization
$O(n^4)$ whenever $\eta \ll n$, and the reduction comes directly from
using indirect diffusion to generate only annihilation indices
associated with occupied orbitals in the input $\eta$-particle state.

\section{Block encoding for structured Hamiltonian} \label{sec:structuredBE}
It follows from the discussions in \Cref{sec:full_encoding} that, for a general second-quantized Hamiltonian, the number of Clifford and T-gates required to construct the $O_A$ oracle circuit typically exceeds that required for the $O_C$ oracle. The majority of Clifford gates are used in the decomposition of the controlled multi-qubit Pauli X gate in the data-lookup oracle. The total number of CNOT gates required is $n^4 \log\left(\frac{n^4}{\epsilon} \right)+o(n^4 \log\left(\frac{n^4}{\epsilon} \right))$. By tuning the parameter \(\lambda\) in the SELECT--SWAP structure, the required \(T\)-gate count is reduced from \(\mathcal{O}(n^4 + \log\left(\frac{n^4}{\epsilon} \right))\) to \(\mathcal{O}\left(\sqrt{n^4 \log\left(\frac{n^4}{\epsilon} \right)} \right) \). This yields an almost quadratic improvement in the scaling with respect to \(L=n^4\). 
This optimization minimizes the fault-tolerant resource cost for the data-lookup oracle, making the gate complexity approach the theoretical lower bound for $T$ gate usage for general state preparation tasks~\cite{gosset2024quantum}.

However, for Hamiltonians with special structures, quantum resources required to implement the block encoding circuit can be further reduced. The reduction in T-gate and T-depth results from the reduction in either the number of distinct coefficients that need to be encoded in the look-up table or the total number of terms in the Hamiltonian \Cref{eq:2quant} itself or both.  For structured Hamiltonians, the expression \Cref{eq:T-general-lambda,eq:Tdepth-general-lambda}, \Cref{eq:eta-T-lambda} and \Cref{eq:eta-Tdepth-lambda} should all be modified to replace $L$ by the number of distinct coefficients, and $m_b$ should still be calculated from the number of non-negligible terms in the Hamiltonian. The optimal $\lambda$ should then be recalculated from the modified expressions.

In this section, we provide quantum resource estimates for implementing the block encoding circuits for second-quantized Hamiltonians with nearest-neighbor interactions, Hamiltonians with a translation-invariance structure, and Hamiltonians in which the two-body term can be factored as products of number operators. 

\subsection{Hamiltonians with nearest-neighbor interactions}
\label{sec:BEnNN}

If the coefficients of the one or two-body terms are negligibly small, it is not necessary to encode these near-zero coefficients in the block encoding circuit of the Hamiltonian. By neglecting these terms, we can further reduce the subnormalization factor of the block encoding scheme and use fewer resources in the data lookup circuit.

Here we discuss how to reduce the gate complexity of block encoding for $\H=\sum_{p,q} h_{pq}\fc{a}{p}\fan{a}{q}$ in which the coefficients $h_{pq}$ decay exponentially with respect to $|p-q|$. To be specific, we assume
\begin{equation}
\label{equ:decay_assum}
    |h_{pq}| \leq Ce^{-\alpha |p-q|}
\end{equation}
for some $\alpha>0, C>0$. If the underlying lattice associated with the Hamiltonian is defined on a torus $\mathbb{T}$, the decay assumption is modified to take the form
\begin{equation}
\label{equ:decay_assum2}
    |h_{pq}| \leq Ce^{-\alpha \min (|p-q|, n-|p-q| )},
\end{equation}
where the distance accounts for the periodic boundary conditions.

%
The rapid magnitude decay of the coefficients $h_{pq}$ with respect to $|p-q|$ allows us to ignore $(p,q)$ pairs that correspond to small $|h_{pq}|$ without sacrificing the accuracy of the block encoding. To construct a block encoding with precision $\epsilon$, 
it suffices to retain only the one-body interaction terms, whose indices satisfy
\begin{equation}
\label{equ:onebody_reduction}
    |p-q|\leq C' \log \left(\frac{n}{\epsilon} \right),
\end{equation}
where $C'$ is a constant independent of $n$ and $\epsilon$. 
The number of $(p,q)$ pairs satisfying \Cref{equ:onebody_reduction} scales as $\mathcal{O}\left(n \log \left(\frac{n}{\epsilon} \right)\right)$. Considerably fewer one-body terms need be considered also if $|h_{pq}| \leq \frac{C}{|p-q|^{\gamma}}$, i.e. $|h_{pq}|$ decays algebraically with respect to $|p-q|$, or $h_{pq}$ is nonzero only for $|p-q|=1$. In these two scenarios, the number of coefficients to be considered is reduced to $\mathcal{O}(n  \left(\frac{1}{\epsilon} \right)^{\gamma})$ and $\mathcal{O}(n )$ respectively. 

Suppose for each index $p$, only $2M-1$ terms in the Hamiltonian satisfying $|p-q|\leq M-1$ are retained. By retaining only these terms, we can reduce the subnormalization factor of the block encoding of such a Hamiltonian to $\mathcal{O}(nM)$ from $\mathcal{O}(n^2)$ by using an indirect diffusion operator that requires fewer Hadamard gates.

\begin{figure}[htbp!]
\centering
\includegraphics[width=0.5\textwidth]{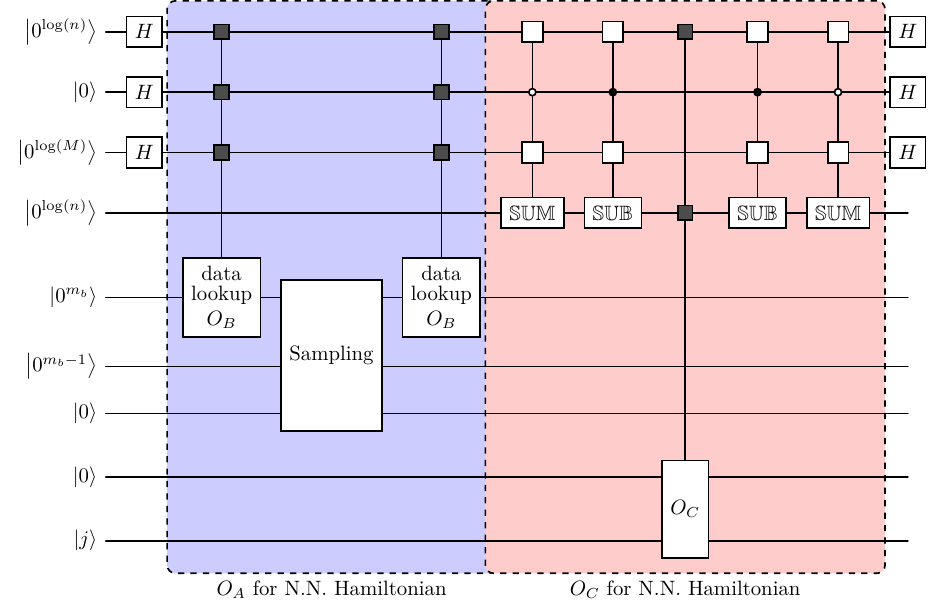}
    \caption{Quantum circuit of block encoding for one-body Hamiltonian with nearest-neighbor interactions.}
    \label{fig: HAM_with_decaying_coef}
\end{figure}

To illustrate how such an indirect diffusion circuit may be constructed, we examine a simplified case where $h_{pq} $ is nonzero only when $|p-q|=1$.
A partial construction of block encoding for this Hamiltonian is described below, and the construction of the quantum circuit is illustrated in \Cref{fig: HAM_with_decaying_coef}.

Intuitively, generating a superposition of $\ket{p}\ket{p-1}$ and $\ket{p}\ket{p+1}$, which represents all index pairs of $h_{p,p-1}$ and $h_{p,p+1}$ requires one qubit to represent a superposition of $\ket{0}$ and $\ket{1}$ in addition to the $\log(n)$ qubits required to represent a superposition of $\ket{p}$ for $p=0,1,...,n-1$ and another $\log(n)$ qubits used to hold superpositions of $\ket{p-1}$ or $\ket{p+1}$.

%

To generate a superposition of $\ket{p}\ket{p+1}$ and $\ket{p}\ket{p-1}$, we can first duplicate the first $\log(n)$ qubits used to hold the superposition of $\ket{p}$ in the second set of $\log(n)$ qubits using CNOT gates. We can then use the superposition of $\ket{0}$ and $\ket{1}$ generated in the third single qubit register to conditionally add or subtract 1 from the second set of $\log(n)$ qubits to yield
%
%
%
\begin{equation}
\label{equ:diffusefortruncation}
    \frac{1}{\sqrt{2n}}\sum_{p=0}^{n-1} \ket{p}\ket{p-1} \ket{0} 
    +  \frac{1}{\sqrt{2n}}\sum_{p=0}^{n-1} \ket{p}\ket{p+1} \ket{1}.
\end{equation}


The superposition of $\ket{p}\ket{p+1}$ and $\ket{p}\ket{p-1}$ can be easily uncomputed using the identical circuit block employed to generate such a superposition. 

When $M > 1$, the superposition states, 
\begin{equation}
\begin{split}
\label{equ:diffusefortruncation2}
    &\frac{1}{\sqrt{2nM}}\sum_{p=0}^{n-1} \sum_{m=0}^{M-1}\ket{p}\ket{p-m} \ket{m} \ket{0}
    \\
    +&  \frac{1}{\sqrt{2nM}}\sum_{p=0}^{n-1} \sum_{m=0}^{M-1}\ket{p}\ket{p+m}\ket{m} \ket{1}.
\end{split}
\end{equation}
can be prepared in a similar way. To generate the state, we first generate a superposition of state

\begin{equation}
    \frac{1}{\sqrt{nM}}\sum_{p=0}^{n-1} \sum_{m=0}^{M-1}\ket{p}\ket{m}
\end{equation}
by applying Hadamard gates to $\log(n)+\log(M)$ qubits. We then use another qubit to prepare a superposition of $\ket{0}$ and $\ket{1}$ states which allows us to generate the superposition of both $\ket{p-m}$ and $\ket{p+m}$ on another quantum register of $\log(n)$ qubits for each fixed $p$ and $m$. Note that for $m=0$ the two branches coincide ($\ket{p-m}=\ket{p+m}=\ket{p}$), so the diagonal term would be counted twice. To avoid this double counting, we store $h_{pp}/2$ at the $m=0$ entries of the data lookup table so that the two branches together encode exactly $h_{pp}$; if the diagonal vanishes, the stored value is simply $0$.

The subnormalization factor resulting from \Cref{equ:diffusefortruncation2} is $2nM$, which is proportional to the number of terms $n(2M-1)$ in the truncated one-body Hamiltonian.  It is significantly less than the corresponding $n^2$ subnormalization factor for block encoding a general one-body Hamiltonian when $M \ll n$. The reduction in the number of terms also translates into a reduction in the number of qubits used to construct the $O_B$ oracle with $\epsilon$ precision in \Cref{eq:ob}, which is $m_b=\log\left( \frac{nM}{\epsilon}\right)$ instead of $m_b=\log\left( \frac{n^2}{\epsilon}\right)$. The number of additional $T$ gates and Clifford gates used to perform conditional summations and subtractions for generating the superposition of $\ket{p}\ket{p-m}$ and $\ket{p}\ket{p+m}$ in \Cref{equ:diffusefortruncation2} is
\begin{equation}
\label{equ:arithcost}\mathcal{O}\left(\log\left(\frac{nM}{\epsilon}\right)+\log(n) \right)=\mathcal{O}\left(\log(n)+\log\left(\frac{1}{\epsilon}\right) \right)
\end{equation}
as discussed in~\cite{gidney2018halving}. This additional cost does not contribute to the main complexity of block encoding asymptotically with respect to $n$, given the sparsity oracle requires $\mathcal{O}(n)$ gate complexity and $\epsilon$ required in scientific problems would typically add a small constant $\log(\frac{1}{\epsilon})$. Moreover, the number of additional gates is relatively small compared to the savings in the optimal gate complexity for implementing the structured \spar \ and \amp \ oracles, $\mathcal{O}\left(\sqrt{n^2 \log\left(n/\epsilon\right)}
-\sqrt{nM \log\left(\frac{nM}{\epsilon}\right)}
\right )$, which can be deduced by substituting $L=nM$ and $L=n^2$ into \Cref{equ:generalBEn_OptimalTgate_depth} and taking the difference. The reduction in the T depth is of the same scaling, which
can also be deduced by substituting $L = nM$ and $L = n^2$ into \Cref{equ:generalBEn_OptimalTgate_depth}.

%

%

More gains in gate complexity can be achieved for Hamiltonians that contain two-body interactions. If the number of two-body interaction terms can be reduced from $L=n^4$ in the general case to $L = n^2M^2$ for the $M$ nearest neighbor case, the subnormalization factor is reduced from $\mathcal{O}(n^4)$ to $\mathcal{O}(n^2M^2)$ and the reduction in $T$ gate complexity can be calculated directly from \Cref{equ:generalBEn_OptimalTgate_depth}. For the two-body interaction, we need to generate a superposition of states of the form $\ket{p}\ket{p\pm m}\ket{q}\ket{q \pm m'}$. We begin this process by preparing a superposition over all $\ket{p}\ket{q}$ states, using $2\log(n)$ Hadamard gates, followed by arithmetic circuits to perform the necessary additions and subtractions between $p$ and $m$, or $q$ and $m$. Compared to the one-body interaction case, the number of arithmetic operations is doubled; therefore, the overall additional cost remains $\mathcal{O}\left(\log\left(\frac{nM}{\epsilon}\right)+\log(n) \right)$. Its asymptotic complexity with respect to $n$ is still negligible in the construction of block encoding. The number of additional gates is also smaller than the savings in the implementation of structured \spar \  and \amp \ oracles. The overall saving is on the order $\mathcal{O}\left(\sqrt{n^4 \log\left(n^4/\epsilon\right)}
-\sqrt{n^2M^2 \log\left(\frac{n^2M^2}{\epsilon}\right)}
\right ),$ which can be deduced by substituting $L=n^2M^2$ and $L=n^4$ into \Cref{equ:generalBEn_OptimalTgate_depth} and taking the difference. The reduction in the T depth is of the same scaling, which
can also be deduced by substituting $L = n^2M^2$ and $L = n^4$ into \Cref{equ:generalBEn_OptimalTgate_depth}.


%



%

\begin{figure*}[htbp!]
\centering
\includegraphics[width=1\textwidth]{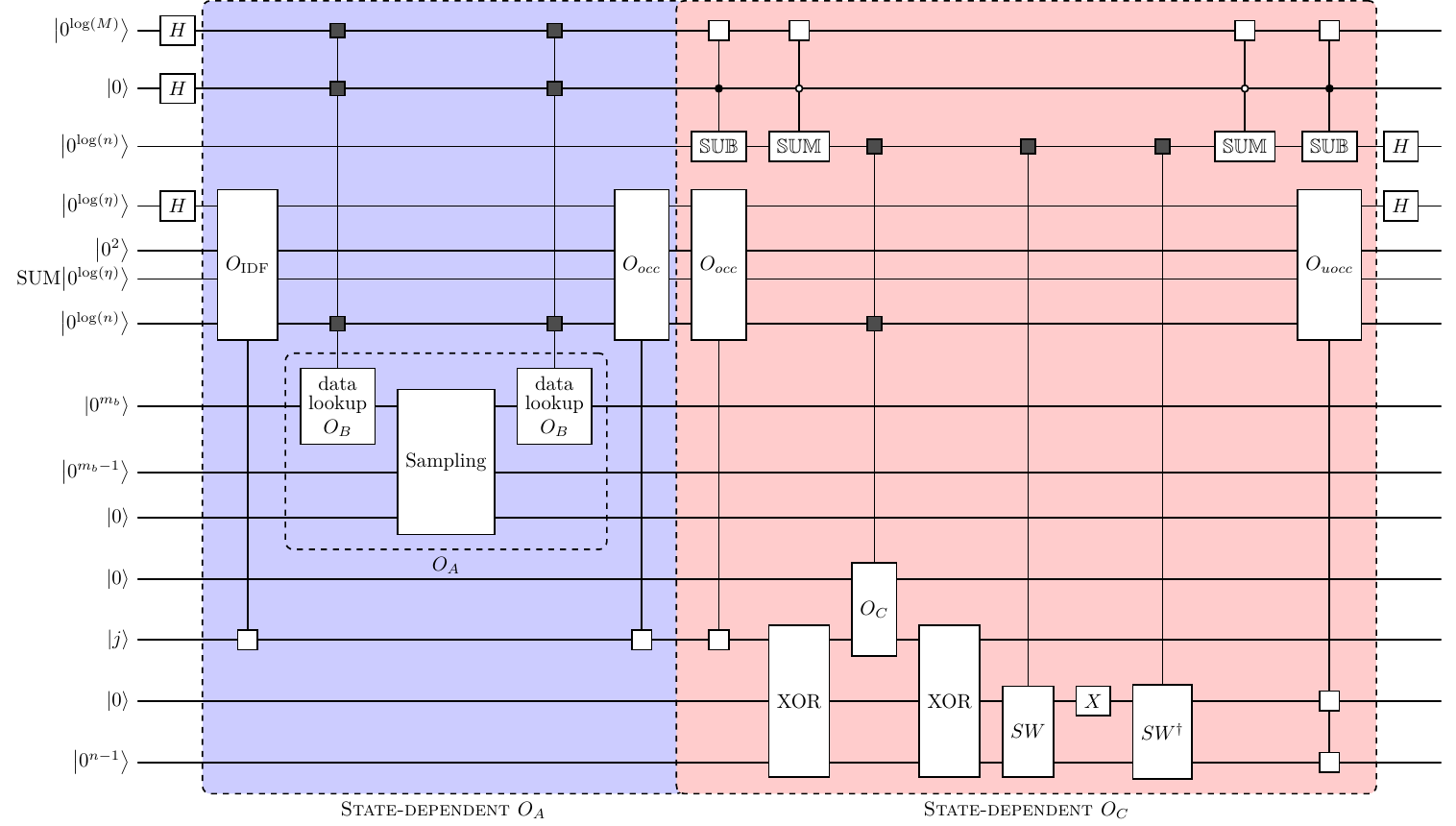}
    \caption{Quantum circuit of block encoding for one-body $\eta$-particle Hamiltonian with nearest-neighbor interactions.}
    \label{fig: NNHeta}
\end{figure*}

If one is interested in block encoding the $\eta$-particle sector of the Hamiltonian with nearest-neighbor interactions, the indirect diffusion scheme introduced in \Cref{sec:eta} can be combined with the method discussed above to prepare the superposition of a subset of $\ket{p}\ket{p-m}$ and $\ket{p}\ket{p+m}$ states for, e.g., the one-body term that does not vanish when $\fc{a}{p}\fan{a}{q}$ is applied to $\ket{j}$'s in which $\eta$ qubits are occupied. The superposition of these states becomes the input to the \amp \ and \spar \ oracles. 

The desired superposition is prepared by first using Hadamard gates and the $O_{\rm IDF}$ oracle discussed in \Cref{sec:input} to obtain

\begin{equation}
\begin{split}
    &\frac{1} {\sqrt{2M\eta}} \sum_{m=0}^{M-1}\sum_{i=0}^{\eta-1}  \ket{m}\ket{0^{\log(n)}}\ket{i}\ket{q_{i}(j)} 
        \ket{0}\ket{j}\\
        +&\frac{1} {\sqrt{2M\eta}} \sum_{m=0}^{M-1}\sum_{i=0}^{\eta-1}  \ket{m}\ket{0^{\log(n)}}\ket{i}\ket{q_{i}(j)} 
\ket{1}\ket{j},
\end{split}
\label{eq:midf}
\end{equation}
where we use the $j$-dependent state $\ket{q_i(j)}$ to represent $\ket{\redrev{\occ{i}{j}}}$ defined in \Cref{eq:occ} to emphasize that it corresponds to valid $q$ indices in the superposition of valid $(p,q)$ pairs. Recall that in the block encoding of an $\eta$-particle Hamiltonian with one-body interaction introduced in \Cref{sec:eta}, the standard procedure begins with the generation of a superposition over all $p$ indices and all valid $q$ indices. Rather than constructing a full superposition over all possible $p$ indices, we here instead produce a superposition over states $\ket{m} = \ket{|p-q|}$ that specify the separation between $p$ and $q$ that guarantees $h_{pq}$ is sufficiently large with respect to a truncation parameter that depends on the overall precision $\epsilon$ of the block encoding circuit. This refinement leverages the locality or decay properties of $h_{pq}$ to reduce circuit complexity, encoding only the necessary $(p,q)$ pairs. As in \Cref{equ:diffusefortruncation2}, the $\ket{0}$ and $\ket{1}$ branches coincide for $m=0$ (both yield $p=q_i$), so the diagonal coefficient stored at the $m=0$ entries of the data lookup table is $h_{q_iq_i}/2$, ensuring the two branches together encode exactly $h_{q_iq_i}$. 
%
The superposition of the $\ket{0}$ and $\ket{1}$ states in \Cref{eq:midf} are generated to perform  conditional additions and subtractions to yield a superposition of  $\ket{q_i(j)-m}$ and $\ket{q_{i}(j)+m}$.

The $O_C$ and $O_A$ oracles take
\begin{equation}
    \begin{split}
    &\frac{1} {\sqrt{2M\eta}} \sum_{m=0}^{M-1}\sum_{i=0}^{\eta-1}  \ket{m}\ket{q_{i}-m}\ket{i}\ket{q_{i}(j)} 
        \ket{0}\ket{j}\\
        &+\frac{1} {\sqrt{2M\eta}} \sum_{m=0}^{M-1}\sum_{i=0}^{\eta-1}  \ket{m}\ket{q_{i}+m}\ket{i}\ket{q_{i}(j)} 
        \ket{1}\ket{j}
\end{split}
\label{eq:qmidf}
\end{equation}
as well as other ancilla qubits as shown in \Cref{fig:onebodyBEeta} as the input to map $\ket{j}$ to $(-1)^{d_{j; q_i \pm m,q_i}}\ket{\mbox{FLIP}(j;q_i \pm m,q_i)}$ (the image of one-body interaction $h_{q_i \pm m, q_i} \fc{a}{q_i\pm m} \fan{a}{q_i}$)  and to encode the selected coefficients via the data lookup and direct sampling scheme discussed in \Cref{sec:sampling}. 

In order to properly uncompute and restore ancilla qubits to $\ket{0}$ states, we need to replicate $\ket{j}$ using the same technique discussed in \cref{sec:state_select_prepare}. In addition, 
conditional addition and subtraction are used to restore  $\ket{q_{i}-m}$ and $\ket{q_{i}+m}$ to $\ket{0^{\log(n)}}$. 

For a one-body Hamiltonian, the nearest-neighbor indirect diffusion produced by \Cref{eq:qmidf} reduces the subnormalization factor from $\mathcal{O}(n^2)$ to $\mathcal{O}(S_\eta)$, where in this case $S_\eta=\Theta(M\eta)$. Because the number of coefficients to be encoded remains $L=nM$, the number of qubits required to encode these coefficients to $\epsilon$ precision in data lookup is $m_b=\log\left( \frac{S_\eta}{\epsilon}\right)$.
It follows from~\cref{eq:eta-T-lambda} and \cref{eq:eta-Tdepth-lambda} that the number of $T$ gates required in the block encoding circuit is  
\begin{equation}
\label{equ:generalBEn_Tgate_count_NN_eta}
    \mathcal{O}\left(n\log(\eta)+\left \lceil\frac{nM}{\lambda}\right\rceil+ \lambda \log\left( \frac{M \eta}{\epsilon}\right)\right),
\end{equation}
and the $T$ depth is 
 \begin{equation}
 \label{equ:general_BEn_Tdepth_count_NN_eta}
     \mathcal{O}\left(n\log(\eta)+\frac{nM}{\lambda}+2\log(\lambda)\right)
 \end{equation}
 with $1\leq \lambda \leq nM$.

 Minimizing the \(\lambda\)-dependent terms in the \(T\)-count gives, up to integer rounding and the constraint \(1 \leq \lambda \leq nM\), the optimal $\lambda$ scales as 
\begin{equation}
    \Theta\left( 
    \sqrt{\frac{nM}{\log\left(M\eta/\epsilon\right)}} \right).
\end{equation}
Substituting this choice into~\cref{equ:generalBEn_Tgate_count_NN_eta} yields the optimized \(T\)-count
\begin{equation}
    \mathcal{O}\left(
        n\log(\eta)
        +
        \sqrt{nM\log\left(\frac{M\eta}{\epsilon}\right)}
    \right).
\end{equation}
The corresponding \(T\)-depth, obtained from~\cref{equ:general_BEn_Tdepth_count_NN_eta}, has the same leading contribution, up to lower-order logarithmic terms.


The circuit that implements the block encoding of this structured Hamiltonian is illustrated in~\Cref{fig: NNHeta}.


\subsection{Hamiltonian with translation invariance}
\label{sec:translation_invariance}


Translation invariance is a key property in many physical systems, which may greatly simplify block encoding constructions. For Hamiltonians with translation invariance, fewer coefficients are to be encoded because they are related by spatial symmetry. This special structure results in both a smaller subnormalization factor and a cheaper data lookup oracle for encoding distinct nonzero coefficients. This feature makes translation-invariant Hamiltonians particularly amenable to resource-saving quantum simulation techniques, similar to the advantages seen in nearest-neighbor interaction models.

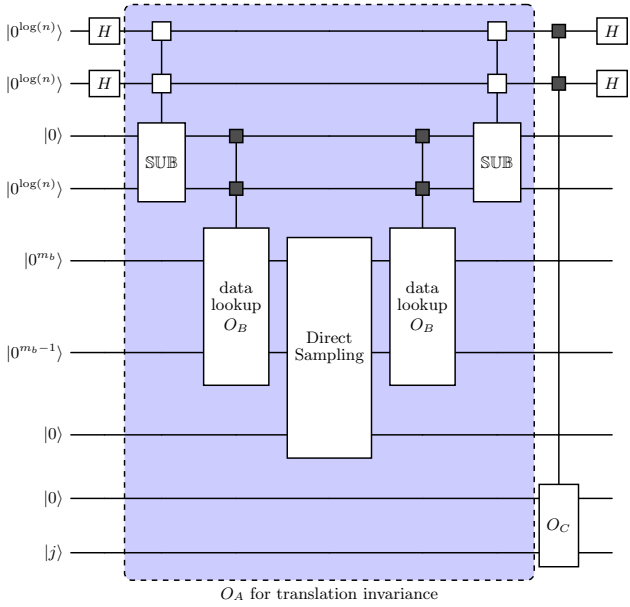
\begin{figure}[htbp!]
\centering
\begin{tikzpicture}
\node[scale=.7]{
   \begin{quantikz}[column sep=0.35cm]
\lstick{$\ket{0^{\log (n)}}$}&\gate{H}&\ctrl[style=U]{1}\gategroup[9,steps=5,style={dashed,rounded corners,fill=blue!20},background,label style={label position=below,anchor=north,yshift=-0.2cm}]{$O_A$ for translation invariance}&&&&\ctrl[style=U]{1}&\ctrl[style=Uf]{1}&\gate{H}\\
\lstick{$\ket{0^{\log (n)}}$}&\gate{H}&\ctrl[style=U]{1}&&&&\ctrl[style=U]{1}&\ctrl[style=Uf]{6}&\gate{H}\\
\lstick{$\ket{0}$}&&\gate[2]{\mathbb{SUB}}&\ctrl[style=Uf]{1}&&\ctrl[style=Uf]{1}&\gate[2]{\mathbb{SUB}}&&\\
\lstick{$\ket{0^{\log (n)}}$}&&&\ctrl[style=Uf]{1}&&\ctrl[style=Uf]{1}&&&\\
\lstick{$\ket{0^{m_b}}$}&&&\gate[2]{\shortstack{data\\
lookup\\$O_B$}}&\gate[3]{\shortstack{Direct\\
Sampling}}&\gate[2]{\shortstack{data\\
lookup\\$O_B$}}&&&\\
\lstick{$\ket{0^{m_b-1}}$}&&&&&&&&\\
\lstick{$\ket{0}$}&&&&&&&&\\
\lstick{$\ket{0}$}&&&&&&&\gate[2]{O_C}&\\
\lstick{$\ket{j}$}&&&&&&&&
\end{quantikz}
};
\end{tikzpicture}
    \caption{Quantum circuit of block encoding for one-body Hamiltonian with translation invariance.}
    \label{fig: TIH_Full}
\end{figure}

To demonstrate how block encoding can be constructed for translation-invariant Hamiltonians, we consider Hamiltonians that contain only one-body interactions, i.e.,
\begin{equation}
    \mathcal{H}= \sum_{p,q=0}^{n-1} T(p-q) \fc{a}{p} \fan{a}{q}.
    \label{eq:hinvar}
\end{equation}
Because there are at most $\mathcal{O}(n)$ nonzero coefficients $T(p-q)$ in \Cref{eq:hinvar}, the $O_A$ oracle can use a data lookup oracle that requires fewer resources. The key to achieve a reduction in complexity is to generate the superposition of $\ket{p-q}$ first by using arithmetic subtraction~\cite{gidney2018halving} and compare oracles~\cite{babbush2018encoding} to yield
%
%
\begin{equation}
\label{equ:TI}
    \frac{1}{n}\sum_{p,q=0}^{n-1} \ket{p} \ket{q}  \ket{p-q}\ket{\text{sgn}(p-q)}.
\end{equation} 

The $O_A$ oracle takes $\ket{p-q}\ket{\text{sgn}(p-q)}$ as the input to retrieve the corresponding $T(p-q)$ value encoded by a direct sampling oracle.

Because the number of distinct coefficients in \eqref{equ:TI} is at most $2n-1$, the $L$ that appears in the T-count estimate \eqref{eq:T-general-lambda} should be replaced by $2n-1$, whereas
$m_b$ is still defined by $m_b = \log\left( \frac{n^2}{\epsilon}\right)$ to achieve $\epsilon$ precision because the Hamiltonian contains $n^2$ terms, with each allowed to have no more than $\epsilon/n^2$ error. It follows that the T-count required to block encode this Hamiltonian is given by 
\begin{equation}
\label{equ:generalBEn_Tgate_count_TI}
    \mathcal{O}\left(n+\left\lceil\frac{n}{\lambda}\right\rceil+ \lambda \log\left( \frac{n}{\epsilon}\right)\right).
\end{equation}
Similarly, it follows from \cref{eq:Tdepth-general-lambda} that the $T$ depth of the block encoding circuit is
 \begin{equation}
\label{equ:general_BEn_Tdepth_count_TI}
\mathcal{O}\left(n+\frac{n}{\lambda}+2\log(\lambda)\right).
 \end{equation}
In this case, the optimal $\lambda$ used in the SELECT--SWAP circuit should be
\begin{equation}
\lambda = \Theta \left( \sqrt{\frac{n}{\log\left(\frac{n}{\epsilon}\right)}}  \right).   
\end{equation}

As a result, the leading order term of the optimal T-count becomes
\begin{equation}
\mathcal{O}\left(\sqrt{n \log \left( \frac{n}{\epsilon}\right)} \right)
\end{equation}
Compared to the T-count obtained from \eqref{eq:mb-general} by setting $L$ to $n^2$, the $T$-count for a translation-invariant Hamiltonian is reduced by
\begin{equation}
    \mathcal{O}\left(
    \sqrt{n^2 \log\left(\frac{n^2}{\epsilon}\right)}
    -
    \sqrt{n \log\left(\frac{n}{\epsilon}\right)}
    \right),
\end{equation}
up to a constant factor. Such a reduction is significantly larger than the additional cost $\mathcal{O}(\log(n)+\log(\frac{1}{\epsilon}))$ shown in \Cref{equ:arithcost}, required to implement the diffusion oracle used to generate the superposition of states in \Cref{equ:TI} using arithmetic subtraction and comparison oracles.
 


%

To restrict the block encoding to the $\eta$-particle sector of the Hamiltonian, we need to combine the techniques discussed above with the indirect diffusion scheme presented in \cref{sec:sampling} to generate
 \begin{equation}
     \frac{1}{\sqrt{n\eta}} \sum_{p=0}^{n-1} \sum_{i=0}^{\eta-1} \ket{p} \ket{i} \ket{q_i} \ket{p-q_i}\ket{\text{sgn}(p-q_i)}\ket{j},
     \label{eq:pmqi}
 \end{equation}
where $q_i(j)$ represents the position of the $i$-th occupied qubit in $\ket{j}$. The superposition represented by \Cref{eq:pmqi} corresponds to a subset of $(p,q)$ pairs that serve as the indices to $T(p-q)$ coefficients associated with valid $a_p^\dagger a_q$'s when applied to a $\eta$-particle state $\ket{j}$.
The $O_A$ oracle then takes $\ket{p-q_i}\ket{\text{sgn}(p-q_i)}$ to retrieve the value of $T(p-q_i)$.

\begin{figure*}[htbp!]
\centering
\includegraphics[width=1\textwidth]{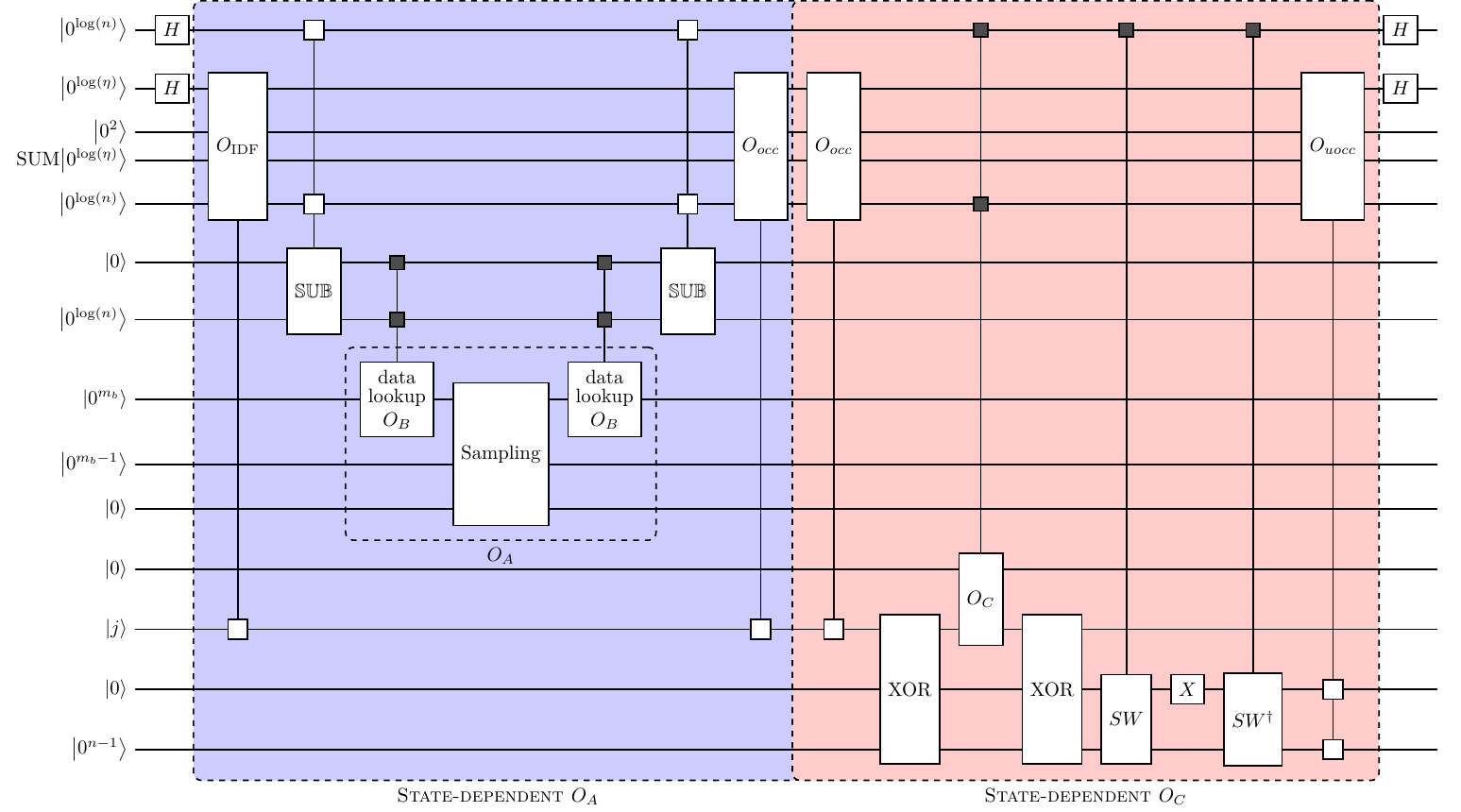}
    \caption{Quantum circuit of block encoding for one-body $\eta$-particle Hamiltonian with translation invariance.}
    \label{fig:eta trans_inv_H}
\end{figure*}

\Cref{fig:eta trans_inv_H} shows the circuit for block encoding a one-body $\eta$-particle Hamiltonian with translation invariance. Notice that,
compared to \Cref{fig:onebodyBEeta}, the additional arithmetic circuit block labeled by SUB is used to generate the superposition of $\ket{p-q_i(j)}$ which is then passed into $O_A$ to retrieve the value of $T(p-q_i)$. 

For this type of Hamiltonian, the number of effective one-body terms for an $\eta$-particle state is $S_{\eta}=\mathcal{O}(n\eta)$. With $L=2n-1$ and 
\begin{equation}
m_b=\log\left(\frac{S_\eta}{\epsilon} \right).
\label{eq:mb4etati}
\end{equation}
%
it follows from \Cref{eq:eta-T-lambda} that the $T$-count for this type of Hamiltonian is 
\begin{equation}
\label{equ:generalBEn_Tgate_count_TI_eta2}
    \mathcal{O}\left(n\log(\eta)+\left \lceil\frac{n}{\lambda}\right\rceil+ \lambda \log\left( \frac{n\eta }{\epsilon}\right)\right)
\end{equation}
and the $T$ depth of the circuit is
\begin{equation}
\label{equ:general_BEn_Tdepth_count_TI_eta2}
    \mathcal{O}\left(n\log(\eta)+\frac{n}{\lambda}+2\log(\lambda)\right),
\end{equation}
 where $1\leq \lambda \leq 2n-1$.  
The optimal $\lambda$ in this case is 
$\Theta \left(\sqrt{\frac{n}{\log\left(\frac{n\eta}{\epsilon}\right)}} \right)$. As a result, the leading order term of the optimal T-count becomes
\begin{equation}
\mathcal{O}\left(\sqrt{n \log\left(\frac{n\eta}{\epsilon}\right)}\right).
\end{equation}

Compared to the general $\eta$-particle block encoding with one-body interaction in~\Cref{sec:eta}, the reduction in optimal gate complexity with this method is $\mathcal{O}\left(\sqrt{n^2 \log\left(\frac{n\eta}{\epsilon}\right)}-\sqrt{n \log\left(\frac{n\eta}{\epsilon}\right)}\right)$, it is significantly larger than the additional cost $\mathcal{O}(\log(n)+\log(\frac{1}{\epsilon}))$ incurred in the subtraction and the comparison oracle.

%

\subsection{Hamiltonians that can be factorized as products of number operators}
\label{sec:number_operator_product}

If the two-body component of the Hamiltonian in \Cref{eq:2quant}%
 only contains terms in which $p=r$ and $q=s$, the Hamiltonian can be simplified to 
\begin{equation}
\H = \sum_{\substack{p,q=0\\p\neq q}}^{n-1} V_{pq} n_{p} n_{q},
\label{eq:Vpq}
\end{equation}
i.e., the two-body term can be written as a sum of products of number operators.

The \spar \ oracle for a product of two number operators can be greatly simplified because no phase or bit flip needs to be considered. Applying the $O_C$ oracle for $n_p n_q$ where $p\neq q$ to $\ket{0} \ket{p} \ket{q}\ket{j}$ yields
\begin{equation}
   \ket{o(j;p,q)} \ket{p} \ket{q}\ket{j},
\end{equation}
where $o(j;p,q)=0$ if $j_p = j_q=1$ and $o(j;p,q)=1$ otherwise. To implement the \spar \ oracle, we employ a SWAP-UP circuit to verify whether both $\ket{j_p} = \ket{1}$ and $\ket{j_q} = \ket{1}$. Following the convention in \Cref{sec:oc}, the validation qubit takes $\ket{0}$ as the input and is first set to $\ket{1}$ by the first $X$ gate within $O_C$; if this condition is satisfied, it is then flipped from $\ket{1}$ back to $\ket{0}$.
This step requires $\mathcal{O}(n)$ $T$ gates and generates a circuit of $\mathcal{O}(\log(n))$ T depth. Because \Cref{eq:Vpq} contains $L=n^2$ terms with potentially all distinct coefficient $V_{pq}$, $m_b=\log(\frac{n^2}{\epsilon})$. Consequently, it follows from  \Cref{eq:T-general-lambda} and \Cref{eq:Tdepth-general-lambda} that the \spar \ oracle and \amp \ oracles require
\begin{equation}
    \mathcal{O}\left(n+\frac{n^2}{\lambda}+\lambda \log\left(\frac{n^2}{\epsilon} \right)\right)
\end{equation}
$T$ gates. The $T$ depth of the circuit is 
\begin{equation}
    \mathcal{O}\left(n+\frac{n^2}{\lambda}+\log(\lambda)\right).
\end{equation}
The optimal $\lambda$ is $\lambda = \Theta \left( \sqrt{\frac{n^2 }{\log \left ( \frac{n^2}{\epsilon}\right)}} \right)$. As a result, the leading order term of the optimal T-count is
\begin{equation}\mathcal{O}\left( \sqrt{n^2\log \left ( \frac{n^2}{\epsilon}\right)}\right).
\end{equation}

To restrict the block encoding to the $\eta$-particle sector of the Hamiltonian, we first generate a superposition of valid index pairs through indirect diffusion for an $\eta$-particle state,
\begin{equation}
    \frac{1}{\sqrt{\eta^2}} \sum_{i_1=0}^{\eta-1} \sum_{i_2=0}^{\eta-1}\ket{i_1}\ket{i_2} \ket{p_{i_1}(j)} \ket{q_{i_2}(j)}\ket{j},
\end{equation}
where we use $p_{i_1}(j)$ to represent the bit index of the $i_1$-th occupied qubit in $\ket{j}$ and $q_{i_2}(j)$ to represent the bit index of the $i_2$-th occupied qubit in $\ket{j}$. We use the SWAP-UP circuit to verify whether both $\ket{j_{p_{i_1}}}=\ket{1}$ and $\ket{j_{q_{i_2}}}=\ket{1}$ hold. The validation qubit takes $\ket{0}$ as the input and is first set to $\ket{1}$ by the first $X$ gate within $O_C$; if the condition is satisfied, we flip it from $\ket{1}$ back to $\ket{0}$ to obtain

\begin{equation*}
\begin{split}
    \frac{1}{\sqrt{\eta^2}} \sum_{i_1=0}^{\eta-1} \sum_{i_2=0}^{\eta-1}\ket{o(j;i_1,i_2)}\ket{i_1}\ket{i_2} \ket{p_{i_1}(j)} \ket{q_{i_2}(j)}\ket{j}.
\end{split}
\end{equation*}
We note that the diagonal terms with $i_1 = i_2$ (equivalently $p_{i_1}(j) = q_{i_2}(j)$), which are excluded in \Cref{eq:Vpq}, pass this validation check since both indices always point to occupied orbitals. Rather than enforcing $i_1 \neq i_2$ with an additional equality check, these terms can simply be removed by setting the corresponding diagonal entries $V_{pp}$ to zero in the lookup table of the \amp\ oracle, at no additional circuit cost. For this product of number operators, the state-dependent label count is $S_\eta=\Theta(\eta^2)$, since both indices must correspond to occupied orbitals in $\ket{j}$. The resulting block encoding of the $\eta$-particle sector requires
\begin{equation}
    \mathcal{O}\left(n\log(\eta)+\frac{n^2}{\lambda}+\lambda \log(\frac{S_\eta}{\epsilon})  \right)
\end{equation}
$T$ gates and generates a circuit of 
\begin{equation}
    \mathcal{O}\left(n\log(\eta)+\frac{n^2}{\lambda}+\log(\lambda)\right)
\end{equation}
T depth, deduced from \Cref{eq:eta-T-lambda} and \Cref{eq:eta-Tdepth-lambda} with $L=n^2$ and $m_b=\log(\frac{S_\eta}{\epsilon})$. The optimal $\lambda$ is $\lambda=\Theta\left (\sqrt{\frac{n^2}{\log\left( \frac{\eta^2}{\epsilon}\right)}} \right)$ and the leading order term of T-count and T-depth is 
\begin{equation}
    \mathcal{O}\left(\sqrt{n^2\log\left( \frac{\eta^2}{\epsilon}\right)}  \right).
\end{equation}

\subsection{Examples}
\label{sec:model}
In this section, we give a few examples of model Hamiltonians that have the special structures described above and show how techniques discussed above can be used to reduce the gate complexity of the block encoding circuits for these  Hamiltonians. 


We also translate the input-model estimates into simulation-level costs and compare them with representative quantum-dynamics simulation approaches for fixed-$\eta$ particle states.  

\subsubsection{Hubbard model}
In this section, we discuss the block encoding for Hubbard model, which serves as a prototypical many-electron system that captures a variety of correlated-electron phenomena. Its Hamiltonian is given by
\begin{equation}
\label{equ:Hhubbard}
\H = - t \sum_{\langle p,q \rangle}^{} \sum_{\sigma\in \{ \uparrow, \downarrow \}}  a_{p,\sigma}^\dagger a_{q,\sigma} + U\sum_{p=0}^{n-1}  n_{p,\uparrow}n_{p,\downarrow},
\end{equation}  
where the notation $\langle p,q \rangle$ implies that the two sites labeled by $p$ and $q$ are adjacent. The parameters $t$ and $U$ are nonzero constants. This model is a linear combination of one special case of nearest-neighbor Hamiltonian as discussed in \Cref{sec:translation_invariance} and product of number operators as discussed in \Cref{sec:number_operator_product}. It will be referred to as Hubbard Model in \Cref{tab:cost2}.

One key feature of the model is that there are only two unique coefficients, corresponding to the one-body and two-body interactions respectively. The data lookup oracle can therefore be greatly simplified. In particular, in the quantum circuit for one-body nearest-neighbor interaction illustrated in \Cref{fig: HAM_with_decaying_coef}, the controlled data lookup oracle $O_B$ can be replaced directly by a sequence of CNOT gates on the quantum register $\ket{0^{m_b}}$.

To construct the amplitude oracle using this simplified approach, we first generate

\begin{equation}
    \frac{1}{\sqrt{2}} \ket{0} \ket{0^{m_b}}\ket{0^{m_b}}+ \frac{1}{\sqrt{2}} \ket{1} \ket{0^{m_b}}\ket{0^{m_b}}.
\end{equation}

The first qubit is used to implement a linear combination of two types of interactions in \Cref{equ:Hhubbard}. The second quantum register, consisting of $m_b$ qubits, is used to load the data corresponding to either $-t$ or $U$. The last quantum register, also with $m_b$ qubits, together with the second register, is employed to perform direct sampling as shown in \Cref{eq:rb}. When the control qubit is in the state $\ket{0}$, the CNOT gates load the binary representation $b_{-t}$ for the amplitude $-t$, whereas when it is in the state $\ket{1}$, they load the binary representation $b_{U}$ for $U$,
\begin{equation}
    \frac{1}{\sqrt{2}} \ket{0} \ket{b_{-t}}\ket{0^{m_b}}+ \frac{1}{\sqrt{2}} \ket{1} \ket{b_{U}}\ket{0^{m_b}}.
\end{equation}
The direct sampling method can then be applied without invoking the data lookup oracle. In particular, we note that the simplified amplitude oracle functions identically for the block encoding of both the full Hamiltonian and the $\eta$-particle sector.

The main $T$ gate and T depth complexity of this simplified amplitude oracle is $\mathcal{O}(m_b)$ where $m_b$ scales as $\mathcal{O}(\log(\frac{n}{\epsilon}))$. Replace the complexity coming from data-lookup oracle in \cref{eq:general_es_complexity}
with this estimate, we can derive that the block encoding for full Hubbard model requires 
\begin{equation}
\mathcal{O}\left(n+\log\left(\frac{n}{\epsilon}\right)\right)
\end{equation}
$T$ gates. The T depth of the circuit has the same scaling. 




Using the fixed-particle construction in \Cref{sec:eta}, the corresponding block encoding for the $\eta$-particle sector requires

\begin{equation}
\mathcal{O}\left(n\log(\eta)+\log\left(\frac{\eta}{\epsilon}\right)\right)
\end{equation}
$T$ gates and T depth of same scaling.

\subsubsection{Nearest neighbor Hubbard model}

In this section, we examine the block encoding circuit for the nearest-neighbor Hubbard model, which serves as a prototypical many-electron system that captures a variety of correlated-electron phenomena. Its Hamiltonian is given by
\begin{equation}
\H = - \sum_{p,q=0}^{n-1} \sum_{\sigma\in \{ \uparrow, \downarrow \}} T(p-q) a_{p,\sigma}^\dagger a_{q,\sigma} + U\sum_{p=0}^{n-1}  n_{p,\uparrow}n_{p,\downarrow},
\end{equation}  
where \( T(p-q) \) is non-zero only for nearest neighbors, i.e., when \( |p - q| \leq M-1 \). This model is one special case as discussed in \Cref{sec:translation_invariance} and will be referred to as the nearest-neighbor Hubbard Model in \Cref{tab:cost2}.

Because the non-zero coefficients of the one-body term are translation-invariant and limited to nearest neighbors on the lattice, we can use techniques discussed in both \Cref{sec:BEnNN} and   \Cref{sec:translation_invariance} to reduce the gate complexity of the block encoding circuit.

Recall the construction in \cref{sec:BEnNN}, a key step is to use \Cref{equ:diffusefortruncation2} to generate a superposition of $\ket{p}\ket{p+m}$ and $\ket{p}\ket{p-m}$.
To take advantage of translation invariance, we do not need to use an amplitude oracle that takes $\ket{p}\ket{p-m}$ as the input. Instead, we can construct an $O_A$ to only take $\ket{m}\ket{0}$ or $\ket{m}\ket{1}$ as the input to retrieve the value $T(p-q)=T(\pm m)$. The number of data items stored in the data lookup oracle is $2M$ instead of $n^2$. Since the Hamiltonian contains $nM$ terms, $m_b$ must be set to $m_b = \log \left(\frac{nM}{\epsilon}\right)$ to ensure $\epsilon$ accuracy of the block encoding. 
It follows from \Cref{eq:T-general-lambda} with $L$ replaced by $2M$ that the T-count for this Hamiltonian is 
\begin{equation} \mathcal{O}\left(n+\frac{M}{\lambda}+\lambda \log\left(\frac{nM}{\epsilon} \right)\right). 
\end{equation} 
The optimal $\lambda$ in this case is $ \Theta\left(\sqrt{\frac{M}{\log\left( \frac{nM}{\epsilon}\right)}}\right) $.

As a result, the leading order of the optimal T-count is $\mathcal{O}\left(\sqrt{M\log\left( \frac{nM}{\epsilon}\right)}\right)$.


For the $\eta$-particle sector of the Hamiltonian, we use $S_\eta = \eta M$ to obtain $m_b = \log\left( \frac{\eta M}{\epsilon}\right)$.
With the number of data stored in the lookup table equals $2M$, it follows from \Cref{eq:eta-T-lambda} and \Cref{eq:eta-Tdepth-lambda} that the T-count for this Hamiltonian is
\begin{equation}   O\!\left(
    n\log\eta
    +
    \frac{M}{\lambda}
    +
    \lambda    \log\frac{\eta M}{\epsilon}
    \right),
\end{equation}
and the T depth for this Hamiltonian is 
\begin{equation}
    O\!\left(
    n\log\eta
    +
    \frac{M}{\lambda}
    +
    2\log\lambda
    \right)
\end{equation}

The optimal $\lambda$ is $\Theta\left( \sqrt{\frac{M}{\log(\frac{\eta M}{\epsilon})}} \right)$ for reducing the T count.
As a result, the leading order of the optimal T-count and the corresponding T-depth is
\begin{equation}
\mathcal{O}\left(n\log(\eta)+\sqrt{ M\log\left( \frac{\eta M}{\epsilon}\right)}\right).
\end{equation} 

\subsubsection{Molecular Electronic Hamiltonian with localized orbitals}

In this section, we show how the techniques presented earlier can be used to construct an efficient block encoding circuit for the electronic structure Hamiltonian of a molecular system in quantized form. We assume molecular orbitals $\{\varphi_{\ell} \}$ can be localized in space, i.e., for each spin-orbital index $\ell$, there exists a vector $\vec{c}_\ell \in \mathbb{R}^3$ (called the center of $\varphi_\ell$) such that, 
%

%

\begin{equation}
\label{equ:lorbital_cond2}
|\varphi_\ell(\vec{r})| \leq \varphi_{\text{max}} \exp\left( -\alpha \|\vec{r} - \vec{c}_\ell\| \right),
\end{equation}
for some $\varphi_{\text{max}}$ which is a finite upper bound of $|\varphi_\ell(\vec{r})|$ for all $\ell$ in $\mathbb{R}^3$.



A direct consequence of orbital localization is that the two electron integral defined by
\begin{equation}
\label{equ:hcoe}
h_{pqrs} = \int_{\mathbb{R}^3} \int_{\mathbb{R}^3} \frac{\varphi_p^*(\vec{r}_1) \varphi_q^*(\vec{r}_2) \varphi_r(\vec{r}_1) \varphi_s(\vec{r}_2)}{\|\vec{r}_1 - \vec{r}_2\|} \, \text{d}\vec{r}_1 \, \text{d}\vec{r}_2,
\end{equation}
which are the coefficients of the two-body interaction terms in the second-quantized Hamiltonian can have a small magnitude when the centers of the localized orbitals are far from each other.
%
%
%
As a result, it is not necessary to encode all these coefficients to obtain a block encoding with $\epsilon$ precision. 

If, for each orbital, the number of orbitals that have sufficient overlap with itself is limited to $M$, then the number of $h_{pqrs}$ we need to consider is limited to $L = n^2 M^2$, which can be significantly less than $L = n^4$ if $M \ll n$. Such a reduction translates directly into a much smaller subnormalization factor as well as the size of $m_b$ required in the data lookup circuit block. In fact, if we can achieve $\epsilon$ accuracy by considering only $L=n^2 M^2$ elements of the $h_{pqrs}$ tensor, $m_b$ can be reduced to 
\begin{equation}
m_b =  \log\left(\frac{n^2M^2}{\epsilon}\right)
\label{eq:mbmol}
\end{equation}
from $m_b =\log\left(\frac{n^4}{\epsilon}\right)$ in the most general case in which no localization of the molecular spin orbitals is considered.

Because the number of electrons for a molecular system is fixed at $\eta$,
the number of two-body interaction terms that contribute to the action of the Hamiltonian on a given $\eta$-particle basis state is further reduced: the annihilation indices must correspond to occupied orbitals, so the state-dependent label count becomes $S_\eta = \Theta(\eta^2 M^2)$, which sets the subnormalization factor to $\mathcal{O}(\eta^2M^2)$ and allows $m_b$ to be reduced to $m_b = \log\left(\frac{\eta^2M^2}{\epsilon}\right)$ from \Cref{eq:mbmol}. We emphasize that, as discussed in \Cref{sec:input}, the data-lookup table itself must still store all $n^2M^2$ retained coefficient entries, since the occupied orbitals depend on the input state $\ket{j}$; the number of stored entries is therefore not reduced by fixing the particle number.

For a system with one-dimensional structure, an $\epsilon$-precision block encoding of the full Hamiltonian can be obtained by choosing $M=\mathcal{O}(\log(\frac{n^2}{\epsilon}))$, so that the truncated Hamiltonian remains $\epsilon$ accurate.
For an $\eta$-particle Hamiltonian, we can choose $M=\mathcal{O}(\log(\frac{\eta^2}{\epsilon}))$ for the same purpose. The resulting block-encoding complexity is summarized in \Cref{tab:cost-combined} under the label Localized.

\subsubsection{Extended Hubbard model}

In this section, we study the block encoding of the extended Hubbard model,
\begin{equation}
\H = - \sum_{p,q=0 }^{n-1} T_{pq} a_{p}^\dagger a_{q} + \sum_{p=0}^{n-1} U_p n_{p} +  \sum_{\substack{p,q=0\\p\neq q}}^{n-1} V_{pq} n_{p} n_{q}.
\end{equation}
This model is referred to as \textit{Factorized} below. The Hamiltonian is a linear combination of a general one-body term, a one-body number-operator term, and a product of number operators. Therefore, the block encoding of $\H$ can be constructed by combining the block encodings for these three types of interactions.

For the full Hamiltonian, the dominant data-lookup cost comes from the arbitrary coefficients $T_{pq}$ and $V_{pq}$. Hence the number of data entries in the lookup oracle is $L=\mathcal{O}(n^2)$, and it is sufficient to choose
\begin{equation}
    m_b=\mathcal{O}\left(\log\left(\frac{n^2}{\epsilon}\right)\right)
\end{equation}
to obtain an $\epsilon$-precision block encoding. It follows from \Cref{eq:T-general-lambda,eq:Tdepth-general-lambda} that the $T$-count for this Hamiltonian is
\begin{equation}
    \mathcal{O}\left(n+\frac{n^2}{\lambda}
    +\lambda\log\left(\frac{n^2}{\epsilon}\right)\right),
\end{equation}
and the corresponding $T$ depth is
\begin{equation}
    \mathcal{O}\left(\frac{n^2}{\lambda}+2\log(\lambda)\right),
\end{equation}
where $1\leq \lambda\leq n^2$. Optimizing the $T$-count gives
\begin{equation}
    \lambda=\Theta\left(\frac{n}{\sqrt{\log(n^2/\epsilon)}}\right),
\end{equation}
and hence the leading order term of the optimal $T$-count is
\begin{equation}
    \mathcal{O}\left(n+n\sqrt{\log\left(\frac{n^2}{\epsilon}\right)}\right).
\end{equation}
With this choice of $\lambda$, the corresponding $T$ depth is
\begin{equation}
    \mathcal{O}\left(n\sqrt{\log\left(\frac{n^2}{\epsilon}\right)}\right).
\end{equation}
This is significantly less than the complexity given for a general second-quantized Hamiltonian in \Cref{eq:general_es_complexity}, where $L=\Theta(n^4)$ and the optimized oracle cost scales as $\mathcal{O}\left(n^2\sqrt{\log(n^4/\epsilon)}\right)$.

The block encoding for the $\eta$-particle Hamiltonian has subnormalization factor $\mathcal{O}(n\eta)$ based on the discussion in \Cref{sec:eta}. 
Substituting $L=\mathcal{O}(n^2)$ and $S_\eta=\mathcal{O}(n\eta)$ into \Cref{eq:eta-T-lambda,eq:eta-Tdepth-lambda} gives
\begin{equation}
    \mathcal{O}\left(n\log(\eta)+\frac{n^2}{\lambda}
    +\lambda\log\left(\frac{n\eta}{\epsilon}\right)\right)
\end{equation}
$T$ gates and
\begin{equation}
    \mathcal{O}\left(n\log(\eta)+\frac{n^2}{\lambda}
    +2\log(\lambda)\right)
\end{equation}
$T$ depth. The optimal choice for reducing the $T$-count is
\begin{equation}
    \lambda=\Theta\left(\frac{n}{\sqrt{\log(n\eta/\epsilon)}}\right),
\end{equation}
which gives the optimized $T$-count
\begin{equation}
    \mathcal{O}\left(n\log(\eta)+n\sqrt{\log\left(\frac{n\eta}{\epsilon}\right)}\right),
\end{equation}
with the same leading scaling for the corresponding $T$ depth. These substitutions give the factorized-model estimate before imposing translation invariance; the translation-invariant factorized specialization below corresponds to the reduced-lookup entries reported in the rows labeled \textit{TI Factorized} in \Cref{tab:cost1-full,tab:cost-combined-eta,tab:cost3}. 

We further consider the extended Hubbard model with translation-invariant one-body and density-density interactions, referred to as \textit{TI Factorized} below, where
\begin{equation}
\begin{split}
    \H =& - \sum_{p,q } T(p-q) a_{p}^\dagger a_{q} + \sum_{p} U_p n_{p} \\ &+  \sum_{p\neq q} V(p-q) n_{p} n_{q}.
\end{split}
\end{equation}
We first discuss the block encoding for the full Hamiltonian. The subnormalization factor remains $\mathcal{O}(n^2)$ and the requirement for $m_b$ remains $\mathcal{O}(\log(\frac{n^2}{\epsilon}))$ in order to obtain $\epsilon$ precision for block encoding. However, following the discussion in \Cref{sec:translation_invariance}, the number of classical data entries stored in the data-lookup oracle is reduced to $L=\mathcal{O}(n)$. Then \Cref{eq:T-general-lambda,eq:Tdepth-general-lambda}  
gives
\begin{equation}
    \mathcal{O}\left(n+\frac{n}{\lambda}
    +\lambda\log\left(\frac{n^2}{\epsilon}\right)\right)
\end{equation}
$T$ gates and
\begin{equation}
    \mathcal{O}\left(\frac{n}{\lambda}+2\log(\lambda)\right)
\end{equation}
$T$ depth. The optimal choice is
\begin{equation}
    \lambda=\Theta\left(\sqrt{\frac{n}{\log(n^2/\epsilon)}}\right),
\end{equation}
which yields an optimized $T$-count
\begin{equation}
    \mathcal{O}\left(n+\sqrt{n\log\left(\frac{n^2}{\epsilon}\right)}\right)
\end{equation}
and corresponding $T$ depth
\begin{equation}
    \mathcal{O}\left(\sqrt{n\log\left(\frac{n^2}{\epsilon}\right)}\right).
\end{equation}


The block encoding for the $\eta$-particle Hamiltonian has a subnormalization factor $\mathcal{O}(n\eta)$ based on the discussion in \Cref{sec:eta}. The overall complexity is obtained through \Cref{eq:eta-T-lambda,eq:eta-Tdepth-lambda} with

\begin{equation}
    L=\mathcal{O}(n), \ m_b = \mathcal{O}\left(\log\left(\frac{n\eta}{\epsilon}\right)\right).
\end{equation}
Substituting these values into \Cref{eq:eta-T-lambda,eq:eta-Tdepth-lambda} gives
\begin{equation}
    \mathcal{O}\left(n\log(\eta)+\frac{n}{\lambda}
    +\lambda\log\left(\frac{n\eta}{\epsilon}\right)\right)
\end{equation}
$T$ gates and
\begin{equation}
    \mathcal{O}\left(n\log(\eta)+\frac{n}{\lambda}+2\log(\lambda)\right)
\end{equation}
$T$ depth. Optimizing $\lambda$ gives
\begin{equation}
    \mathcal{O}\left(n\log(\eta)+\sqrt{n\log\left(\frac{n\eta}{\epsilon}\right)}\right)
\end{equation}
for both the leading $T$-count and the corresponding $T$ depth. These substitutions give the translation-invariant factorized entries reported in the rows labeled \textit{TI Factorized} in \Cref{tab:cost1-full,tab:cost-combined-eta,tab:cost3}.

\section{Comparison with other input models}
\label{sec:compare}

This section compares the resource estimates of the block-encoding circuits developed in this work with representative existing input models and simulation constructions. The closest baseline is the LCU-based qubitization input model of Babbush et al.~\cite{babbush2018encoding} for second-quantized electronic-structure Hamiltonians. We also include the Trotter-based low-depth simulation construction of Kivlichan et al.~\cite{kivlichan2018quantum}; that work does not provide a full PREPARE/SELECT input model, so its qubit and subnormalization entries are marked as not applicable. Finally, we compare with the SWAP-UP-based fermionic \sel construction of Wan~\cite{wan2021exponentially}; the SELECT--SWAP lookup primitive used for coefficient data follows Low et al.~\cite{low2018trading}.

We compare qubit count, subnormalization factor, $T$-gate count, and $T$ depth. The first comparison treats electronic-structure Hamiltonians in the full Hilbert space. We then discuss the fixed-$\eta$ particle sector, followed by Hubbard-type examples and simulation-level costs.

\subsection{Electronic-structure Hamiltonian in the full Hilbert space}

We begin with the general second-quantized electronic Hamiltonian in~\Cref{eq:2quant}, without imposing a fixed-particle constraint. For a generic two-body Hamiltonian, the number of formal interaction terms is $L=\Theta(n^4)$. As derived in~\Cref{eq:general_es_complexity}, the present construction has qubit count, $T$ count, and $T$ depth
\begin{equation}
  \mathcal{O}\!\left(n^2\sqrt{\log\!\left(\frac{n^4}{\epsilon}\right)}\right),
\end{equation}
while the subnormalization factor remains $\alpha=\mathcal{O}(n^4)$ in the full Hilbert space. The improvement over a term-by-term LCU input model is therefore not a smaller full-space normalization, but a lower oracle-realization cost obtained by replacing linear coefficient loading with SELECT--SWAP lookup and direct sampling.

For translation-invariant factorized Hamiltonians, the number of distinct coefficients is much smaller. In this case, the symmetry-aware lookup construction gives qubit and $T$-count scaling $\mathcal{O}(n+\sqrt{n\log(n^2/\epsilon)})$ and $T$ depth $\mathcal{O}(\sqrt{n\log(n^2/\epsilon)})$, while preserving the same $\mathcal{O}(n^2)$ subnormalization scaling as the comparable Babbush et al. input model. For localized electronic Hamiltonians, the present construction gives an explicit end-to-end PREPARE/SELECT-style cost model: the qubit, $T$-count, and $T$-depth costs scale as $\tilde{\mathcal{O}}(n\sqrt{\log^3(n^2/\epsilon)})$, with subnormalization $\mathcal{O}(n^2\log^2(n^2/\epsilon))$.

\begin{table*}[htbp!]
\centering
\def\arraystretch{1.75}
\begin{tabular}{c|c|c|c|c|c}
  Model & Reference & Qubit  & Subnormalization factor & $T$ count & $T$ depth  \\
  \hline
  \multirow{2}{*}{General} 
    & Babbush et al.~\cite{babbush2018encoding} 
      & $\mathcal{O}(n+\log(\tfrac{n^4}{\epsilon}))$ 
      & $\mathcal{O}(n^4)$ 
      & $\mathcal{O}(n^4+\log(\tfrac{n^4}{\epsilon}))$ 
      & $\mathcal{O}(n^4)$ \\
  \cline{2-6} 
    & \textbf{This work} 
      & $\mathcal{O}(n^2\sqrt{\log(\tfrac{n^4}{\epsilon})})$  
      & $\mathcal{O}(n^4)$ 
      & $\mathcal{O}(n^2\sqrt{\log(\tfrac{n^4}{\epsilon})})$ 
      & $\mathcal{O}(n^2\sqrt{\log(\tfrac{n^4}{\epsilon})})$ \\
  \Xhline{1.5pt}
  \multirow{2}{*}{Factorized}
    & Kivlichan et al.~\cite{kivlichan2018quantum} 
      & N/A & N/A 
      & $\mathcal{O}(n^2\log(\tfrac{1}{\epsilon}))$ 
      & $\mathcal{O}(n)$ \\
  \cline{2-6}
    & \textbf{This work}
      & $\mathcal{O}(n+n\sqrt{\log(\tfrac{n^2}{\epsilon})})$
      & $\mathcal{O}(n^2)$
      & $\mathcal{O}(n+n\sqrt{\log(\tfrac{n^2}{\epsilon})})$
      & $\mathcal{O}(n\sqrt{\log(\tfrac{n^2}{\epsilon})})$ \\
  \Xhline{1.5pt}
  \multirow{2}{*}{TI Factorized} 
    & Babbush et al.~\cite{babbush2018encoding} 
      & $\mathcal{O}(n+\log(\tfrac{n^2}{\epsilon}))$ 
      & $\mathcal{O}(n^2)$ 
      & $\mathcal{O}(n+\log(\tfrac{n^2}{\epsilon}))$ 
      & $\mathcal{O}(n)$ \\
  \cline{2-6} 
    & \textbf{This work} 
      & $\mathcal{O}(n+\sqrt{n \log(\tfrac{n^2}{\epsilon})})$ 
      & $\mathcal{O}(n^2)$ 
      & $\mathcal{O}(n+\sqrt{n\log(\tfrac{n^2}{\epsilon})})$ 
      & $\mathcal{O}(\sqrt{n \log(\tfrac{n^2}{\epsilon})})$ \\
  \Xhline{1.5pt}
  \multirow{2}{*}{Localized} 
    & Wan~\cite{wan2021exponentially} 
      & $\mathcal{O}(n)$  
      & $\mathcal{O}(n^2)$ 
      & $\mathcal{O}(n)+\text{PREP}$  
      & $\mathcal{O}(n)+\text{PREP}$ \\
  \cline{2-6}
    & \textbf{This work} 
      & $\tilde{\mathcal{O}}(n\sqrt{\log^3(\tfrac{n^2}{\epsilon})})$  
      & $\mathcal{O}(n^2 \log^2(\tfrac{n^2}{\epsilon}))$ 
      & $\tilde{\mathcal{O}}(n\sqrt{\log^3(\tfrac{n^2}{\epsilon})})$ 
      & $\tilde{\mathcal{O}}(n\sqrt{\log^3(\tfrac{n^2}{\epsilon})})$ \\
\end{tabular}
\caption{Complexity comparison of input models for electronic-structure Hamiltonians in the full Hilbert space.}
\label{tab:cost1-full}
\end{table*}

\subsection{Complexity comparison for Hamiltonian represented in an $\eta$-particle sector}
We next compare complexity in the fixed-particle sector, again using Babbush et al.~\cite{babbush2018encoding} as the baseline where applicable.
In~\Cref{tab:cost-combined-eta}, we report the cost of the circuits developed in this work for implementing a block encoding of the Hamiltonian~\Cref{eq:2quant} restricted to the $\eta$-particle subspace. Such a restriction changes the scaling in all four resource columns.

The main difference from the full-space construction is that the relevant normalization is controlled by the number $S_\eta$ of state-dependent labels generated by indirect diffusion, rather than by the total number $L$ of formal Hamiltonian coefficients stored in the data-lookup oracle. For a general two-body electronic Hamiltonian, $L=\Theta(n^4)$, while the two annihilation indices must be chosen from the occupied orbitals of the input $\eta$-particle basis state. Hence $S_\eta=\Theta(n^2\eta^2)$, and the full-space $n^4$ factor in~\Cref{tab:cost1-full} is replaced by $n^2\eta^2$ in the subnormalization factor and in the precision dependence of the fixed-$\eta$ block encoding. For structured Hamiltonians the same principle gives the corresponding entries in~\Cref{tab:cost-combined-eta}: for factorized or free-fermion one-body structure the relevant state-dependent label count scales as $S_\eta=O(n\eta)$, while locality further reduces the number of coefficient labels that must be prepared.

This point distinguishes our construction from existing fixed-$\eta$ approaches. Standard LCU/QROM input models block encode the full second-quantized Hamiltonian, so even when the simulated state lies in the $\eta$-particle sector, the subnormalization factor is not generally certified to scale with $S_\eta$; in particular, it is not clear from those constructions how to obtain the $O(n\eta)$ scaling for free-fermion or one-body fixed-$\eta$ dynamics, or the $O(n^2\eta^2)$ scaling for a general two-body Hamiltonian, without changing the input model. The fixed-$\eta$ input model of Tong, An et al.~\cite{tong2021fast} gives an $O(\eta)$ normalization for Hamiltonians that are linear combinations of number operators, but it does not provide a direct block encoding for general off-diagonal second-quantized one- and two-body interactions.

Several recent works instead reduce the subnormalization factor in a symmetry-restricted sector by modifying the Hamiltonian before applying an LCU construction~\cite{cortes2024assessing,loaiza2023block,loaiza2023reducing}. For a fixed-particle sector, these methods subtract a symmetry-preserving shift such as $f(\sum_i n_i)$ or, more generally, $f(S)$ with $[f(S),\sum_i n_i]=0$, and then optimize the parameters of $f$ to reduce the resulting coefficient $\ell_1$ norm,
\begin{equation}
    H - f\!\left(\sum_i n_i\right) \quad \text{or} \quad H-f(S).
\end{equation}
Such preprocessing can be very effective for particular instances, but the final subnormalization factor depends on the performance of the classical optimization and does not by itself give a general theoretical guarantee of $S_\eta$ scaling. By contrast, to our knowledge, the present construction is the \textit{first} explicit block encoding that directly targets a fixed-particle sector of a general second-quantized Hamiltonian and achieves optimal $\eta$-particle-sector subnormalization scaling from the circuit construction itself, without Hamiltonian shifts or instance-dependent optimization.

\subsection{Relation to compact and succinct fermion encodings}
\label{sec:compact-encoding-comparison}

Kirby et al.~\cite{kirby2021compactencoding} simulate second-quantized Hamiltonians in a compact encoding whose register stores the ordered list of occupied modes, $\mathcal{O}(\eta\log n)$ system qubits in a fixed-$\eta$ fermionic sector, assuming a constant number of interaction forms with coherently evaluable coefficient functions; Carolan and Schaeffer~\cite{carolan2024succinct} compress further, to within $\mathcal{O}(\eta)$ qubits of the information-theoretic minimum $\lceil\log_2\binom{n}{\eta}\rceil$, with low-depth particle-preserving rotations. Both encodings remain second-quantized. The enumerator oracle in Sec.~V of Ref.~\cite{kirby2021compactencoding} is functionally analogous to the state-dependent label generation by $O_{\rm IDF}$ (which includes $O_{occ}$) together with the validation and fermionic update in $O_C$. Its matrix-element oracle in Sec.~VI performs the value-processing roles of $O_A$ and $O_C$, but returns a binary value rather than encoding a coefficient as an amplitude. The essential difference is where the occupied-mode addresses live: in compact encoding the list is the register, so enumeration over occupied modes is available by construction, whereas here $O_{occ}$ computes an address transiently in ancilla space and uncomputes it, leaving the persistent register in the standard occupation basis.

For the unstructured coefficients of two-body interaction terms, the number of qubits required for the quantum lookup table dominates the overall qubit usage when the SELECT-SWAP optimization is applied for $T$ count. If the objective is to reduce $T$ count, the SELECT--SWAP lookup at its optimal block size (\Cref{eq:lambda-opt-general}) already uses $\Theta(\sqrt{L m_b})=\Theta(n^2\sqrt{\log(n/\epsilon)})$ ancilla qubits for $L=\Theta(n^4)$, whether clean or dirty; this exceeds the $n$-qubit system register itself by a factor of $\tilde{\Theta}(n)$, and with it any qubit saving that compact encoding can offer. This comparison concerns total logical workspace at this operating point, including any borrowed dirty qubits; choosing a smaller $\lambda$ trades a higher $T$ count for fewer workspace qubits. Moreover, the elementary ordered list uses $\Theta(\eta\log n)$ system qubits and yields an asymptotic reduction relative to the occupation string when $\eta\log n=o(n)$; at constant filling, it uses $\Theta(n\log n)$ system qubits, which can be less compact compared to the method proposed in this paper. 

Compact or succinct encodings can be attractive when coefficient preparation is inexpensive and qubit availability is restrictive. For the compact encoding, the condition $\eta=o(n/\log n)$ ensures an asymptotic saving in system qubits, because $\eta\log n=o(n)$. An advantage in total circuit depth over the present construction requires a comparison that includes coefficient preparation and subnormalization. Ref.~\cite{kirby2021compactencoding} characterizes direct versus compact encoding as a space--time tradeoff; a complete comparison of these representations, covering system qubits, clean and dirty ancillas, $T$ count and depth, coefficient access, and subnormalization, is left for future work.


\begin{table*}[htbp!]
\centering

\resizebox{\textwidth}{!}{%
\def\arraystretch{1.75}
\begin{tabular}{c|c|c|c|c}
  Model &  Qubit & Subnormalization factor & T count & T depth  \\
  \hline
  {General} & $\mathcal{O}(n^2\sqrt{\log(\tfrac{n^2 \eta^2}{\epsilon})})$ 
      & $\mathcal{O}(n^2\eta^2)$ 
      & $\mathcal{O}(n^2\sqrt{\log(\tfrac{n^2 \eta^2}{\epsilon})})$ 
      & $\mathcal{O}(n^2\sqrt{\log(\tfrac{n^2 \eta^2}{\epsilon})})$ \\
  \Xhline{1.5pt}
  \hline
   Factorized & $\mathcal{O}(n+n\sqrt{\log(\tfrac{n\eta}{\epsilon})})$
      & $\mathcal{O}(n\eta)$ 
      & $\mathcal{O}(n\log(\eta)+n\sqrt{\log(\tfrac{n\eta}{\epsilon})})$
      & $\mathcal{O}(n\log(\eta)+n\sqrt{\log(\tfrac{n\eta}{\epsilon})})$ \\
  \Xhline{1.5pt}
   TI Factorized & $\mathcal{O}(n+\sqrt{n \log(\tfrac{n\eta}{\epsilon})})$
      & $\mathcal{O}(n\eta)$
      & $\mathcal{O}(n\log(\eta)+\sqrt{n \log(\tfrac{n\eta}{\epsilon})})$ 
      & $\mathcal{O}(n\log(\eta)+\sqrt{n \log(\tfrac{n\eta}{\epsilon})})$ \\
  \Xhline{1.5pt}
 Localized & $\tilde{\mathcal{O}}(n+\eta \sqrt{\log^3(\tfrac{\eta^2}{\epsilon})})$ 
      & $\mathcal{O}(\eta^2 \log^2(\tfrac{\eta^2}{\epsilon}))$ 
      & $\tilde{\mathcal{O}}(n\log(\eta)+\eta \sqrt{\log^3(\tfrac{\eta^2}{\epsilon})})$ 
      & $\tilde{\mathcal{O}}(n\log(\eta)+\eta \sqrt{\log^3(\tfrac{\eta^2}{\epsilon})})$ \\
\end{tabular}
} 
\caption{Complexity of the block encoding circuit for electronic structure Hamiltonians within an $\eta$-particle subspace. Here $n$ is the number of spin orbitals, $\eta$ is the number of particles, and $\epsilon$ is the target precision. 
$\tilde{\mathcal{O}}$ hides polylogarithmic factors in $n$ and $\epsilon$. 
PREP refers to the cost of the \prep oracle, and N/A means not applicable.}
\label{tab:cost-combined-eta}
\end{table*}

 \begin{table*}[htbp!]
    \centering
    \def\arraystretch{1.5}
    \begin{tabular}{c|c|c|c|c|c}
      Space &Reference & Qubit & Subnormalization factor &T count & T depth  \\
      \cline{1-6}
      \multirow{3}{*}{Full} & 2018 Babbush et al. \cite{babbush2018encoding} & $\mathcal{O}(n+\log(\frac{n}{\epsilon}))$ &  $\mathcal{O}(n)$ & $\mathcal{O}(n+\log(\frac{n}{\epsilon}))$ & $\mathcal{O}(n)$  
      \\
      \cline{2-6}
        &2018 Kivlichan et al. \cite{kivlichan2018quantum} & N/A  & N/A & $\mathcal{O}(n^2\log(\frac{1}{\epsilon}))$ & $\mathcal{O}(n)$ 
      \\
      \cline{2-6}
       &\textbf{This paper} & $\mathcal{O} (n+\log(\frac{n}{\epsilon}))$ &  $\mathcal{O}(n)$ & $\mathcal{O}(n+\log(\frac{n}{\epsilon}))$ & $\mathcal{O}(\log(n))$      
      \\
      \Xhline{1.5pt}
      \multirow{3}{*}{$\eta$-particle} &2021 Tong, An et al. \cite{tong2021fast} & $\mathcal{O}(n+\log(\frac{n}{\epsilon}))$ &  $\mathcal{O}(\eta)$ & $\mathcal{O}(n \log(n) \log(\frac{1}{\epsilon}))$ & $\mathcal{O}(n)$ 
           \\
      \cline{2-6}
        &\textbf{This paper} & $\mathcal{O} (n+ \log(\frac{\eta}{\epsilon}))$ &  $\mathcal{O}(\eta)$ & $\mathcal{O}(n\log(\eta)+ \log(\frac{\eta}{\epsilon}))$ & $\mathcal{O}(n\log(\eta))$  
    \end{tabular}
    \caption{Comparison of Input model cost for Hubbard Model. In the table, $n$ is the number of spin orbitals in the system, and $\eta$ is the number of particles for the states we consider in the paper. Each row corresponds to the input model constructed in a paper. Note that we use $N/A$ to indicate that the cost is not applicable for the paper~\cite{kivlichan2018quantum} as the paper focuses on the quantum simulation with trotterization without a clear construction of the input model. }
    \label{tab:cost2}
\end{table*}

\begin{table*}[htbp!]
    \centering
    \def\arraystretch{1.5}
    \begin{tabular}{c|c|c|c}
       Model  & Reference & T count & Method
      \\
      \hline
      \multirow{2}{*}{General} 
      & 2018 Babbush et al. \cite{babbush2018encoding}&  $\mathcal{O}(n^8+n^4\log(\frac{n^4}{\epsilon}))$  &  Qubitization
      \\
      \cline{2-4}
       & \textbf{This paper} & $\mathcal{O}(n^2\eta^2[n\log(\eta)+n^2\sqrt{\log(\frac{n^2\eta^2}{\epsilon})}])$ 
       &  Qubitization
      \\
      \Xhline{1.5pt}
      \multirow{2}{*}{Factorized} & 2018 Kivlichan et al. \cite{kivlichan2018quantum} &  $ n\eta^{1+\frac{1}{p}}  \cdot n^2 \log(\frac{1}{\epsilon}) $  &  Trotterization
      \\
      \cline{2-4}
       & \textbf{This paper} & $\mathcal{O}(n\eta[ n\log(\eta)+n\sqrt{ \log(\frac{n\eta}{\epsilon})}])$
       &  Qubitization
      \\
      \Xhline{1.5pt}
      \multirow{2}{*}{TI Factorized} 
      & 2018 Babbush et al. \cite{babbush2018encoding}&  $\mathcal{O}(n^3+n^2\log(\frac{1}{\epsilon}))$  & Qubitization
      
       \\
\cline{2-4}

       & \textbf{This paper} & $\mathcal{O}(n\eta[ n\log(\eta)+\sqrt{n \log(\frac{n\eta}{\epsilon})}])$ &  Qubitization
      \\
      \Xhline{1.5pt}
      \multirow{2}{*}{Localized} & 2018 Babbush et al. \cite{babbush2017exponentially} & $\mathcal{O}(\frac{\eta^2n^3 \log(1/\epsilon)}{\log\log(1/\epsilon)} )$  &  Truncated Taylor series 
   \\
      \cline{2-4}
       & \textbf{This paper} &

       $\tilde{\mathcal{O}}(\eta^2[n\log(\eta)+n \log(\frac{\eta^2}{\epsilon})]) $
       &  Qubitization
    \end{tabular}
    \caption{Comparison of $T$ gate cost for quantum dynamics simulation for second-quantized electronic-structure Hamiltonians. For the rows labeled \textbf{This paper}, the SELECT--SWAP parameters have been chosen to minimize the leading $T$ gate count. We neglect the simulation time $t$ in the complexity analysis, since all the bounds scale almost linearly with respect to $t$.}
    \label{tab:cost3}
\end{table*}


\subsection{Comparison for Hubbard Models}

\Cref{tab:cost2} compares the input-model costs for the Hubbard model in \Cref{equ:Hhubbard}, both in the full Hilbert space and in the $\eta$-particle sector. Because the model contains only two distinct coefficients, the data-lookup oracle degenerates to a constant-size circuit, and the dominant cost comes from the sparsity oracle. In the full Hilbert space, our construction matches the qubit count, subnormalization factor, and $T$ count of the input model of Babbush et al.~\cite{babbush2018encoding}, while reducing the $T$ depth from $\mathcal{O}(n)$ to $\mathcal{O}(\log(n))$, owing to the logarithmic-depth SWAP-UP networks used in the sparsity oracle. The Trotter-based construction of Kivlichan et al.~\cite{kivlichan2018quantum} is listed for reference, although it does not provide an explicit input model. In the $\eta$-particle sector, the input model of Tong, An et al.~\cite{tong2021fast} attains the same $\mathcal{O}(\eta)$ subnormalization factor for Hamiltonians expressed as linear combinations of number operators, although no explicit circuit construction is given there. Our construction achieves this scaling within a framework that also accommodates off-diagonal one- and two-body interactions, and reduces the $T$ count from $\mathcal{O}(n\log(n)\log(\frac{1}{\epsilon}))$ to $\mathcal{O}(n\log(\eta)+\log(\frac{\eta}{\epsilon}))$, with a comparable qubit count.

\subsection{Comparison for quantum dynamics simulation}

\Cref{tab:cost3} compares the end-to-end $T$-gate complexity for quantum simulation after combining the cost of the block-encoding input model with the corresponding normalization factor. The main message is that the improvements in the input-model construction translate directly into lower simulation-level resource estimates, particularly when the Hamiltonian has additional structure or when the dynamics are restricted to a fixed $\eta$-particle sector. Since qubitization-based simulation has complexity proportional to the product of the subnormalization factor and the cost of implementing the block-encoding oracle, reductions in either quantity can lead to substantial savings in the final $T$ count.

For a general second-quantized electronic-structure Hamiltonian, the previous generic input model of Babbush et al. leads to a scaling of order $\mathcal{O}(n^8+n^4\log(n^4/\epsilon))$. In contrast, the construction in this work gives a cost of
\[
\mathcal{O}\left(n^2\eta^2\left[n\log(\eta)+n^2\sqrt{\log\left(\frac{n^2\eta^2}{\epsilon}\right)}\right]\right).
\] 
This expression makes explicit the dependence on the particle number $\eta$, rather than treating all $n^4$ two-body terms as uniformly active in the full Fock space. When $\eta \ll n$, this particle-number dependence can yield a significant asymptotic improvement. The SELECT--SWAP parameter has already been chosen to minimize the leading $T$-gate count, which gives the square-root lookup contribution shown in \Cref{tab:cost3}.

For the factorized model, the present construction gives the simulation cost
\[
\mathcal{O}\left(n\eta\left[n\log(\eta)+n\sqrt{\log\left(\frac{n\eta}{\epsilon}\right)}\right]\right),
\]
where the lookup size is reduced from the generic two-body value to $\mathcal{O}(n^2)$ because the two-body term is a product of number operators. This provides the ``This paper'' comparison row for the factorized model in \Cref{tab:cost3}.

For translation-invariant factorized Hamiltonians, the table shows a further reduction because the number of distinct coefficients is much smaller than in the fully general case. The cost from Babbush et al.~\cite{babbush2018encoding} scales as $\mathcal{O}(n^3+n^2\log(1/\epsilon))$, while the present construction gives
\[
\mathcal{O}\left(n\eta\left[n\log(\eta)+\sqrt{n\log\left(\frac{n\eta}{\epsilon}\right)}\right]\right).
\]
Here the improvement comes from exploiting both translation-invariant structure and the fixed-particle constraint. The resulting complexity depends on the effective number of particle-index combinations, rather than the full set of orbital-index combinations. This demonstrates that the proposed input model is not merely a generic replacement for coefficient loading, but can also take advantage of algebraic structure in the Hamiltonian.

The localized case shows a similar pattern. Compared with the truncated-Taylor-series configuration-interaction scaling $\mathcal{O}(\eta^2 n^3\log(1/\epsilon)/\log\log(1/\epsilon))$ reported by Babbush et al.~\cite{babbush2017exponentially}, the present qubitization-based construction gives
\[
\tilde{\mathcal{O}}\left(\eta^2\left[n\log(\eta)
+n\log\left(\frac{\eta^2}{\epsilon}\right)\right]\right).
\]
This bound reflects the fact that locality reduces the number of relevant interaction terms, while the $\eta$-particle restriction further limits the number of nonzero transitions that must be coherently prepared. The result is a simulation cost that is governed by the physically active part of the Hamiltonian rather than by the worst-case full-space representation.

Overall, \Cref{tab:cost3} shows that the proposed block-encoding construction leads to simulation-level advantages in three complementary ways: it reduces the effective number of terms through particle-number restriction, exploits structure such as factorization, translation invariance, or locality when available, and uses the optimized SELECT--SWAP tradeoff in the data-lookup oracle. Thus, the improvement is not limited to the implementation of the input oracle itself; it propagates to the total fault-tolerant $T$-gate complexity for Hamiltonian simulation.

\section{Conclusion and future work}
\label{sec:conclusion}
In this paper, we construct block encodings for a general second-quantized Hamiltonian with one-body and two-body interactions. 
Instead of applying multi-controlled gates in the sparsity oracle or amplitude oracle as illustrated in~\cite{babbush2018encoding,liu2025efficient,du2024hamiltonian}, we use the SWAP-UP gate~\cite{wan2021exponentially} in the sparsity oracle and data lookup with direct sampling in the amplitude oracle.
To the best of our knowledge, this is the first time the SWAP-UP gate and data lookup techniques are used in this way.


For a general quantum many-body system with one-body and two-body interactions, we achieve a quadratic improvement in $T$ gate count and improve the constant factor for Clifford gates for the number of interaction terms. 
Furthermore, we develop a state-dependent sparsity oracle and a state-dependent amplitude oracle such that the block encoding has a smaller subnormalization factor for $\eta$-particle block encoding. We further adapt the framework to structured Hamiltonians. In particular, we consider Hamiltonians with translation invariance and nearest-neighbor interactions.

%
When locality or coefficient decay permits an $\epsilon$-accurate truncation, fewer coefficient entries need to be retained in the data-lookup oracle, thereby reducing its implementation cost. Combining the resulting block encodings with qubitization, we derive asymptotic $T$-gate costs for simulating time evolution under time-independent Hamiltonians. For localized-orbital models, \Cref{tab:cost3} compares these costs with those of the configuration-interaction algorithm of Babbush et al.~\cite{babbush2017exponentially}.

Several directions remain open. Our constructions already exploit translation invariance, locality, and the localized-orbital decay model considered above. Extending them to broader decay laws may yield further reductions. Dedicated block encodings of matrix-product operators are already available~\cite{nibbi2024block,dumitrescu2026matrix}. It remains open to determine when a matrix-product operator restricted to a preserved invariant subspace admits efficient state-dependent $O_C$ and $O_A$ oracles in the formalism developed here, and whether matrix-product-operator compression and subspace restriction can jointly reduce the subnormalization factor and circuit cost. A complementary direction is to encode a fixed-$\eta$ basis state using the ordered positions of the $\eta$ nonzero bits in its occupation string. Related compact encodings have been developed in first-quantized and succinct second-quantized settings, including arbitrary-basis first-quantized chemistry constructions~\cite{babbush2017exponentially,carolan2024succinct,georges2025quantum}; adapting our block-encoding and data-lookup oracles to such representations remains open.

The general resource analysis for the direct-sampling construction of the amplitude oracle $O_A$ assumes uniformly bounded coefficients, while the structured cases retain significant entries within the same uniform-diffusion/direct-sampling framework after truncation. Gosset, Kothari, and Wu showed that an arbitrary quantum state can be prepared up to error $\epsilon$ by a Clifford+$T$ circuit with optimal $T$-count when ancillary qubits are allowed~\cite{gosset2024quantum}. Arithmetic-free and function-structured state-preparation methods provide related alternatives~\cite{sanders2019black,mcardle2022quantum,lemieux2024quantum}. It would be interesting to determine whether these state-synthesis methods can be adapted to construct the state-dependent amplitude oracle $O_A$ for broader classes of structured Hamiltonians. Generic controlled state preparation, implementing $\ket{j}\ket{0}\mapsto\ket{j}\ket{\psi_j}$, is known but has exponential circuit size in the worst case~\cite{yuan2023optimal}. The open question here is whether the structured dependence of the transition amplitudes on $\ket{j}$ and the fixed-$\eta$ restriction can be exploited to construct $O_A$ efficiently. Developing a state-dependent analogue of alias sampling for constructing $O_A$ is another open problem.

Recent works have introduced new approaches to block encoding and quantum simulation, including general methods for reducing ancilla overhead~\cite{vasconcelos2025methods}, low-ancilla approximate block encodings derived from Hamiltonian simulation~\cite{zhang2026lowancilla}, structured LCU circuits~\cite{della2025efficient} and block encodings of Pauli sums without ancilla state preparation~\cite{schillo2026tare}, block-encoding-free QSVT~\cite{chakraborty2025qsvtwithoutblockencoding}, and quantum oracle sketching for constructing quantum-access oracles from streaming classical samples~\cite{zhao2026exponential}. It would be interesting to investigate whether indirect diffusion, state-dependent preparation, or other techniques developed in this work can be combined with these approaches to improve quantum simulation within invariant subspaces such as the fixed-$\eta$ sector.

More broadly, the fixed-$\eta$ construction suggests a sector-adapted block-encoding principle: when an operator preserves an invariant subspace and the nonzero transitions within that subspace can be enumerated efficiently, the diffusion, sparsity, and amplitude oracles may be constructed directly for that sector rather than for the full Hilbert space. Potential applications include symmetry sectors of graph Laplacians~\cite{yang2025dictionary}; fixed-boson-number sectors of number-conserving bosonic Hamiltonians, for which signal-processing block encodings provide a complementary starting point~\cite{kane2025block}; fixed-pair-number or seniority sectors of superconducting pairing Hamiltonians~\cite{liu2025efficient}; and fixed-magnetization or total-spin sectors of spin Hamiltonians. Establishing precise conditions under which such restrictions reduce the subnormalization factor and circuit cost---and, when the coefficient data share the symmetry, the data-lookup cost---is an important open problem.

Finally, recent spectral-amplification methods for low-energy problems~\cite{low2025fast,king2025quantum} motivate investigating whether they can be combined with our fixed-$\eta$ block encodings for low-energy simulation or energy estimation.

\begin{acknowledgments}
This work is supported by the U.S. Department of Energy, Office of Science, Accelerated Research in Quantum Computing Centers, Quantum Utility through Advanced Computational Quantum Algorithms, grant No. DE-SC0025572 under Contract No. DE-AC02-05CH11231 (LL, CY, SZ). The authors acknowledge additional support from U. S. Department of Energy, Office of Science, National Quantum Information Science Research Centers, Quantum Systems Accelerator. The authors thank Di Fang, Bert de Jong, Christopher Francis Kane and Yu Tong for their helpful comments and suggestions. Part of the work was performed while DL, CY, SZ, and LL were visiting the Institute for Pure and Applied Mathematics (IPAM), which is supported by the National Science Foundation (Grant No. DMS-1925919). D.L. acknowledges support from the U.S. Department of Energy (DOE) under Contract No. DE-AC02-05CH11231, through the Office of Advanced Scientific Computing Research Accelerated Research for Quantum Computing Program, MACH-Q project. SZ is supported by the National Science Foundation CAREER award (grant CCF-1845125). We acknowledge the use of Fable 5 and ChatGPT 5.5 Pro to check the manuscript for typos and inconsistencies.  
\end{acknowledgments}


\bibliographystyle{apsrev4-2}
\bibliography{bibo} 

\appendix

\section{Algorithmic complexity of block encoding with tunable parameter}
The resource estimates for block encoding second-quantized Hamiltonians are summarized in \Cref{tab:costfull} for the full Hamiltonian and in \Cref{tab:app_cost1} for the $\eta$-particle sector. The parameters $\lambda_i, \tilde{\lambda}_i$ for $i=1,2,3,4$ are tunable parameters in the SELECT-SWAP oracle used in the block encoding.

\begin{table*}[htbp!]
  \centering
  \scriptsize
  \def\arraystretch{1.75}

  \begin{subtable}{1\textwidth}
    \centering
    \begin{tabular}{c|c|c|c|c|c}
      Model & Reference & Qubit  & Subnormalization factor & T count & T depth  \\ 
      \hline 
      \multirow{2}{*}{General} & Babbush et al. (2018)~\cite{babbush2018encoding} & $\mathcal{O}(n+\log(\tfrac{n^4}{\epsilon}))$ & $\mathcal{O}(n^4)$ & $\mathcal{O}(n^4+\log(\tfrac{n^4}{\epsilon}))$ & $\mathcal{O}(n^4)$ \\
      \cline{2-6} 
       & \textbf{This paper} & $\mathcal{O}(n+\lambda_1\log(\tfrac{n^4}{\epsilon}))$ & $\mathcal{O}(n^4)$ & $\mathcal{O}(\tfrac{n^4}{\lambda_1}+\lambda_1\log(\tfrac{n^4}{\epsilon}))$ & $\mathcal{O}(\tfrac{n^4}{\lambda_1}+\log\lambda_1)$ \\
      \Xhline{1.5pt}
      \multirow{2}{*}{Factorized} & Kivlichan et al.(2018)~\cite{kivlichan2018quantum} & N/A & N/A & $\mathcal{O}(n^2\log(\tfrac{1}{\epsilon}))$ & $\mathcal{O}(n)$ \\
      \cline{2-6}
       & \textbf{This paper} & $\mathcal{O}(n+\lambda_2\log(\tfrac{n^2}{\epsilon}))$ & $\mathcal{O}(n^2)$ & $\mathcal{O}(n+\tfrac{n^2}{\lambda_2}+\lambda_2\log(\tfrac{n^2}{\epsilon}))$ & $\mathcal{O}(\tfrac{n^2}{\lambda_2}+\log\lambda_2)$ \\
      \Xhline{1.5pt}
      \multirow{2}{*}{TI Factorized} & Babbush et al.(2018)~\cite{babbush2018encoding} & $\mathcal{O}(n+\log(\tfrac{n^2}{\epsilon}))$ & $\mathcal{O}(n^2)$ & $\mathcal{O}(n+\log(\tfrac{n^2}{\epsilon}))$ & $\mathcal{O}(n)$ \\
      \cline{2-6} 
       & \textbf{This paper} & $\mathcal{O}(n+\lambda_3\log(\tfrac{n^2}{\epsilon}))$ & $\mathcal{O}(n^2)$ & $\mathcal{O}(n+\tfrac{n}{\lambda_3}+\lambda_3\log(\tfrac{n^2}{\epsilon}))$ & $\mathcal{O}(\tfrac{n}{\lambda_3}+\log\lambda_3)$ \\
      \Xhline{1.5pt}
      \multirow{2}{*}{Localized} & Wan (2021)~\cite{wan2021exponentially} & $\mathcal{O}(n)$ & $\mathcal{O}(n^2)$ & $\mathcal{O}(n)+\text{PREP}$ & $\mathcal{O}(n)+\text{PREP}$ \\
      \cline{2-6} 
       & \textbf{This paper} & $\tilde{\mathcal{O}}(n+\lambda_4\log(\tfrac{n^2}{\epsilon}))$ & $\mathcal{O}(n^2\log^2(\tfrac{n^2}{\epsilon}))$ & $\tilde{\mathcal{O}}(\tfrac{n^2}{\lambda_4}\log^2(\tfrac{n^2}{\epsilon})+\lambda_4\log(\tfrac{n^2}{\epsilon}))$ & $\mathcal{O}(n+\tfrac{n^2}{\lambda_4}\log^2(\tfrac{n^2}{\epsilon})+\log\lambda_4)$  
    \end{tabular}
    \caption{Full Hilbert space (no particle-number constraint).}
    \label{tab:costfull}
  \end{subtable}

  \vspace{1em}

  \begin{subtable}{\textwidth}
    \centering
    \resizebox{\textwidth}{!}{%
    \def\arraystretch{1.75}
    \begin{tabular}{c|c|c|c|c|c}
      Model & Reference & Qubit & Subnormalization factor & T count & T depth  \\
      \hline
      \multirow{2}{*}{General} & Babbush et al.(2018)~\cite{babbush2018encoding} & $\mathcal{O}(n+\log(\tfrac{n^4}{\epsilon}))$ & $\mathcal{O}(n^4)$ & $\mathcal{O}(n^4+\log(\tfrac{n^4}{\epsilon}))$ & $\mathcal{O}(n^4)$ \\
      \cline{2-6}
       & \textbf{This paper} & $\mathcal{O}(n+\tilde{\lambda}_1\log(\tfrac{n^2\eta^2}{\epsilon}))$ & $\mathcal{O}(n^2\eta^2)$ & $\mathcal{O}(n\log\eta+\tfrac{n^4}{\tilde{\lambda}_1}+\tilde{\lambda}_1\log(\tfrac{n^2\eta^2}{\epsilon}))$ & $\mathcal{O}(n\log\eta+\tfrac{n^4}{\tilde{\lambda}_1}+\log\tilde{\lambda}_1)$ \\
      \Xhline{1.5pt}
      \multirow{2}{*}{Factorized} & Kivlichan et al.(2018)~\cite{kivlichan2018quantum} & N/A & N/A & $\mathcal{O}(n^2\log(\tfrac{1}{\epsilon}))$ & $\mathcal{O}(n)$ \\
      \cline{2-6}
       & \textbf{This paper} & $\mathcal{O}(n+\tilde{\lambda}_2\log(\tfrac{n\eta}{\epsilon}))$ & $\mathcal{O}(n\eta)$ & $\mathcal{O}(n\log\eta+\tfrac{n^2}{\tilde{\lambda}_2}+\tilde{\lambda}_2\log(\tfrac{n\eta}{\epsilon}))$ & $\mathcal{O}(n\log\eta+\tfrac{n^2}{\tilde{\lambda}_2}+\log\tilde{\lambda}_2)$ \\
      \Xhline{1.5pt}
      \multirow{2}{*}{TI Factorized} & Babbush et al.(2018)~\cite{babbush2018encoding} & $\mathcal{O}(n+\log(\tfrac{n^2}{\epsilon}))$ & $\mathcal{O}(n^2)$ & $\mathcal{O}(n+\log(\tfrac{1}{\epsilon}))$ & $\mathcal{O}(n)$ \\
      \cline{2-6}
        & \textbf{This paper} & $\mathcal{O}(n+\tilde{\lambda}_3\log(\tfrac{n\eta}{\epsilon}))$ & $\mathcal{O}(n\eta)$ & $\mathcal{O}(n\log\eta+\tfrac{n}{\tilde{\lambda}_3}+\tilde{\lambda}_3\log(\tfrac{n\eta}{\epsilon}))$ & $\mathcal{O}(n\log\eta+\tfrac{n}{\tilde{\lambda}_3}+\log\tilde{\lambda}_3)$ \\
      \Xhline{1.5pt}
      \multirow{2}{*}{Localized} & Wan (2021)~\cite{wan2021exponentially} & $\mathcal{O}(n)$ & $\mathcal{O}(n^2)$ & $\mathcal{O}(n)+\text{PREP}$ & $\mathcal{O}(n)+\text{PREP}$ \\
      \cline{2-6}
       & \textbf{This paper} & $\tilde{\mathcal{O}}(n+\tilde{\lambda}_4\log(\tfrac{\eta^2}{\epsilon}))$ & $\mathcal{O}(\eta^2\log^2(\tfrac{\eta^2}{\epsilon}))$ & $\tilde{\mathcal{O}}(n\log\eta+\tfrac{n^2}{\tilde{\lambda}_4}\log(\tfrac{\eta^2}{\epsilon})+\tilde{\lambda}_4\log(\tfrac{\eta^2}{\epsilon}))$ & $\mathcal{O}(n\log\eta+\tfrac{n^2}{\tilde{\lambda}_4}\log(\tfrac{\eta^2}{\epsilon})+\log\tilde{\lambda}_4)$  
    \end{tabular}
    }
    \caption{Fixed $\eta$-particle subspace.}
    \label{tab:app_cost1}
  \end{subtable}

  \caption{Comparison of input-model cost for second-quantized electronic Hamiltonians with precision $\epsilon$. 
  Here $n$ is the number of spin orbitals and $\eta$ is the number of particles in the states considered. 
  The parameters $\lambda_i$ and $\tilde{\lambda}_i$ are free integers used in~\cite{low2018trading}, with ranges 
  $\lambda_1,\tilde{\lambda}_1 \in [1,n^4]$,
  $\lambda_2,\tilde{\lambda}_2 \in [1,n^2]$,
  $\lambda_3,\tilde{\lambda}_3 \in [1,n]$, 
  $\lambda_4 \in [1,n^2\log^2(\tfrac{n^2}{\epsilon})]$, and 
  $\tilde{\lambda}_4 \in [1,n^2\log^2(\tfrac{\eta^2}{\epsilon})]$. 
  Each row corresponds to the input model constructed in a paper. 
  We use $N/A$ to indicate that the cost is not applicable for~\cite{kivlichan2018quantum}, as that work focuses on Trotter-based simulation without an explicit input-model construction. 
  PREP denotes the cost of the \prep oracle, since~\cite{wan2021exponentially} mainly analyzes the \sel oracle.}
  \label{tab:cost-combined}
\end{table*}

\end{document}